\documentclass[aps,pra,twocolumn,showpacs,notitlepage,superscriptaddress,letterpaper, nofootinbib]{revtex4-1}

\usepackage{graphicx, amsmath, amssymb, amsfonts, pifont, bm, cancel, bbold}
\usepackage[usenames]{color}
\usepackage{subfigure}
\usepackage[normalem]{ulem} 
\usepackage{hyperref}
\usepackage{threeparttable}
\usepackage{array}
\usepackage{upgreek}
\usepackage{multirow}
\usepackage{enumerate}
\newcolumntype{P}[1]{>{\centering\arraybackslash}p{#1}}
\newcolumntype{M}[1]{>{\centering\arraybackslash}m{#1}}

\newcommand{\bel}{\begin{align*}}
\newcommand{\tamam}{\end{align*}}

\newcommand{\ket}[1]{|#1\rangle}

\newcommand{\ud}{\mathrm{d}}

\newcommand{\beginsupplement}{
    \setcounter{equation}{0}
    \renewcommand{\theequation}{S\arabic{equation}}
    \setcounter{table}{0}
    \renewcommand{\thetable}{S\arabic{table}}
    \setcounter{figure}{0}
    \renewcommand{\thefigure}{S\arabic{figure}}
    \newcounter{SIfig}
    \renewcommand{\theSIfig}{S\arabic{SIfig}}}

\begin{document}

\title{ 
A scalable superconducting quantum simulator with long-range connectivity based on a photonic-bandgap metamaterial
}
\author{Xueyue~Zhang}
\thanks{These authors contributed equally to this work.}
\author{Eunjong~Kim}
\thanks{These authors contributed equally to this work.}
\affiliation{Thomas J. Watson, Sr., Laboratory of Applied Physics and Kavli Nanoscience Institute, California Institute of Technology, Pasadena, California 91125, USA.}
\affiliation{Institute for Quantum Information and Matter, California Institute of Technology, Pasadena, California 91125, USA.}
\author{Daniel K.~Mark}
\affiliation{Center for Theoretical Physics, Massachusetts Institute of Technology, Cambridge, MA 02139, USA}
\author{Soonwon Choi}
\affiliation{Center for Theoretical Physics, Massachusetts Institute of Technology, Cambridge, MA 02139, USA}
\author{Oskar~Painter}
\email{opainter@caltech.edu}
\homepage{http://copilot.caltech.edu}
\affiliation{Thomas J. Watson, Sr., Laboratory of Applied Physics and Kavli Nanoscience Institute, California Institute of Technology, Pasadena, California 91125, USA.}
\affiliation{Institute for Quantum Information and Matter, California Institute of Technology, Pasadena, California 91125, USA.}
\affiliation{AWS Center for Quantum Computing, Pasadena, California 91125, USA.}

\date{\today}


\begin{abstract} 
Synthesis of many-body quantum systems in the laboratory can help provide further insight into the emergent behavior of quantum materials \cite{georgescu2014quantum, carusotto2020photonic, altman2021quantum}, whose properties may provide improved methods for energy conversion, signal transport, or information processing \cite{cava2021introduction}. While the majority of engineerable many-body systems, or quantum simulators, consist of particles on a lattice with local interactions, quantum systems featuring long-range interactions \cite{bruzewicz2019trapped, browaeys2020many, lahaye2009physics} are particularly challenging to model and interesting to study due to the rapid spatio-temporal growth of quantum entanglement and correlations \cite{jurcevic2014quasiparticle, richerme2014non}. Here, we present a scalable quantum simulator architecture based on a linear array of superconducting qubits locally connected to an extensible photonic-bandgap metamaterial. The metamaterial acts both as a quantum bus mediating qubit-qubit interactions \cite{douglas2015quantum}, and as a readout channel for multiplexed qubit-state measurement. As an initial demonstration, we realize a 10-qubit simulator of the one-dimensional Bose-Hubbard model with in situ tunability of both the hopping range and the on-site interaction. We characterize the Hamiltonian of the system using a measurement-efficient protocol based on quantum many-body chaos \cite{mark2022inprep}. Further, we study the many-body quench dynamics of the system, revealing through global bit-string statistics \cite{boixo2018characterizing} the predicted crossover from integrability to ergodicity as the hopping range increases. The metamaterial quantum bus architecture presented here can be extended to two-dimensional lattice systems \cite{gonzalez2015subwavelength} and used to generate a wide range of qubit interactions \cite{douglas2015quantum, hung2016quantum}, expanding the accessible Hamiltonians for analog quantum simulation and increasing the flexibility in implementing quantum circuits for gate-based computations \cite{bravyi2010tradeoffs, baspin2021quantifying, delfosse2021bounds, tremblay2022constant}.

\end{abstract}

\maketitle

Realizing a scalable architecture for quantum computation and simulation is a central goal in the field of quantum information science, with numerous physical platforms under active investigation \cite{browaeys2020many, schafer2020tools, bruzewicz2019trapped, chatterjee2021semiconductor, kjaergaard2020superconducting, flamini2018photonic}.
While architectures with nearest-neighbor (NN) coupling between quantum particles on a lattice are prevalent, quantum systems with long-range interactions can realize a richer set of computational tasks and physical phenomena \cite{molmer1999multiparticle, korenblit2012quantum, vaidya2018tunable, wright2019benchmarking, manovitz2020quantum, periwal2021programmable, bluvstein2022quantum}.
For instance, in the case of gate-based quantum computation, coupling beyond the nearest-neighbor level enables non-local gate operations between qubits which can reduce the overhead of quantum algorithms and lift the restrictions on code rate and distance of local-interaction-based quantum error-correcting codes \cite{bravyi2010tradeoffs,baspin2021quantifying, delfosse2021bounds, tremblay2022constant}. In the case of analog quantum simulation \cite{altman2021quantum}, the inclusion of long-range interactions can alter the behavior of otherwise integrable many-body systems, resulting in quantum chaotic dynamics, at the root of such topics as quantum thermalization \cite{nandkishore2015thermquantstat} and quantum information scrambling \cite{joshi2020quantum}. Furthermore, control over the range of lattice connectivity grants access to different physical regimes and the crossover between them, such as in many-body quantum phase transitions \cite{islam2013emergence, landig2016quantum, ebadi2021quantum} and the hydrodynamics of non-equilibrium quantum states \cite{joshi2022observing}. 

For engineered quantum systems consisting of interacting quantum particles on a lattice, it is often challenging to scale to larger lattice sizes while maintaining a high degree of lattice connectivity and single-site control. One common approach, developed for trapped-ion and neutral-atom systems, is to use resonant modes of either vibrational \cite{molmer1999multiparticle} or optical \cite{Zheng:2000} cavities as a quantum bus for mediating interactions between the internal states of atoms across the lattice. The cavity bus can realize all-to-all coupling via a single mode of the cavity, tailored long-range interactions using multiple cavity modes \cite{korenblit2012quantum, islam2013emergence, vaidya2018tunable}, or even programmable complex coupling graphs which are distinct from the underlying physical lattice geometry \cite{manovitz2020quantum, periwal2021programmable}.
Similar schemes have been employed in superconducting quantum circuits \cite{Majer:2007,zhong2019violating}, realising systems as large as 20 qubits with all-to-all coupling via a common microwave cavity \cite{guo2021observation}.
The cavity-bus approach, however, eventually faces scaling issues as well. For a fixed cavity volume, as the number of lattice sites is increased and the physical spacing between sites is reduced, parasitic near-field coupling between neighboring lattice sites becomes a challenge, as does single-site control of quantum particles due to cross-talk. On the other hand, increasing the cavity size to accommodate more lattice sites results in frequency-crowding effects stemming from the increase in the spectral density of cavity modes \cite{zhong2019violating}.

An alternative approach for connecting quantum particles on a lattice is to construct a quantum bus from an intrinsically extensible structure, such as a waveguide \cite{roy2017colloquium, sheremet2021waveguide}. Along this direction, engineered photonic-bandgap waveguides have been proposed as a quantum bus that simultaneously protects quantum particles from radiative damping through the waveguide while allowing for long-range lattice connectivity \cite{douglas2015quantum}. 
The waveguide-bus concept has been investigated in the context of many-body simulation with cold atoms coupled to engineered nanophotonic waveguides \cite{douglas2015quantum, chang2018colloquium, bello2022spin}, and recent experiments have explored qubit-photon bound states in superconducting quantum circuits with microwave photonic-bandgap waveguides \cite{liu2017quantum, mirhosseini2018superconducting, sundaresan2019interacting, scigliuzzo2021extensible, kim2021quantum}. 
However, the realization of a scalable many-body quantum simulator, with single-site quantum-particle control and a high level of lattice connectivity, has remained an open challenge.

In this work, we demonstrate a scalable many-body quantum simulator consisting of a one-dimensional (1D) lattice of superconducting transmon qubits coupled to a common metamaterial waveguide. This system provides both tunable long-range connectivity between qubits and full single-site control and state measurement of individual qubits. The metamaterial waveguide acts both as a bus for mediating long-range interactions between qubits, and as a hardware-efficient Purcell filter enabling multiplexed, rapid readout of the qubit states with high fidelity. When qubits are tuned inside a bandgap of the metamaterial waveguide, this system realizes an extended version of the Bose-Hubbard model with tunable hopping range and on-site interaction. We characterize the system properties using conventional single- and two-qubit measurements, in combination with a sample-efficient Hamiltonian learning protocol based on the emergent statistics under many-body dynamics of qubit-state measurement outcomes across the lattice (global bit-strings) \cite{mark2022inprep}. The many-body characterization reveals not only the long-range nature of the hopping, but also information on the hopping phases of the extended Bose-Hubbard Hamiltonian. Finally, we investigate a crossover from integrable to ergodic many-body dynamics versus the hopping range, and characterize the crossover using two-point correlators and the fluctuations in the probabilities of global bit-string measurement outcomes. 

\begin{figure*}[t!]
\begin{center}
\includegraphics[width=1\textwidth]{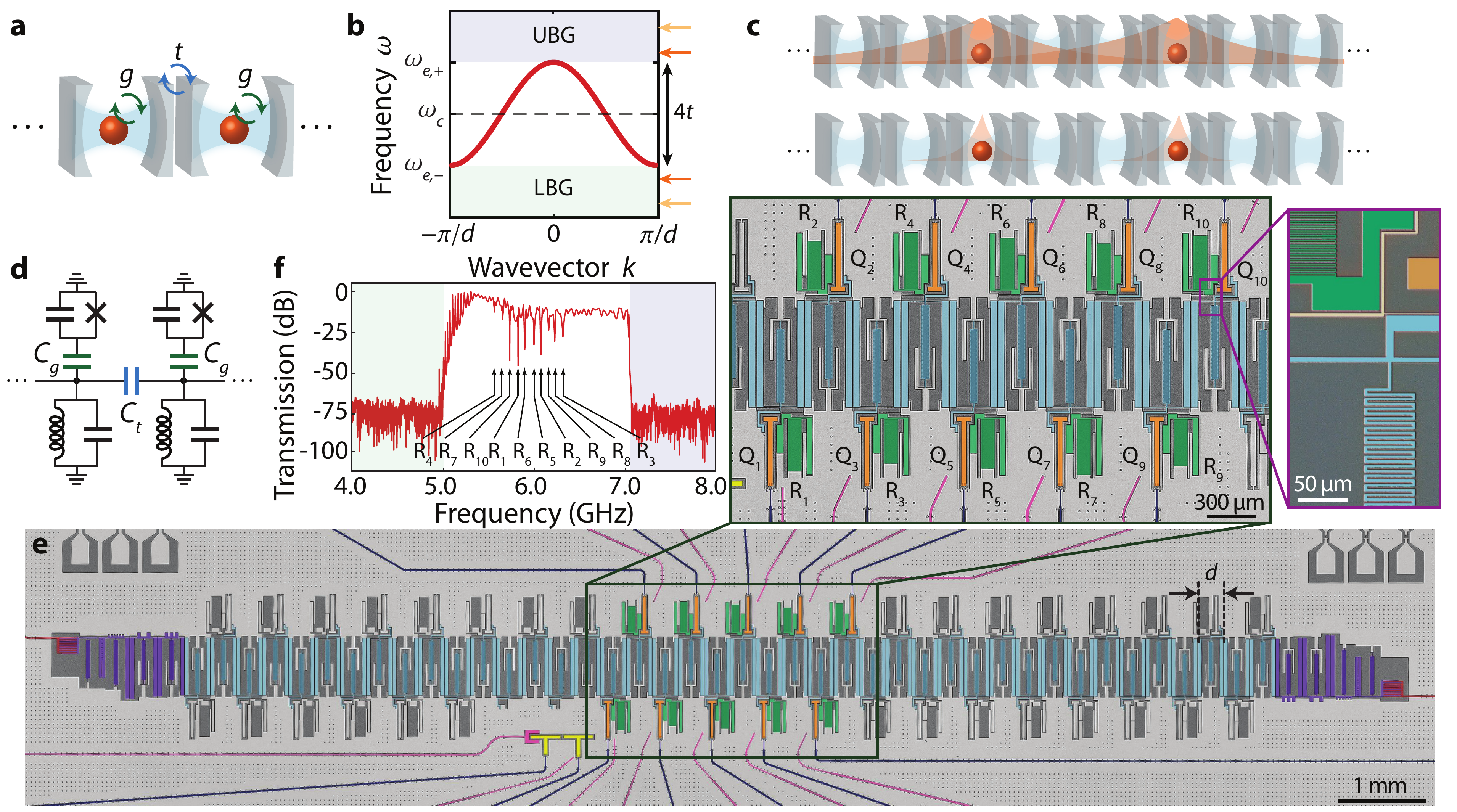}
\caption{\textbf{Metamaterial-based quantum simulator.} 
\textbf{a}, Schematic showing a 1D array of coupled cavities with nearest-neighbor coupling $t$. Each cavity is occupied by a quantum emitter (orange ball) with coupling $g$ to the cavity. 
\textbf{b}, Dispersion relation of the coupled cavity array in panel \textbf{a} with a passband between $\omega_{e, \pm}$ centered at $\omega_c$ (bandwidth of $4t$). The LBG (UBG) below (above) the passband is shaded in green (purple). 
\textbf{c}, Top (Bottom): Cartoon of two emitter-photon bound states at small (large) detuning $|\Delta|$, indicated by dark (light) orange arrows in panel \textbf{b}, exhibiting an extended (restricted) spatial range and large (small) photonic component in the bound states. 
\textbf{d}, Electrical circuit realization of panel \textbf{a} with capacitively coupled LC resonators and transmon qubits corresponding to the cavity array and the quantum emitters, respectively. The coupling capacitors are color coded in accordance with panel \textbf{a}.
\textbf{e}, Optical micrograph (false colored) of the fabricated quantum simulator with 42 metamaterial resonators (lattice constant $d=292\,\upmu$m) colored blue connected to input-output ports (red) via tapering sections (purple). Ten qubits (Q$_i$, colored orange), controlled by individual charge drive lines (pink) and flux bias lines (dark blue), and their readout resonators (R$_i$, colored green) couple to the ten inner unit cells of the metamaterial waveguide with a zoomed-in view in the left inset. Detailed view of the coupling region is shown in the right inset. Two auxiliary qubits (yellow) are not used in this experiment.
\textbf{f}, Transmission spectrum through the metamaterial waveguide (red curve) with black arrows indicating the ten resonances of the readout resonators R$_i$. }
\label{fig:FigureN1} 
\end{center}
\end{figure*}

\hfill \break 
\noindent
\textbf{Metamaterial-based quantum simulator}

\noindent
The backbone of the many-body quantum simulator in this work is a metamaterial waveguide formed from a chain of lumped-element LC microwave resonators. 
This metamaterial waveguide can be described by a generic model, illustrated in Fig.~\ref{fig:FigureN1}a, of a 1D cavity array with NN coupling $t$ \cite{hartmann2006strongly, angelakis2007photon, calajo2016atom}. 
The corresponding dispersion relation (see Fig.~\ref{fig:FigureN1}b) is given by $\omega_k = \omega_c + 2t \cos{(kd)}$, exhibiting a passband centered around the cavity frequency $\omega_c$ with a bandwidth of $4t$, where $k$ is the wavevector and $d$ is the lattice constant of the array. 
The bandgap at frequencies below $\omega_{e,-}=\omega_c - 2t$ (above $\omega_{e,+}=\omega_c + 2t$) is denoted as the lower (upper) bandgap, abbreviated as LBG (UBG).
Inside the bandgaps, the off-resonant coupling between a bare quantum emitter and the waveguide modes gives rise to an emitter-photon bound state \cite{john1990quantum} whose photonic tail is localized around the emitter. 
The localization follows a spatial profile $(\mp 1)^{\Delta x}e^{-|\Delta x|/\xi}$ in the LBG/UBG \cite{calajo2016atom}, where $\Delta x$ is the displacement in the number of unit cells from the emitter and $\xi$ is the localization length controlled by the detuning $\Delta$ between the band-edge frequency and the transition frequency of the bound state. 
The overlap of two bound states results in photon-mediated coupling with a range covering multiple unit cells, i.e., long-range coupling, exhibiting a greater strength and a more extended range $\xi$ at a smaller detuning $|\Delta|$, as displayed in Fig.~\ref{fig:FigureN1}c.

The implementation of the metamaterial waveguide is shown in Figs.~\ref{fig:FigureN1}d and e, consisting of a 42-unit-cell array of  capacitively coupled lumped-element microwave resonators. 
The metamaterial waveguide is equipped at both ends with engineered tapering sections, designed to reduce the impedance mismatch to external 50-$\Omega$ input-output ports at frequencies lying within the passband of the metamaterial waveguide \cite{ferreira2021collapse, kim2021quantum}. 
Each of the middle ten metamaterial resonators (unit cells labeled by $i=1$--$10$) couples to a transmon qubit \cite{koch2007charge} serving as the quantum emitter. 
Individual addressing of each qubit is achieved by excitation (XY control) from a charge drive line and frequency tuning (Z control) from a flux bias line. 
Dispersive qubit readout \cite{Schuster:2005} is enabled by capacitively coupling each qubit Q$_i$ to a compact readout resonator R$_i$, which itself is then coupled to the metamaterial resonator of the same unit cell. 
The entire metamaterial and transmon qubit system (the device) is fabricated using evaporated thin-film aluminum on a high-resistivity silicon substrate, with fabrication procedures detailed in Refs.~\cite{mirhosseini2018superconducting, ferreira2021collapse}. Further details of the device modelling and the experimental setup used to measure and test the device, are discussed in Secs.~\ref{sec:SI} and \ref{sec:SII} of the Supplementary Information, respectively.



This realization of the device enables qubit readout utilizing the passband of the metamaterial waveguide with built-in protection against Purcell decay channels \cite{houck2008controlling} (see Sec.~\ref{sec:SIII} of the Supplementary Information). 
The transmission spectrum through the metamaterial waveguide, displayed in Fig.~\ref{fig:FigureN1}f, shows a passband ranging from $\omega_{e,-}/2\pi\approx 5.01\,$GHz to $\omega_{e,+}/2\pi \approx 7.08\,$GHz with ripples smaller than 8$\,$dB near the center.
The extinction ratio of the transmission between that measured in the passband and that measured in the bandgaps is greater than 65$\,$dB with a sharp transition in the transmission occurring within 100$\,$MHz of the band-edges. 
In the middle of the passband, resonances associated with the readout resonators are observed between 5.574$\,$GHz and 6.328$\,$GHz. 
The average decay rate of ten readout resonators is $\overline{\kappa_{\mathrm{R}_i}}/2\pi=11.8\,$MHz, enabling fast, high-fidelity multiplexed readout while maintaining a low level of readout crosstalk. For details of readout methods and characterization, refer to Sec.~\ref{sec:RO} of the Supplementary Information.

\hfill \break 
\noindent
\textbf{Bose-Hubbard model with long-range hopping}

\begin{figure*}[t!]
\begin{center}
\includegraphics[width=1\textwidth]{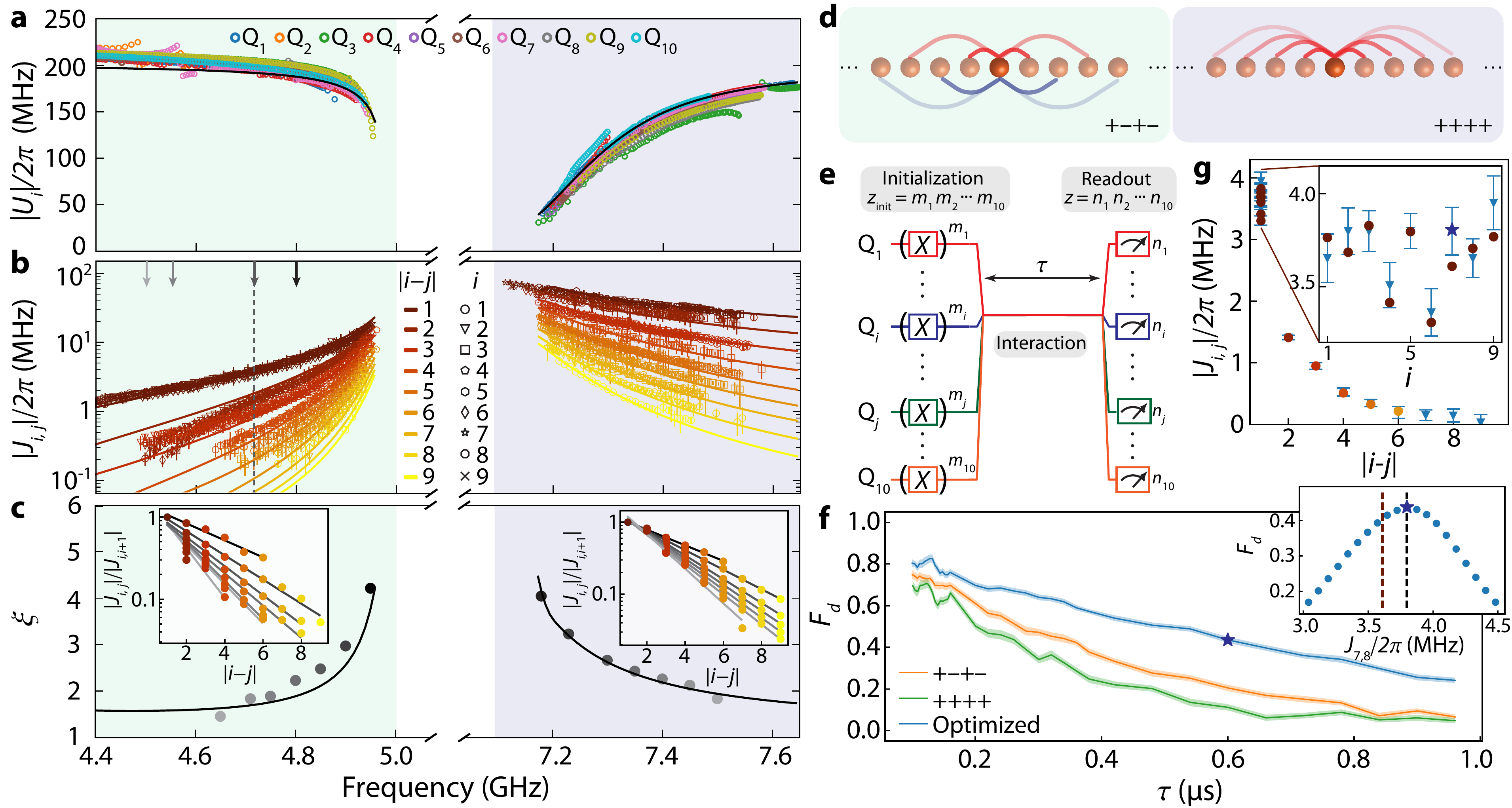}
\caption{\textbf{Hamiltonian learning.} 
\textbf{a}, Magnitude of on-site interaction $U_i$ plotted as a function of frequency $\omega_{01}$ with measured values indicated with colored circles. 
\textbf{b}, Hopping amplitude $|J_{i,j}|$ between sites Q$_i$ and Q$_j$ versus frequency. Experimental data are shown as markers with errorbars indicating a standard deviation. Colors represent the separation $|i-j|$ between sites. Four gray-scale arrows specify frequencies in Fig.~\ref{fig:FigureN3}.
\textbf{c}, Localization length $\xi$ extracted by fitting the exponential decay of measured hopping rate (results of polynomial fitting of datapoints in panel \textbf{b}) as a function of distance $|i-j|$ (insets) at a few different frequencies. The darkness of a marker matches a fitting curve in the inset at the same frequency. In panels \textbf{a}-\textbf{c}, theory curves (solid) are obtained from numerical calculations using an identical circuit model and LBG/UBG is shaded in green/purple.
\textbf{d}, Left (Right): cartoon illustrating hopping from a site inside the LBG (UBG) with positive/negative sign represented in red/blue and the amplitude represented by opacity. The signs of hopping are alternating (all positive) in the LBG (UBG), denoted as ${+}{-}{+}{-}$ (${+}{+}{+}{+}$). 
\textbf{e}, Pulse sequence for many-body evolution where ten sites Q$_1$--Q$_{10}$ are initialized at their idle frequencies by using $\pi$-pulses on the local drive lines ($X$ gate) to prepare an initial bit-string $z_\textrm{init}=m_1\,m_2\, \cdots \, m_{10}$, where $m_i \in \{0,1\}$. Then the ten sites are tuned to resonance for evolution time $\tau$ during the interaction stage, followed by a site-resolved single-shot readout at their idle frequencies to obtain a final bit-string $z=n_1\,n_2\, \cdots \, n_{10}$ where $n_i \in \{0,1\}$. 
\textbf{f}, Many-body fidelity estimator $F_d$ at $\omega_{01}/2\pi=4.72\,$GHz versus evolution time $\tau$ assuming $J_{i,j}$'s from two-qubit measurement indicated by the dashed line in panel \textbf{b} with alternating signs ${+}{-}{+}{-}$ (orange) and all positive signs ${+}{+}{+}{+}$ (green), and from the numerical optimization discussed in the main text (blue). The shading on each curve corresponds to the standard deviation of the mean $F_d$ for 40 randomly chosen $z_\textrm{init}$'s in the five-excitation sector. Inset: $F_d$ at $\tau=0.6\,\upmu$s as a function of $J_{7,8}$ with optimized parameters on the remaining $J_{i,j}$'s, where the brown (black) dashed line indicates $J_{7,8}$ extracted from panel \textbf{b} (from the numerical optimization). 
\textbf{g}, Comparison of $|J_{i,j}|$ from two-qubit measurements (colored circles) and from numerical optimization (blue triangles with errorbars for 68\% confidence interval), corresponding to the orange curve and the blue curve in panel \textbf{f}, respectively. The measured hopping amplitudes are shown for all possible qubit pairs at distance $|i-j|=1$ (inset shows measured nearest-neighbor hopping amplitudes $|J_{i,i+1}|$ and their optimized values for $1\le i \le 9$), while for larger distances $|i-j|>1$ only the average values are indicated. The navy stars represent the maximum $F_d$ at $\tau=0.6\,\upmu$s in panel \textbf{f} and the optimized $J_{7,8}$ in the insets of panels \textbf{f} and \textbf{g}.}
\label{fig:FigureN2} 
\end{center}
\end{figure*}

\noindent
The spatially extended bound-state excitations, formed between transmon-qubit excitations and waveguide photons of the metamaterial-waveguide bus, creates a lattice of interacting microwave photons \cite{houck2012chip, carusotto2013quantum, carusotto2020photonic}.
This quantum system is described by an extended version of the 1D Bose-Hubbard model \cite{cazalilla2011one} with tunable long-range hopping and on-site interaction. 
Specifically, each bound state formed from qubit Q$_i$, inheriting the level structure of an anharmonic oscillator from a transmon qubit \cite{koch2007charge}, serves as a bosonic site with local site energy $\epsilon_i = \omega_{01,i}$ and the on-site interaction $U_i=\omega_{12,i}-\omega_{01,i}$. Here, $\omega_{01,i}$ and $\omega_{12,i}$ are the transition frequencies of the bound state on site Q$_i$ from its ground state $\ket{0}$ to the first excited state $\ket{1}$ and that from the first to the second excited state $\ket{2}$, respectively. In addition, the long-range hopping $J_{i,j}$ is enabled by the overlap between a pair of qubit-photon bound states on sites Q$_i$ and Q$_j$. The Hamiltonian of this model that captures the basic processes mentioned above can be written as
\begin{equation}
\label{eq:Ham}
    \hat{H}/\hbar = \sum_{i,j} J_{i,j}\hat{b}_i^{\dagger}\hat{b}_j + \sum_i\frac{U_i}{2}\hat{n}_i(\hat{n}_i - 1) + \sum_i \epsilon_i\hat{n}_i,
\end{equation}
where $\hat{b}_i^{\dagger}$ ($\hat{b}_i$) is the creation (annihilation) operator and $\hat{n}_i \equiv \hat{b}_i^{\dagger}\hat{b}_i$ is the number operator on site Q$_i$. The parameters of the Hamiltonian realized in this simulator can be learned through experiments enabled by the precise, single-site-level control over qubits.

We measure the on-site interaction $U_i$ (Fig.~\ref{fig:FigureN2}a) by performing spectroscopy of $\omega_{12}$ after initializing Q$_i$ in its first excited state $\ket{1}$.
From within either bandgap, $|U_i|$ decreases as $\omega_{01}$ approaches the closest band-edge due to dressing from the passband modes of the metamaterial waveguide \cite{sundaresan2019interacting}, i.e., the Lamb shift.
In the UBG, a wide tuning range of $|U_i|$ is achievable from the strong hybridization between the $\ket{1}$-$\ket{2}$ transition and the band-edge modes at $(\omega_{01} - \omega_{e, +})/2\pi < 300\,$MHz.
The magnitude of hopping $|J_{i,j}|$ is measured from vacuum Rabi oscillations between sites Q$_i$ and Q$_j$ by initializing one site with a $\pi$-pulse and tuning $\omega_{01}$ of both sites on resonance for a duration $\tau$ with fast flux pulses. 
As shown in Fig.~\ref{fig:FigureN2}b, for a fixed distance $|i-j|$, $|J_{i,j}|$ increases with a decreasing $|\Delta|$, resulting from larger photonic components of the bound states.
Compared to the LBG, the UBG exhibits larger $|J_{i,j}|$ at the same $|\Delta|$, owing to a stronger coupling $g$ of the bare qubits to the metamaterial at higher frequencies and the breakdown of the tight-binding cavity array model in the circuit realization of the metamaterial in Fig.~\ref{fig:FigureN1}d (see Sec.~\ref{sec:SI} of the Supplementary Information). 
At a specific $\omega_{01}$, $|J_{i,j}|$ decreases exponentially as a function of distance $|i-j|$, as shown in Fig.~\ref{fig:FigureN2}c, resembling the profile of the photonic tail in a qubit-photon bound state. From fitting the exponential decay curve we extract the localization length $\xi$, which ranges from $\xi = 1.4$ to $4.2$, with the largest localization length occurring at the smallest achievable band-edge detunings. For even smaller detunings, the eigenstate of the two interacting bound states merges into the passband and becomes radiative to the waveguide.

\hfill\break
\noindent
\textbf{Many-body Hamiltonian learning}

\noindent
Beyond the above single- and two-qubit measurements, we perform in situ many-body characterization of Hamiltonian parameters \cite{choi2021emergent, cotler2021emergent, mark2022inprep} which are otherwise hard to access. 
For example, the sign of the hopping term $J_{i,j}$ inherits the spatial profile  of the photonic component of the bound states. In the case of the bound states in the UBG, the sign of the hopping terms are all uniform (positive), whereas for bound states in the LBG the hopping terms alternate sign as the distance between lattice sites increases by one (see Fig.~\ref{fig:FigureN2}d). 
This is because the photonic component of the bound state behaves as a defect mode inside the bandgap, exhibiting a spatial profile resembling the wavevector at the nearest band-edge, $k=0$ at the upper band-edge and $k=\pi/d$ at the lower bend-edge.
Although insignificant in measurements involving only two lattice sites, the sign of the hopping terms does alter the many-body dynamics of the system. 
Here, we utilize a many-body fidelity estimator $F_d$ proposed in Ref.~\cite{mark2022inprep} to reveal this information. This fidelity estimator, which closely tracks the true many-body fidelity, is obtained for ergodic quench evolution of simple initial states (see Sec.~\ref{sec:SVI} of the Supplementary Information).

We follow the sequence described in Fig.~\ref{fig:FigureN2}e to perform the many-body quench evolution. The sequence consists of preparing a set of five randomly chosen sites in their first excited state, followed by using flux pulses to align $\omega_{01}$ of all ten sites for time $\tau$, and then finally performing site-resolved single-shot measurement on all lattice sites to obtain a ten-bit string $z=n_1\,n_2\,\cdots \,n_{10}$.
The many-body fidelity estimator $F_d$ is calculated by comparing bit-string statistics of repeated measurements with numerical simulation of the evolution assuming a set of Hamiltonian parameters in Eq.~\eqref{eq:Ham}. The maximum $F_d$ is achieved at the parameter values closest to the Hamiltonian realized in the experiment.
The fast repetition rate of this experiment enables us to perform a large number of measurements ($1.6\times10^{5}$ in total), reducing statistical error and increasing sensitivity to small Hamiltonian parameter variations (see Sec.~\ref{sec:SV} of the Supplementary Information for details of qubit control, pulse sequence, and repetition rate).

Figure~\ref{fig:FigureN2}f compares $F_d$ at $\omega_{01}/2\pi=4.72\,$GHz using three different parameter sets for $J_{i,j}$: a first set with amplitudes derived from the two-qubit experiments in Fig.~\ref{fig:FigureN2}b assuming alternating signs (Fig.~\ref{fig:FigureN2}d, left), a second set with the same amplitudes as the first but all positive signs (Fig.~\ref{fig:FigureN2}d, right), and a third set of optimized parameter values that maximize $F_d$. The optimized hopping terms are restricted to be real-valued, with independent $J_{i,i+1}$ for each $i=1-9$ qubits and $J_{i,j}$ for each distance $|i-j| > 1$ (all qubit pairs of the same distance having the same $J_{i,j}$). 
An alternating sign of $J_{i,j}$ with distance is favored, yielding a higher many-body fidelity compared to hopping terms with all positive signs. This is further evidenced by the alternating signs of the resulting optimized parameter set. Although we find small differences between the set of optimized hopping amplitudes and those from the two-qubit experiments with alternating signs (see Fig.~\ref{fig:FigureN2}g), $F_d$ of the optimized parameter set is markedly better. The sensitivity of $F_d$ to the hopping terms is highlighted in the inset of Fig.~\ref{fig:FigureN2}f, where the variation of the fidelity versus $J_{7,8}$ is shown. For details of the $F_d$ calculation and parameter optimization, refer to Sec.~\ref{sec:SVI} of the Supplementary Information.

\hfill \break 
\noindent
\textbf{Ergodic many-body dynamics with long-range hopping}

\begin{figure}[t!]
\begin{center}
\includegraphics[width=0.48\textwidth]{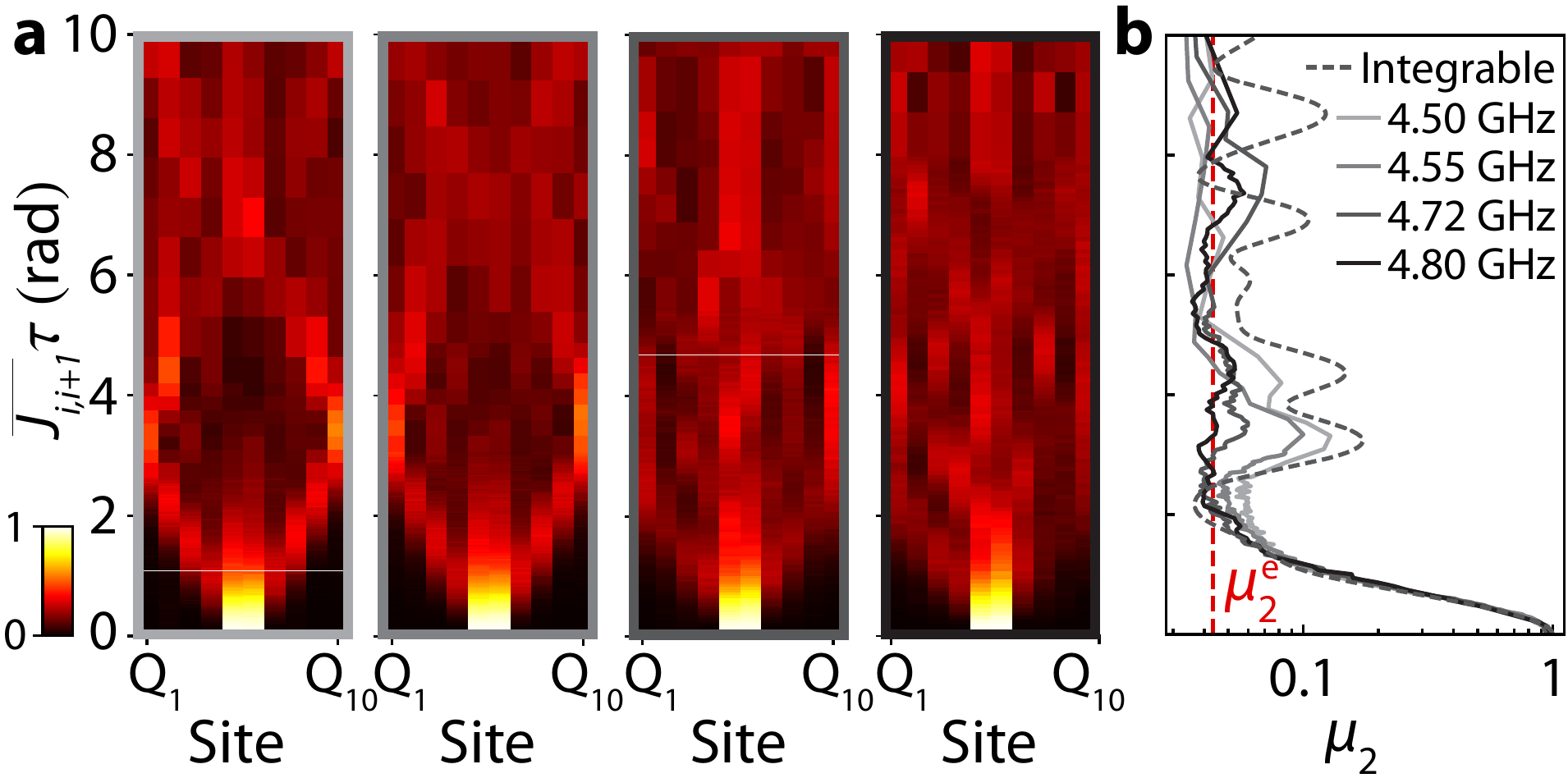}
\caption{\textbf{Two-particle quantum walk with increasing hopping range.} 
\textbf{a}, Evolution of the population $\langle \hat{n}_i\rangle$ on sites Q$_1$--Q$_{10}$ as a function of normalized evolution time $\overline{J_{i,i+1}}\tau$. The system is initialized in $z_\textrm{init}=0000110000$ and the evolution occur at $\omega_{01}/2\pi=4.50$\,GHz, 4.55\,GHz, 4.72\,GHz, and 4.80\,GHz from left to right. 
\textbf{b}, The second moment $\mu_2$ as a function of normalized evolution time $\overline{J_{i,i+1}}\tau$. Results calculated from the data in panel \textbf{a} are shown in solid curves with gray scales corresponding to frames in panel \textbf{a} and arrows in Fig.~\ref{fig:FigureN2}b. Result from numerical simulation of the integrable Hamiltonian is shown as the dotted curve and $\mu_2^\mathrm{e}$ for a generic ergodic system is indicated by the red dashed line.}
\label{fig:FigureN3}
\end{center}
\end{figure}

\begin{figure*}[t!]
\begin{center}
\includegraphics[width=1\textwidth]{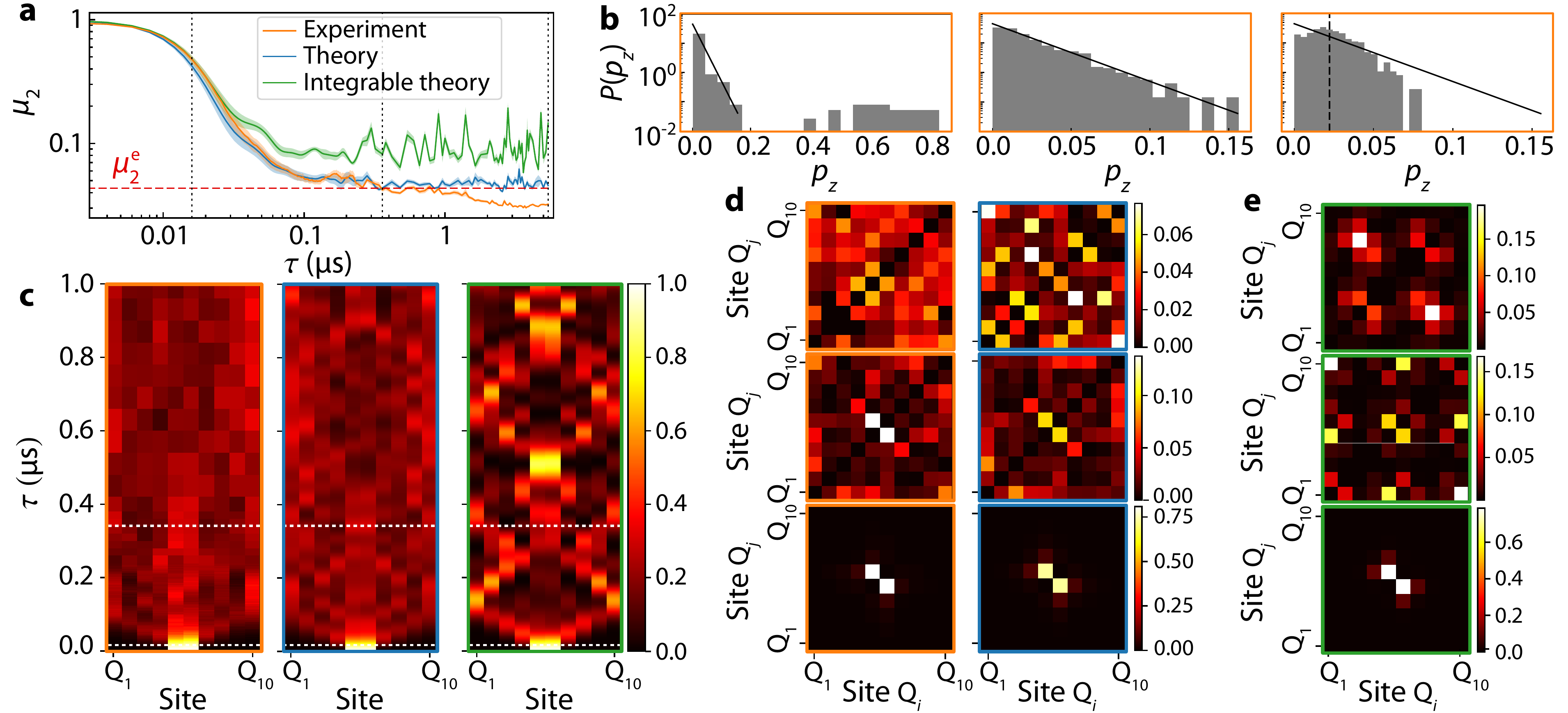}
\caption{\textbf{Ergodic many-body dynamics with long-range hopping at 4.72$\,$GHz.} 
\textbf{a}, Second moment $\mu_2$ as a function of evolution time $\tau$ in our system from the experiment (orange) and the theory with the optimized parameter set in Fig.~\ref{fig:FigureN2}g (blue), compared to theoretical predictions of the integrable model (green). The shading on each curve corresponds to a standard deviation of the mean second moment for 20 randomly chosen initial bit-strings $z_\textrm{init}$ in the two-particle sector, and the red dashed line represents the ergodic value $\mu_2^\mathrm{e}$.
\textbf{b}, Density histogram $P(p_z)$ of the distribution of experimental bit-string probabilities $\{p_z\}$ with the 20 initializations $z_\textrm{init}$'s at evolution times $\tau=16$\,ns, 360\,ns, and 5.4\,$\upmu$s from left to right (indicated by the dotted lines in panel \textbf{a}). The solid lines show the PT distribution and the dashed line in the right plot shows the value $p_z=1/D$ of a classical uniform distribution.
\textbf{c}, Evolution of the population $\langle \hat{n}_i\rangle$ on sites Q$_1$--Q$_{10}$ as a function of time $\tau$ with $z_\textrm{init}=0000110000$ in the cases of experiment, theory, and integrable theory from left to right. The white dashed lines at the bottom (in the middle) indicates $\tau=16\,$ns (360\,ns).
\textbf{d}--\textbf{e}, Two-site correlator $\langle \hat{n}_i \hat{n}_j\rangle$ with $z_\textrm{init}=0000110000$ at evolution times $\tau=16\,$ns, 360\,ns, and 5.4\,$\upmu$s from bottom to top in the cases of experiment (left column of panel \textbf{d}), theory (right column of panel \textbf{d}), and integrable theory (panel \textbf{e}).}
\label{fig:FigureN4} 
\end{center}
\end{figure*}

\noindent
Having demonstrated the tunability of Hamiltonian parameters, we now utilize the platform to study the effect of long-range hopping on the many-body dynamics.
Specifically, the ergodicity of the 1D Bose-Hubbard model in the hardcore limit ($|U/J| \gg 1$) depends on the range of hopping, exhibiting integrable behavior with NN hopping, and chaotic behavior with long-range hopping. 
We study this crossover with various hopping ranges and investigate the resulting dynamics using both conventional one- and two-site correlators, and the statistics of the global bit-strings resulting from qubit-state measurement outcomes across the lattice. This latter techniques is particularly useful in identifying the universal signatures of ergodicity, and subsequent deviations at long evolution times due to decoherence.

The crossover between integrable and ergodic dynamics can be qualitatively visualized by a two-particle quantum walk \cite{peruzzo2010quantum, yan2019strongly, gong2021quantum} with initial excitations on sites Q$_5$ and Q$_6$ using the sequence shown in Fig.~\ref{fig:FigureN2}e.
We show in Fig.~\ref{fig:FigureN3}a the measured quantum walk as a function of normalized evolution time $\overline{J_{i,i+1}}\tau$ at a few different $\omega_{01}$'s indicated by arrows in Fig.~\ref{fig:FigureN2}b (corresponding numerical simulation is provided in Sec.~\ref{sec:SVII} of the Supplementary Information), where $\overline{J_{i,i+1}}$ is the average NN hopping rate.
The excitation wave packets smear over the system when $\omega_{01}$ is close to the band-edge frequency.
More quantitatively, this trend can be probed by computing the probability $p_z$ of measuring a certain bit-string $z$ in the two-excitation sector at evolution time $\tau$. 
For a generic ergodic Hamiltonian, the second moment $\mu_2 \equiv \sum_z p_z^2$ \cite{boixo2018characterizing}, which reflects the probability fluctuations, converges to $\mu_2^\text{e}=2/(D+1)$ after initial evolution \cite{mark2022inprep} due to the chaotic nature of its quantum dynamics ($D=45$ is the dimension of the two-excitation Hilbert space).
No such convergence is expected in an integrable Hamiltonian due to revivals associated with ballistic propagation of wave packets  \cite{aharonov1993quantum}. 
As an example, we show in Fig.~\ref{fig:FigureN3}b the results from the spin-$1/2$ XY model \cite{Kardar:2007} obtained from modifying the Hamiltonian in Eq.~\eqref{eq:Ham} by keeping only NN hopping terms in the hardcore limit.
When $\omega_{01}$ is closer to the band-edge, the measured second moment deviates from the simulated integrable result and converges to $\mu_2^\mathrm{e}$ at an earlier normalized evolution time $\overline{J_{i,i+1}}\tau$ consistent with the breaking of integrability due to the extended hopping range. 



To further highlight the effect of long-range hopping, we compare the many-body dynamics of the experimentally measured system (Experiment) at $\omega_{01}/2\pi = 4.72\,\mathrm{GHz}$, to both a theoretical calculation using the optimal learned Hamiltonian parameters with long-range hopping (Theory) and a theoretical calculation using a model based on the integrable Hamiltonian mentioned above (Integrable theory). Decoherence is not included in either of the theory calculations. 
We show in Fig.~\ref{fig:FigureN4}a the time evolution of the second moment $\mu_2$ averaged over 20 randomly chosen two-excitation initial states for all three cases. 
At a short time ($\tau=16\,\mathrm{ns}$), the excitations remain in their initial sites for each of the cases. This is visualized for a quantum walk with initial excitations on sites Q$_5$ and Q$_6$ in Fig.~\ref{fig:FigureN4}c (evolution of population $\langle \hat{n}_i \rangle$) and in the bottom panels of Fig.~\ref{fig:FigureN4}d and e (two-site correlator $\langle \hat{n}_i \hat{n}_j \rangle$).
The histogram $P(p_z)$ of experimentally measured bit-string probabilities $\{p_z\}$ at this early evolution stage (Fig.~\ref{fig:FigureN4}b, left) shows a distribution with a tail of large $p_z$ values. This is associated with an insufficient scrambling of the initially localized quantum information. 
At an intermediate time ($\tau=360\,\mathrm{ns}$), the excitations are shown to aggregate on a few sites in the case of the integrable theory (Fig.~\ref{fig:FigureN4}e, middle), whereas the excitations are more spread out over the entire 1D lattice in the experiment and in the learned Hamiltonian model with long-range hopping (middle panels of Fig.~\ref{fig:FigureN4}d). This spreading is not uniform, though, and forms a ``speckle" pattern with site-to-site fluctuation associated with quantum interference \cite{arute2019quantum}. 
This can be quantitatively analyzed through the histogram $P(p_z)$ of bit-string probabilities $\{p_z\}$. The measured bit-string probabilities for the experimental system at intermediate times (Fig.~\ref{fig:FigureN4}b, middle) follow the Porter-Thomas (PT) distribution \cite{porter1956fluctuations, boixo2018characterizing}, a strong signature of a quantum system uniformly exploring all states of its Hilbert space \cite{neill2018blueprint}, and deeply connected to quantum chaos, computational complexity, and many-body benchmarking protocols \cite{boixo2018characterizing, choi2021emergent, cotler2021emergent, mark2022inprep}. 
The second moment $\mu_2$ of the PT distribution is equal to $\mu_2^\mathrm{e}$, to which the experimentally measured system data and the learned Hamiltonian model with the long-range hopping converge at intermediate times as shown in Fig.~\ref{fig:FigureN4}a.
After a long evolution time ($\tau=5.4\,\upmu$s), the speckle pattern in the two-site correlator of the experimental system begins to wash out in comparison to the theory model with long-range hopping (see top panels of Fig.~\ref{fig:FigureN4}d). At these long times, the histogram of measured $\{p_z\}$ for the experimental system (Fig.~\ref{fig:FigureN4}b, right) deviates from the PT distribution, narrowing substantially, and approaching a uniform distribution corresponding to a completely decohered, maximally mixed state. 
This is also reflected in Fig.~\ref{fig:FigureN4}a, where for long times the measured value of $\mu_2$ for the experimental system eventually begins to decrease below the ergodic value of $\mu_2^\mathrm{e}$ \cite{arute2019quantum}. 
Additional numerical simulations of $\mu_2$ and $P(p_z)$ for ergodic and integrable systems can be found in Sec.~\ref{sec:SVIII} of the Supplementary Information.

\hfill \break 
\noindent
\textbf{Conclusion and outlook}


\noindent
In conclusion, we demonstrate a many-body quantum simulator based on a one-dimensional lattice of transmon qubits connected together using a superconducting metamaterial waveguide. 
The metamaterial exhibits photonic bandgaps that protect qubit-photon bound states from decay, and a transmission passband which is used for high-fidelity multiplexed qubit-state readout.
Furthermore, the metamaterial plays the role of a scalable photonic bus to mediate tunable long-range coupling between qubit-photon bound states. 
This system of interacting bound states realizes a Bose-Hubbard model with long-range hopping. We characterize the system using conventional single- and two-qubit measurements along with a sample-efficient many-body Hamiltonian learning protocol.
Lastly, we study the many-body quench dynamics of the system versus the range of the lattice hopping, revealing the ergodic nature of the extended Bose-Hubbard model, distinct from its nearest-neighbor-coupling counterpart.

Building on the many-body characterization methods employed in this work, the metamaterial-based quantum simulator is well-suited to exploring a variety of subtle aspects of quantum chaos and many-body dynamics. These include the study of many-body dephasing \cite{kiendl2017many,kaplan2020many} and pre-thermalization \cite{neyenhuis2017observation, tang2018thermalization}, and the emergence of universal randomness in projected state ensembles \cite{cotler2021emergent, choi2021emergent}. In addition, the widely tunable range of parameters of this system enables the study of the extended Bose-Hubbard model in regimes such as the Mott insulator phase ($|U/J| \gg 1$) and near quantum criticality ($|U/J| \approx 1$) at the Kosterlitz-Thouless (KT) phase transition \cite{kuhner1998phases}.
The distinct hopping signs of the upper and lower bandgap of the metamaterial quantum bus provide a means to investigate the effects of frustration on quantum many-body phases, such as changes in the critical value of $|U/J|$ at the KT transition \cite{kim2021superconducting}.
Various long-range couplings between qubits may also be introduced, including chiral coupling through engineering of the dispersion of the waveguide bus \cite{bello2022spin}, and couplings with a tailorable power-law decay profile \cite{douglas2015quantum} or with geometric phases \cite{hung2016quantum} using lattice-site-dependent modulation of the transmon qubit transition frequency. 
The metamaterial can also act as a dissipative bath in the passband regime, giving rise to correlated decay \cite{kim2021quantum}, which can be used for reservoir engineering and dissipative stabilization of many-body states \cite{ma2019dissipatively}. 
Two-dimensional (2D) lattices of qubits embedded in 2D metamaterials can be fabricated using flip-chip technology \cite{rosenberg20173d,foxen2017qubit}, creating opportunities for quantum simulation on novel lattices, such as those with flat bands \cite{hung2016quantum, kollar2019hyperbolic} or topological properties \cite{owens2021chiral}.
Finally, besides quantum simulation, the metamaterial architecture presented in this work can also be used for quantum computation. Specifically, incorporating tunable couplers \cite{chen2014qubit, yan2018tunable, kounalakis2018tuneable} between qubits and the metamaterial bus can realize high-fidelity long-range gates, paving the way for the implementation of error-correcting codes with reduced resource overhead, such as quantum low-density parity-check codes \cite{baspin2021quantifying, delfosse2021bounds, tremblay2022constant}, on superconducting quantum processors.

\begin{acknowledgments}
The authors thank Alexey Gorshkov, Alejandro Gonz\'{a}lez-Tudela, Darrick Chang, Olexei Motrunich, Ruichao Ma, Fernando Brand\~{a}o, Gil Refael, Srujan Meesala, Vinicius Ferreira, Gihwan Kim, Andreas Butler, and Zhaoyi Zheng for helpful discussions. We appreciate MIT Lincoln Laboratories for the provision of traveling-wave parametric amplifiers used for both spectroscopic and time-domain measurements in this work, and the AWS Center for Quantum Computing for the Eccosorb filters installed in the cryogenic setup for infrared filtering. We also thank the Quantum Machines team for technical support and discussions on the Quantum Orchestration Platform. This work was supported by the AFOSR Quantum Photonic Matter MURI (grant FA9550-16-1-0323), the DOE-BES Quantum Information Science Program (grant DE-SC0020152), the Institute for Quantum Information and Matter, an NSF Physics Frontiers Center (grant PHY-1125565) with support of the Gordon and Betty Moore Foundation, the Kavli Nanoscience Institute at Caltech, and the AWS Center for Quantum Computing. D.~K.~M. acknowledges support from the NSF QLCI program (2016245) and the DOE Quantum Systems Accelerator Center (contract no. 7568717).
  
\end{acknowledgments}


%



\clearpage

\beginsupplement
\begin{center}
    \textbf{\large Supplementary Information}
\end{center}
\section{Theoretical modeling of the quantum simulator}
\label{sec:SI}
In this section, we describe the theoretical modeling of the metamaterial-based quantum simulator used in the main text. 
\subsection{Analytical modeling}\label{sec:analytical-modeling-mm}
In the analytical modeling, we discuss the Hamiltonian description of our quantum simulator by mapping its basic circuit model onto a tight-binding array of coupled cavities with locally coupled qubits under various approximations. We derive the effective Hamiltonian in the qubit subspace, resulting in an analytical form of the metamaterial-mediated coupling between qubits described in the main text.
\subsubsection{Approximate canonical quantization of the metamaterial waveguide}
\begin{figure}[tbhp]
\begin{center}
\includegraphics[width=0.48\textwidth]{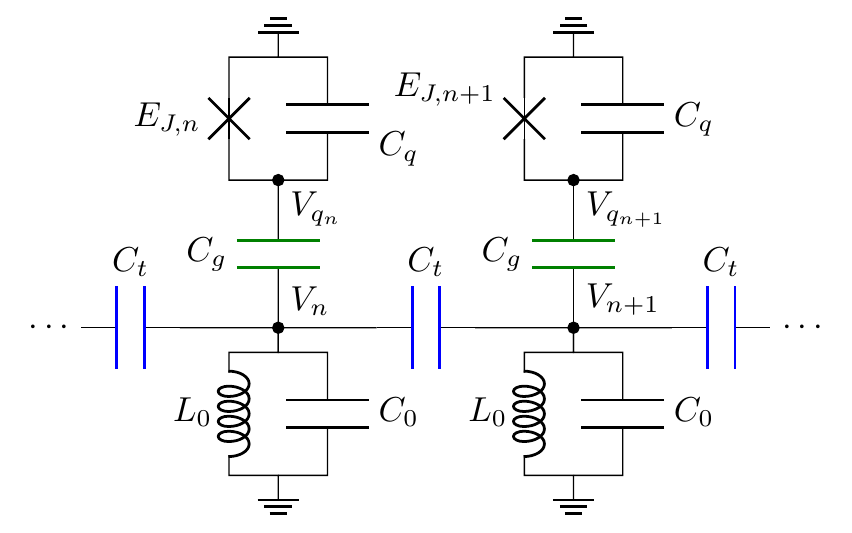}
\caption{\textbf{Basic circuit model of the metamaterial-based quantum simulator.} 
The metamaterial waveguide is described by an array of LC resonators with inductance $L_0$ and capacitance $C_0$ capacitively coupled to nearest neighbors with a capacitance $C_t$ (colored blue). Superconducting transmon qubits, each represented as a parallel circuit of a Josephson junction and a capacitor, are coupled to each metamaterial resonator site with a capacitance $C_g$ (colored green). The Josephson energy and the capacitance of the qubit at the $n$th site is given by $E_{J, n}$ and $C_q$, respectively.}
\refstepcounter{SIfig}\label{fig:mmschematic-analytic} 
\end{center}
\end{figure}

The metamaterial waveguide consists of an array of inductor-capacitor (LC) resonators with inductance $L_0$ and capacitance $C_0$ coupled with capacitance $C_t$ illustrated in Fig.~\ref{fig:mmschematic-analytic}. We denote the flux variable of each node of the metamaterial as $\Phi_n(t)\equiv \int_{-\infty}^{t} \ud t' \: V_n (t')$. The Lagrangian in the position space reads
\begin{align}
    \mathcal{L} = \sum_{n = 0}^{N} \left[\frac{C_t}{2}\left(\dot{\Phi}_{n+1} - \dot{\Phi}_{n}\right)^2 + \frac{C_0}{2}\left(\dot{\Phi}_{n}\right)^2 - \frac{\Phi_{n}^2}{2L_0}\right] \label{eq:lagn-metamaterial-position}
\end{align}
where $\Phi_{n=0}=\Phi_{n=N+1} \equiv 0$. Starting from this Lagrangian, the canonical quantization of the metamaterial waveguide can be performed approximately in the position space. The node charge variable $Q_{n}$ conjugate to the node flux variable $\Phi_{n}$ is evaluated as
\begin{align}
    Q_{n} = \frac{\partial\mathcal{L}}{\partial \dot{\Phi}_{n}} 
    = C_{w0} \dot{\Phi}_{n} - C_t \left(\dot{\Phi}_{n+1} + \dot{\Phi}_{n-1}\right),
\end{align}
where $C_{w0} = C_0 + 2 C_t$ is the effective self-capacitance of LC resonators forming the metamaterial waveguide. Using a vector notation $\boldsymbol{\Phi} \equiv (\Phi_{1}, \Phi_{2}, \cdots, \Phi_{N})^\top$, $\boldsymbol{Q} \equiv (Q_{1}, Q_{2}, \cdots, Q_{N})^\top$, the linear relation between node charges and voltages can be written in a compact form $\boldsymbol{Q} = \boldsymbol{C} \dot{\boldsymbol{\Phi}}$, with the capacitance matrix $\boldsymbol{C}$ given by
\begin{align}
\boldsymbol{C} 
 = C_{w0}\boldsymbol{I} - C_t \boldsymbol{J}_{1}.
\end{align}
Here, $\boldsymbol{I}$ is an $N\times N$ identity matrix and $\boldsymbol{J}_{k}$ is a matrix with components $[\boldsymbol{J}_{k}]_{n,n'} = \delta_{n,k+n'} + \delta_{n+k, n'}$, i.e., unity on the $k$th off-diagonal. Also, the Lagrangian in Eq.~\eqref{eq:lagn-metamaterial-position} can be rewritten as $\mathcal{L} = \frac{1}{2}\dot{\boldsymbol{\Phi}}^\top \boldsymbol{C}\dot{\boldsymbol{\Phi}} - \frac{1}{2L_0}\boldsymbol{\Phi}^\top \boldsymbol{\Phi}$ using this vector notation. The Legendre transformation $H=\sum_{n}Q_{n}\dot{\Phi}_{n} - \mathcal{L}$ \cite{Goldstein:2002} gives the Hamiltonian of the metamaterial waveguide
\begin{align}
    H = \frac{1}{2}\boldsymbol{Q}^\top \boldsymbol{C}^{-1}\boldsymbol{Q} + \frac{1}{2L_0}\boldsymbol{\Phi}^\top \boldsymbol{\Phi}.
\end{align}
The first-order approximation to the inverse of capacitance matrix is given by
\begin{align}
\label{eq:Cinv_approx}
    \boldsymbol{C}^{-1}&= \left[C_{w0}(\boldsymbol{I} - r \boldsymbol{J}_{1})\right]^{-1} = C_{w0}^{-1}(\boldsymbol{I} + r \boldsymbol{J}_{1} + r^{2} \boldsymbol{J}_{1}^{2} + \cdots) \nonumber\\
    & \approx \frac{1}{C_{w0}}\boldsymbol{I} + \frac{C_t}{C_{w0}^2}\boldsymbol{J}_{1} + \mathcal{O}(r^2),
\end{align}
where $r = C_t / C_{w0} <1$ is the ratio of the coupling capacitance to the self capacitance, which is assumed to be small. Using this, the Hamiltonian of the metamaterial waveguide is evaluated as
\begin{align}
    H \approx \sum_{n=1}^{N} \left(\frac{Q_{n}^2}{2 C_{w0}} + \frac{\Phi_{n}^2}{2 L_{0}} + \frac{C_t}{C_{w0}^2}Q_{n}Q_{n+1} \right) \label{eq:approx-Hamiltonian-first-order}
\end{align}
up to first order in $r$, where $Q_{N+1}\equiv 0$. Note that higher-order approximations to the inverse of capacitance matrix can be calculated using the matrix relations
$$\boldsymbol{J}_1^{2} \approx 2\boldsymbol{I} + \boldsymbol{J}_{2}, \ \boldsymbol{J}_1^{3} \approx 3\boldsymbol{J}_{1} + \boldsymbol{J}_{3}, \ 
\boldsymbol{J}_1^{4} \approx 6\boldsymbol{I} + 4\boldsymbol{J}_{2} + \boldsymbol{J}_{4},\  \cdots,$$
under which the Hamiltonian exhibits long-range coupling beyond nearest neighbors. Here, the magnitude of charge-charge coupling between a pair of resonators at sites $(i, j)$ scales as $\sim r^{|i - j|}$, exponentially decaying with the distance $|i-j|$ between resonators.

We promote the flux and charge variables to quantum operators by imposing the canonical commutation relation $[\hat{\Phi}_{n}, \hat{Q}_{n'}]=i\hbar \delta_{n,n'}$. The annihilation (creation) operator $\hat{a}_n$ ($\hat{a}_n^\dagger$), defined with 
\begin{align}
    \hat{\Phi}_{n} = \sqrt{\frac{\hbar Z_w}{2}}\left(\hat{a}_{n} + \hat{a}_{n}^\dagger\right),\
    \hat{Q}_{n} = \frac{1}{i}\sqrt{\frac{\hbar }{2Z_w}}\left(\hat{a}_{n} - \hat{a}_{n}^\dagger\right)\label{eq:canonical-flux-charge-ops}
\end{align}
where $Z_{w0}=\sqrt{L_0 / C_{w0}}$, satisfies the commutation relation $[\hat{a}_{n},\hat{a}_{n'}^\dagger]=\delta_{n,n'}$. Substituting Eq.~\eqref{eq:canonical-flux-charge-ops} into Eq.~\eqref{eq:approx-Hamiltonian-first-order}, we obtain an approximate second-quantized Hamiltonian of the metamaterial under the rotating-wave approximation (RWA), corresponding to the Hamiltonian of an array of nearest-neighbor-coupled cavities \cite{hartmann2006strongly, angelakis2007photon, calajo2016atom} illustrated in Fig.~\ref{fig:FigureN1}a of the main text. The Hamiltonian is written as
\begin{align}
    \hat{H} = \hbar\sum_{n=1}^{N}\left[ \omega_{c} \hat{a}_{n}^\dagger \hat{a}_{n} + t\left(\hat{a}_{n}^\dagger \hat{a}_{n+1} + \hat{a}_{n+1}^\dagger \hat{a}_{n}\right)\right],
\end{align}
where the cavity frequency $\omega_c$ and the coupling $t$ between neighboring cavities are given by
\begin{align}
    \omega_{c} = \frac{1}{\sqrt{L_0(C_0 + 2C_t)}}, \quad t = \frac{C_t}{2(C_0 + 2C_t)}\omega_{c}.\label{eq:cavity-freq-coupling-t}
\end{align}
\subsubsection{Coupling of superconducting qubits to the metamaterial waveguide}
We assume coupling a transmon qubit (in the schematic form of a Josephson junction with energy $E_J$ and parallel capacitance $C_q$) to each metamaterial resonator site via a capacitance $C_g$ as illustrated in Fig.~\ref{fig:mmschematic-analytic}. Following the procedures similar to the above, the full Hamiltonian of the qubit-metamaterial system can be obtained up to first order in the coupling capacitances $C_{g}$ and $C_{t}$, written as
\begin{align}
\label{eq:HamNN}
    \hat{H} &\approx \sum_{n}\bigg( \frac{\hat{Q}_{n}^2}{2C_{w\Sigma}} + \frac{\hat{\Phi}_{n}^2}{2L_0} + \frac{C_t}{C_{w\Sigma}^2} \hat{Q}_{n}\hat{Q}_{n+1} \nonumber\\
    &\qquad\quad  + \frac{\hat{Q}_{q_n}^2}{2C_{q\Sigma}}- E_{J,n}\cos{\frac{2\pi\hat{\Phi}_{q_n}}{\Phi_0}} + \frac{C_{g}}{C_{w\Sigma}C_{q\Sigma}}\hat{Q}_{n}\hat{Q}_{q_n}\bigg).
\end{align}
Here, $\hat{\Phi}_{n}$ ($\hat{Q}_n$) and $\hat{\Phi}_{q_n}$ ($\hat{Q}_{q_n}$) are the node flux (charge) operators of the metamaterial resonator and the qubit at the $n$th unit cell, respectively, and $\Phi_0=h/2e$ is a magnetic flux quantum. The self-capacitance of metamaterial resonators and qubits are renormalized to $C_{w\Sigma} = C_{w0} + C_{g}$ and $C_{q\Sigma}=C_{q} + C_{g}$, respectively, redefining the parameters of the cavity array in Eq.~\eqref{eq:cavity-freq-coupling-t} into
\begin{align}
    \omega_{c} = \frac{1}{\sqrt{L_0(C_0 + C_g + 2C_t)}}, \quad t = \frac{C_t}{2(C_0 + C_g + 2C_t)}\omega_{c}. \tag{\ref*{eq:cavity-freq-coupling-t}$'$} \label{eq:cavity-freq-coupling-t-with-qubits}
\end{align}
Introducing the annihilation and the creation operators following a procedure similar to that of  Eq.~\eqref{eq:canonical-flux-charge-ops} and restricting the subspace of qubits to their lowest two energy levels, we obtain the Hamiltonian of the metamaterial-qubit system in a second-quantized form
\begin{align}
    \hat{H} &= \hbar\sum_{n}\bigg[ \omega_{c} \hat{a}_n^\dagger \hat{a}_n + t\left(\hat{a}_{n}^\dagger \hat{a}_{n+1} + \hat{a}_{n+1}^\dagger \hat{a}_{n}\right) \nonumber\\
    &\qquad \qquad + \frac{\omega_{q_n}}{2}\hat{\sigma}_n^{z} + g_{n}\left(\hat{a}_n^\dagger\hat{\sigma}_n^{-}+\hat{a}_n\hat{\sigma}_n^{+}\right)\bigg],\label{eq:quantized-Hamiltonian-real-space}
\end{align}
where
\begin{align}
g_{n}= \frac{C_{g}}{2\sqrt{C_{w\Sigma}C_{q\Sigma}}}\sqrt{\omega_{c}\omega_{q_n}}    
\end{align}
is the coupling between a qubit and a metamaterial resonator at the $n$th unit cell. Here, $\hat{\sigma}_n^{\alpha}$ ($\alpha\in\{\pm, x, y, z\}$) denotes the Pauli operator of qubit at the $n$th unit cell. We can use the annihilation operators $\hat{a}_k = \frac{1}{\sqrt{N}}\sum_{n}e^{-iknd}\hat{a}_n$ in the momentum space (satisfying the commutation relation $[\hat{a}_{k},\hat{a}_{k'}^\dagger]=\delta_{k,k'}$) to rewrite the Hamiltonian of the metamaterial waveguide in terms of its normal modes, where $d$ is the lattice constant. In this case, the Hamiltonian is given by
\begin{align}
    \hat{H}/\hbar &= \sum_{k}\omega_k \hat{a}_k^\dagger \hat{a}_k +  \sum_{n}\frac{\omega_{q_n}}{2}\hat{\sigma}_n^{z} \nonumber\\
    &\qquad+\sum_{n, k} \left(g_{k,n}\hat{a}_k^\dagger \hat{\sigma}_n^{-} + g_{k,n}^*\hat{a}_k \hat{\sigma}_n^{+} \right).\label{eq:quantized-Hamiltonian}
\end{align}
Here, $\omega_k = \omega_c + 2 t \cos{(kd)}$ is the dispersion relation of the metamaterial waveguide (plotted in Fig.~\ref{fig:FigureN1}b of the main text) up to first order in $C_{t}/C_{w\Sigma}$ and $g_{k,n} \equiv g_{n} e^{-iknd} / \sqrt{N}$ is the coupling of a qubit at site $n$ to a metamaterial mode with wavevector $k$.

\subsubsection{Effective Hamiltonian in the dispersive limit}
The effective Hamiltonian $\hat{H}_\mathrm{eff}$ of the system can be calculated by performing the Schrieffer-Wolff transformation  $$\hat{U} = \exp{\left(\sum_{k,n}\frac{{g_{k,n}^*}\hat{a}_k \hat{\sigma}_{n}^{+} - g_{k,n} \hat{a}_k^\dagger \hat{\sigma}_{n}^{-}}{\Delta_{n,k}}\right)}$$ on the original Hamiltonian in Eq.~\eqref{eq:quantized-Hamiltonian}, where $\Delta_{n,k}=\omega_{q_n} - \omega_k$ is the detuning of the $n$th qubit from the metamaterial mode at wavevector $k$. The result of this transformation is given by
\begin{align}
    \hat{H}_\mathrm{eff}/\hbar &= \sum_{k} \omega_k \hat{a}_k^\dagger\hat{a}_{k} + \sum_{n}\frac{\omega_{q_n}}{2}\hat{\sigma}_{n}^{z} \nonumber
    \\&\qquad +\sum_{n,n'} J_{n,n'} \hat{\sigma}_{n}^{+} \hat{\sigma}_{n'}^{-} + \sum_{k,k',n} K_{k,k',n}\hat{\sigma}_{n}^{z}  \hat{a}_{k}^\dagger \hat{a}_{k'},
\end{align}
where $K_{k,k',n}$ denotes coupling between a pair of metamaterial modes $(k, k')$ dependent on the state of the $n$th qubit, giving rise to qubit-state-dependent shift of the metamaterial  given by
\begin{align}
    {K}_{k,k',n} = \frac{g_{k,n}g_{k',n}^*}{2}\left(\frac{1}{\Delta_{n,k}} + \frac{1}{\Delta_{n,k'}}\right),
\end{align}
and the exchange interaction $J_{n,n'}$ between a qubit pair $(n, n')$ is written as
\begin{align}
    J_{n,n'} = \sum_{k}\frac{g_{k,n'}g_{k,n}^*}{2}\left(\frac{1}{\Delta_{n,k}} + \frac{1}{\Delta_{n',k}}\right).
\end{align}
We focus on this exchange interaction, the interaction between qubits mediated by the virtual photons of the metamaterial, by evaluating the sum
\begin{align}
    J_{n,n'} &= \sum_{k}\frac{g_{n}g_{n'}}{N}\frac{e^{ik(n-n')d}}{2}\nonumber\\
    &\ \qquad\times\left[\frac{1}{\Delta_n - 2t\cos{(kd)}} + \frac{1}{\Delta_{n'} - 2t\cos{(kd)}}\right]\label{eq:tunneling}
\end{align}
where $\Delta_{n} \equiv \omega_{q_n} - \omega_c$ is the detuning of the $n$th qubit from the bare cavity frequency $\omega_c$. In the discrete model consisting of a finite number of unit cells $N$, the wavevectors of the metamaterial mode are equally spaced by $\Delta k = {2\pi}/{Nd}$. In the continuum limit, the summation on the right-hand side of Eq.~\eqref{eq:tunneling} is cast into
\begin{align*}
    & \frac{g_{n}g_{n'}}{2} \frac{d}{2\pi} \int_{-\pi / d}^{\pi/d} \ud k\left[\frac{e^{ik(n-n')d}}{\Delta_n - 2t\cos{(kd)}} + \frac{e^{ik(n-n')d}}{\Delta_{n'} - 2t\cos{(kd)}}\right] \\
    &= - \frac{g_{n}g_{n'}}{4\pi}\frac{1}{2t} \left[I\left(n-n', a_{n}\right) + I\left(n-n', a_{n'}\right)\right],
\end{align*}
where $a_n = -{\Delta_n}/{2J}$ and $I$ is the integral function defined and evaluated as
\begin{align}
    I(n, a) &\equiv \int_{-\pi}^{\pi} \ud k \frac{e^{ikn}}{a + \cos{k}}\nonumber\\
    &=\left\{\begin{array}{lr}
        \frac{2\pi }{\sqrt{a^2 - 1}}  (-1)^{|n|}e^{-|n|/\lambda} , & \text{if } a > 1\\
         -\frac{2\pi }{\sqrt{a^2 - 1}} e^{-|n|/\lambda}, & \text{if } a < -1
        \end{array}\right..
\end{align}
Here, $\lambda = 1/\mathrm{arccosh}(|a|) = 1 / \ln{\left(|a| + \sqrt{a^2 - 1}\right)}$. In the following, we describe the behavior of metamaterial-mediated coupling $J_{n,n'}$ between qubits in the bandgap regime.

\begin{enumerate}[\itshape(a)] 
\item \textit{Lower bandgap.} When qubits are tuned to the lower bandgap, i.e., $\Delta_n = \omega_{q_n} - \omega_c < -2t$ and $a_n > 1$, the exchange interaction $J_{n,n'}$ between the qubits mediated by virtual photons of the metamaterial waveguide is calculated as
\begin{align}
J_{n,n'}
&=-(-1)^{|n-n'|}\frac{g_{n}g_{n'}}{2}\nonumber\\
&\ \quad\times \left(\frac{e^{-|n-n'|/\lambda_n}}{\sqrt{\Delta_n^2 - 4t^2}}+ \frac{e^{-|n-n'|/\lambda_{n'}}}{\sqrt{\Delta_{n'}^2 - 4t^2}} \right),
\end{align}
where $\lambda_n$ is the localization length given by
\begin{align}
    \frac{1}{\lambda_{n}} = {\mathrm{arccosh}\left(\frac{|\Delta_{n}|}{2J}\right)}.\label{eq:localization-length}
\end{align}
Note that the diagonal components $J_{n,n}=-g_{n}^2/\sqrt{\Delta_{n}^2-4t^2}$ corresponds to the Lamb shift and takes negative values inside the lower bandgap as the metamaterial modes are located at frequencies higher than that of the qubit. Considering the off-diagonal components, if the qubits at sites $n$ and $n'$ are resonant with one another ($\Delta_n=\Delta_{n'}$), the coupling $J_{n,n'}$ falls off exponentially with the distance $|n-n'|$ at a length scale $\lambda_n$ determined by the detuning  $\Delta_n$ of qubits and the tunneling rate $t$. In addition, the coupling $J_{n,n'}$ alternates sign with a factor $(-1)^{|n-n'|}$ due to the fact that localized modes inside the lower bandgap have quasi-wavevectors $k$ with the real part of $\pi/d$. This is because the lowest frequency of the band takes place at $k=\pi/d$ according to the dispersion relation, and any localized modes inside the lower bandgap must have a real part of the wavevector that is an analytic continuation of the lowest-frequency point.
%
\item \textit{Upper bandgap.} When qubits are tuned to the upper bandgap, $\Delta_n = \omega_{q_n} - \omega_c  > 2t$ and $a_n < -1$, the exchange interaction $J_{n,n'}$ is evaluated as
\begin{align}
J_{n,n'}
=\frac{g_{n}g_{n'}}{2}\left(\frac{e^{-|n-n'|/\lambda_n}}{\sqrt{\Delta_n^2 - 4t^2}}+ \frac{e^{-|n-n'|/\lambda_{n'}}}{\sqrt{\Delta_{n'}^2 - 4t^2}} \right),
\end{align}
with the localization length $\lambda_{n}$ having a form identical to Eq.~\eqref{eq:localization-length}. The diagonal components $J_{n,n}=g_n^2/\sqrt{\Delta_n^2 - 4t^2}$ (i.e., the Lamb shift) takes positive values inside the upper bandgap since the metamaterial modes are located at frequencies lower than that of the qubits. Similar to the case of lower bandgap, the coupling $J_{n,n'}$ between qubits at sites $n$ and $n'$ inside the upper bandgap falls off exponentially with the distance $|n-n'|$ at a length scale $\lambda_n$. However, the alternating sign factor is not present in the upper bandgap case due to the fact that the quasi-wavevector inside the upper bandgap has real part of zero. Therefore, the coupling $J_{n,n'}$ mediated by the metamaterial waveguide is always positive inside the upper bandgap.
\end{enumerate}

\begin{figure*}[t!]
\begin{center}
\includegraphics[width=0.9\textwidth]{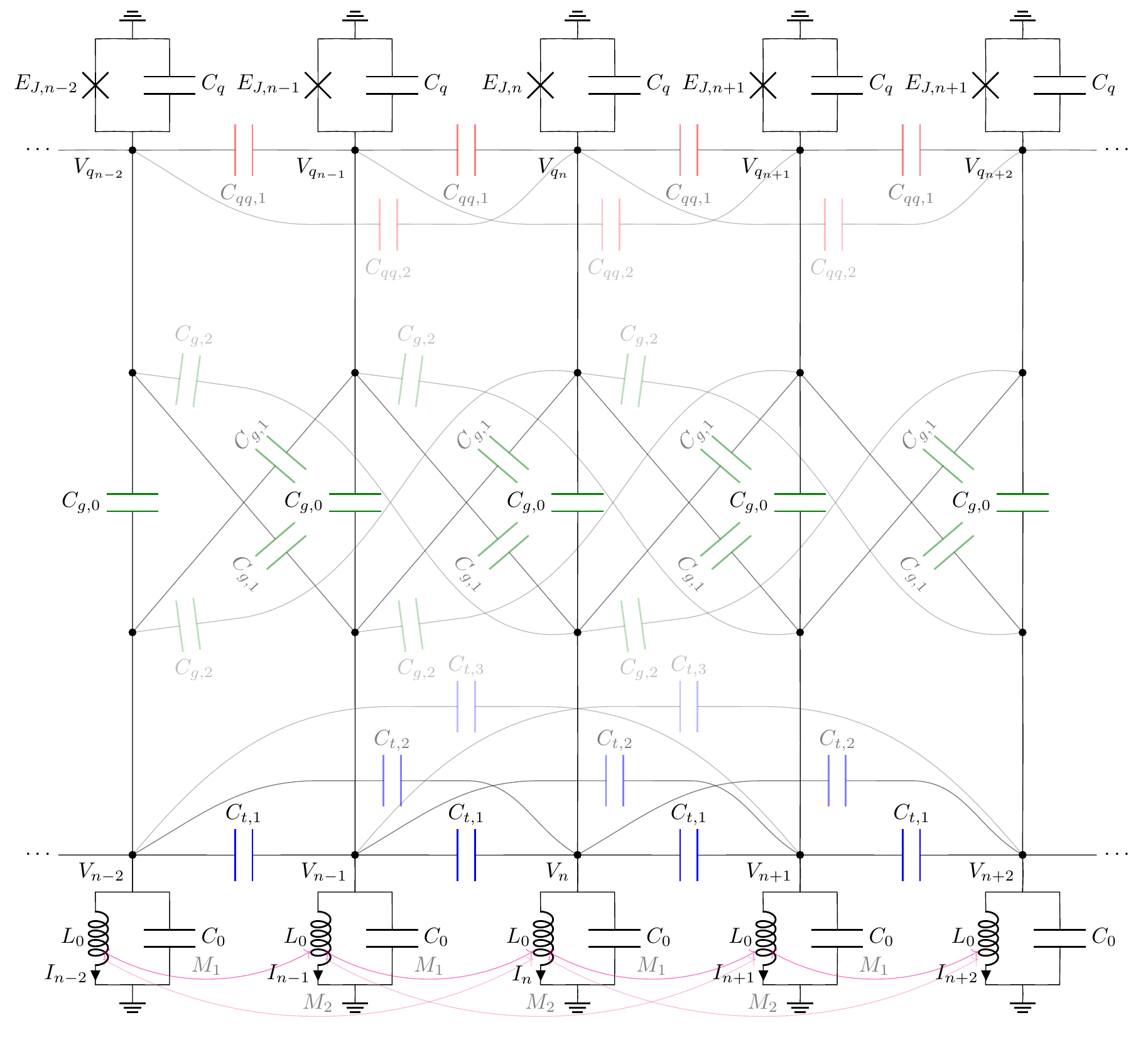}
\caption{\textbf{General circuit model of the metamaterial-based quantum simulator for numerical modeling.} $L_0$ ($C_0$) is the self-inductance (self-capacitance) of metamaterial resonators and $E_{J,n}$ ($C_q$) is the Josephson energy (capacitance) of qubit coupled to the $n$th unit cell. The capacitance between metamaterial resonators (qubits) separated by a distance $x=1, 2, \cdots$ is denoted as $C_{t,x}$ ($C_{qq,x}$), colored blue (red). The mutual inductance between inductors of metamaterial resonators at a distance $x=1, 2, \cdots$, colored magenta, is denoted as $M_x$. The distributed coupling of a qubit to the metamaterial is specified by capacitance $C_{g,x}$ between a qubit to a metamaterial resonator at a distance $x=0, 1, 2, \cdots$, colored green. Opaque elements in the figure represent parasitic capacitive and mutual inductive contributions, processes of which are shown only up to second order.}
\refstepcounter{SIfig}\label{fig:mmschematic-numerical} 
\end{center}
\end{figure*}

\subsection{Numerical modeling}
While the analytical modeling discussed in Sec.~\ref{sec:analytical-modeling-mm} is useful for understanding the basic processes in our quantum simulator, the presence of long-range coupling between metamaterial resonators (due to large coupling capacitance values) and parasitic coupling mechanisms in the realized device complicate the picture, causing our experimental data to deviate significantly from the simplest analytical theory. To resolve this, we come up with a theory for numerical modeling that allows us to find realistic parameters of the device.
\subsubsection{Derivation of Hamiltonian}
We assume a general circuit model illustrated in Fig.~\ref{fig:mmschematic-numerical}, extended from the one used in the analytical modeling in Fig.~\ref{fig:mmschematic-analytic} which consists of an array of LC resonators with inductance $L_0$ and capacitance $C_0$ forming the metamaterial waveguide and qubits with Josephson energy $E_{J,n}$ and capacitance $C_q$. Here, the capacitance $C_{t,1}\equiv C_{t}$ between nearest-neighboring metamaterial resonators and capacitance $C_{g,0}\equiv C_{g}$ between a qubit and a metamaterial resonator at each unit cell give rise to the desired couplings in the system. In addition, this model includes parasitic capacitance $C_{t,x}$ ($x=2, 3, \cdots$) and mutual inductance $M_{x}$ ($x=1, 2, \cdots$)  between metamaterial resonators not limited to nearest neighbors, parasitic long-range capacitance between qubits $C_{qq,x}$ ($x=1, 2, \cdots$), and distributed capacitive coupling between qubits and metamaterial resonators represented by capacitance $C_{g,x}$  ($x=1, 2, \cdots$). 

The capacitive part $\mathcal{L}_{C}$ of the Lagrangian contains terms that are quadratic in time-derivatives of the node flux variables $\Phi_{n}(t)\equiv\int_{-\infty}^{t}\ud t'\: V_{n}(t')$ and $\Phi_{q_n}(t)\equiv\int_{-\infty}^{t} \ud t'\: V_{q_n}(t')$, written as
\begin{align}
 \mathcal{L}_{C} &= \sum_{n} \bigg[\frac{C_0}{2}\dot{\Phi}_{n}^{2} + \sum_{x >0} \frac{C_{t,x}}{2} (\dot{\Phi}_{n + x} - \dot{\Phi}_{n})^2 \nonumber\\
 &\quad\qquad + \frac{C_{q}}{2}\dot{\Phi}_{q_n}^{2} + \sum_{x>0}\frac{C_{qq,x}}{2} (\dot{\Phi}_{q_{n+x}} - \dot{\Phi}_{q_{n}})^2 \nonumber \\
 &\quad\qquad +\sum_{x} \frac{C_{g,|x|}}{2}(\dot{\Phi}_{q_n} - \dot{\Phi}_{n + x})^2 \bigg].
\end{align}
The node charge variables $Q_n = \partial{\mathcal{L}_C}/{\partial\dot{\Phi}_n}$ and $Q_{q_n} = \partial{\mathcal{L}_C}/{\partial\dot{\Phi}_{q_n}}$ canonically conjugate to the flux variables are evaluated as
\begin{subequations}
\begin{align}
    Q_{n} 
    &= \left(C_0 + \sum_{x\neq 0} C_{t,|x|} + \sum_{x} C_{g,|x|}\right)\dot{\Phi}_{n} \nonumber \\
     &\qquad - \sum_{x \neq 0} C_{t,|x|} \dot{\Phi}_{n + x}  - \sum_{x}C_{g,|x|} \dot{\Phi}_{q_{n+x}},\label{eq:qn-phidotn-a}\\
    Q_{q_n} 
    &= \left(C_q + \sum_{x\neq 0} C_{qq,|x|} + \sum_{x} C_{g,|x|}\right)\dot{\Phi}_{q_n} \nonumber \\
     &\qquad - \sum_{x\neq 0} C_{qq,|x|} \dot{\Phi}_{q_{n + x}} - \sum_{x}C_{g,|x|} \dot{\Phi}_{n+x}.\label{eq:qn-phidotn-b}
\end{align}
\end{subequations}
Equations~\eqref{eq:qn-phidotn-a}-\eqref{eq:qn-phidotn-b} can be rewritten in a compact form by introducing a vector of node charge variables $\boldsymbol{Q} = (Q_1, Q_2, \cdots, Q_{q_1}, Q_{q_2}, \cdots)^\top$ and a vector of node flux variables $\boldsymbol{\Phi} = ({\Phi}_1, {\Phi}_2, \cdots, {\Phi}_{q_1}, {\Phi}_{q_2}, \cdots)^\top$, giving $\boldsymbol{Q} = \boldsymbol{C}\boldsymbol{\dot{\Phi}}$ where
\begin{equation}
    \boldsymbol{C} = 
    \begin{pmatrix}
        C_{w\Sigma} & -C_{t,1} & -C_{t,2} &  \cdots & -C_{g,0} & -C_{g,1} & \cdots \\
        -C_{t,1} & C_{w\Sigma} & -C_{t,1} &  \cdots & -C_{g,1} & -C_{g,0} & \cdots \\
        -C_{t,2} & -C_{t,1} & C_{w\Sigma} &  \cdots & -C_{g,2} & -C_{g,1} & \cdots \\
        \vdots & \vdots & \vdots & \ddots & \vdots & \vdots & \ddots  \\
        -C_{g,0} & -C_{g,1} & -C_{g,2} & \cdots & C_{q\Sigma} & -C_{qq,1} & \cdots \\
        -C_{g,1} & -C_{g,0} & -C_{g,1} &  \cdots & -C_{qq,1} & C_{q\Sigma} & \cdots \\
        \vdots & \vdots & \vdots &  \ddots & \vdots & \vdots & \ddots
    \end{pmatrix}.
\end{equation}
Here, the effective self-capacitance $C_{w\Sigma}$ ($C_{q\Sigma}$) of a metamaterial resonator (qubit) is given by
\begin{subequations}
\begin{align}
    C_{w\Sigma}&= C_0 + \sum_{x \neq 0}C_{t,|x|} + \sum_{x}C_{g,|x|}, \\
    C_{q\Sigma}&= C_q +  \sum_{x\neq 0}C_{qq,|x|} + \sum_{x} C_{g,|x|}.
\end{align}
\end{subequations}
Note that the capacitance matrix $\boldsymbol{C}$ is symmetric, satisfying $\boldsymbol{C}^\top=\boldsymbol{C}$.

The inductive part $\mathcal{L}_L$ of the Lagrangian reads
\begin{align}
    \mathcal{L}_{L} &= \sum_n\left(-\frac{1}{2}L_0 I_n^2 - \sum_{x > 0}M_x I_n I_{n+x}\right)\nonumber\\
    &= -\frac{1}{2}\sum_n I_n \left( L_0 I_n + \sum_{x\neq 0} M_{|x|} I_{n+|x|}\right),
    \label{eq:Lagrangian-inductive}
\end{align}
where $I_n$ is the current flowing through the inductor of the $n$th metamaterial resonator. The node flux $\Phi_n$ and current $I_n$ satisfies the relation
\begin{align}
    \Phi_n = L_0 I_n + \sum_{x\neq 0}M_{|x|} I_{n + x},
\end{align}
which can be compactly written in terms of a vector of node flux variables of metamaterial resonators $\boldsymbol{\Phi}_w \equiv (\Phi_1, \Phi_2, \cdots)^\top$ and that of current variables $\boldsymbol{I}_w \equiv (I_1, I_2, \cdots)^\top$ as $\boldsymbol{\Phi}_w = \boldsymbol{L}_{w} \boldsymbol{I}_{w}$, where
\begin{equation}
    \boldsymbol{L}_w = \begin{pmatrix}
        L_0 & M_1 & M_2 & M_3 & \cdots \\
        M_1 & L_0 & M_1 & M_2 & \cdots \\
        M_2 & M_1 & L_0 & M_1 &  \cdots \\
        M_3 & M_2 & M_1 & L_0 & \cdots \\
        \vdots & \vdots & \vdots & \vdots & \ddots 
    \end{pmatrix}
\end{equation}
is a symmetric matrix, i.e., $\boldsymbol{L}_w^\top = \boldsymbol{L}_w$. This allows us to rewrite Eq.~\eqref{eq:Lagrangian-inductive} as 
\begin{equation}
    \mathcal{L}_L = -\frac{1}{2}\boldsymbol{I}_w^\top \boldsymbol{\Phi}_w = - \frac{1}{2} \boldsymbol{\Phi}_w^\top \boldsymbol{L}_w^{-1}\boldsymbol{\Phi}_w.
\end{equation}

The last part of the Lagrangian $\mathcal{L}_{J}$ comes from the Josephson junctions forming qubits
\begin{equation}
    \mathcal{L}_{J} = \sum_n E_{J,n} \cos{\frac{2\pi\Phi_{q_n}}{\Phi_0}}.
\end{equation}

The Lagrangian of the system is given by sum of the three contributions mentioned above, i.e., $\mathcal{L}=\mathcal{L}_C + \mathcal{L}_L + \mathcal{L}_J$. The Hamiltonian can be obtained by the Legendre transformation 
\begin{align}
    H &= \sum_{n} \left(Q_n\dot{\Phi}_n  + Q_{q_n}\dot{\Phi}_{q_n}\right) \nonumber\\
    &= \frac{1}{2}\boldsymbol{Q}^\top \boldsymbol{C}^{-1}\boldsymbol{Q} + \frac{1}{2}\boldsymbol{\Phi}_w^\top \boldsymbol{L}_w^{-1} \boldsymbol{\Phi}_w - \sum_{n}E_{J,n}\cos{\frac{2\pi\Phi_{q_n}}{\Phi_0}}.
\end{align}
Finally, we expand the cosine potential of the Josephson junctions according to
\begin{equation}
    \cos{\frac{2\pi\Phi_{q_n}}{\Phi_0}} \approx 1 - \frac{1}{2}\left(\frac{2\pi\Phi_{q_n}}{\Phi_0}\right)^2 + \frac{1}{24}\left( \frac{2\pi\Phi_{q_n}}{\Phi_0}\right)^4,
\end{equation}
which gives the final form of the Hamiltonian
\begin{align}
    H &= \frac{1}{2}\boldsymbol{Q}^\top \boldsymbol{C}^{-1}\boldsymbol{Q}+ \frac{1}{2}\boldsymbol{\Phi}_w^\top \boldsymbol{L}_w^{-1} \boldsymbol{\Phi}_w \nonumber\\
    &\quad + \sum_{n}\frac{E_{J,n}}{2}\left(\frac{2\pi\Phi_{q_n}}{\Phi_0}\right)^2 - \sum_{n} \frac{E_{J,n}}{24}\left( \frac{2\pi\Phi_{q_n}}{\Phi_0}\right)^4 \label{eq:numerical-model-Hamiltonian}.
\end{align}
\subsubsection{Second quantization}
We promote the node flux and charge variables to quantum operators by imposing the canonical commutation relation $[\hat{\Phi}_n, \hat{Q}_{n'}] = [\hat{\Phi}_{q_n}, \hat{Q}_{q_n'}]=i\hbar \delta_{n,n'}$. We also decompose the inverse capacitance and inductance matrices into their diagonal ($D$) and off-diagonal ($O$) parts, i.e., $\boldsymbol{C}^{-1}=(\boldsymbol{C}^{-1})_{D} + (\boldsymbol{C}^{-1})_{O}$ and $\boldsymbol{L}^{-1}_{w}=(\boldsymbol{L}^{-1}_{w})_{D} + (\boldsymbol{L}^{-1}_{w})_{O}$. Then the Hamiltonian in Eq.~\eqref{eq:numerical-model-Hamiltonian} can be rearranged into $\hat{H} = \hat{H}_0 + \hat{V}$, where
\begin{subequations}
\begin{align}
    \hat{H}_0 &= \sum_{n} \Bigg[ \frac{(\boldsymbol{C}^{-1})_{n,n}}{2}\hat{Q}_n^2 + \frac{(\boldsymbol{L}^{-1}_{w})_{n,n}}{2}\hat{\Phi}_n^2 \nonumber \\
    &\qquad\qquad + \frac{(\boldsymbol{C}^{-1})_{q_n,q_n}}{2}\hat{Q}_{q_n}^2 + \frac{E_{J,n}}{2}\left(\frac{2\pi\hat{\Phi}_{q_n}}{\Phi_0}\right)^2 \nonumber \\
    &\qquad\qquad -\frac{E_{J,n}}{24}\left( \frac{2\pi\hat{\Phi}_{q_n}}{\Phi_0}\right)^4 \Bigg]\label{eq:numerical-model-Hamiltonian-diagonal}
\end{align}
contains components associated with the diagonal part of the inverse matrices and the cross coupling terms are described by
\begin{align}
    \hat{V} = \frac{1}{2}\hat{\boldsymbol{Q}}^\top (\boldsymbol{C}^{-1})_{O} \hat{\boldsymbol{Q}} + \frac{1}{2}\hat{\boldsymbol{\Phi}}_w^\top (\boldsymbol{L}^{-1}_w)_{O} \hat{\boldsymbol{\Phi}}_w.\label{eq:numerical-model-Hamiltonian-offdiag}
\end{align}
\end{subequations}

The second quantization can be performed by writing the canonical flux and charge operators in terms of annihilation and creation operators by noting the form of $\hat{H}_0$ in Eq.~\eqref{eq:numerical-model-Hamiltonian-diagonal}:
\begin{subequations}
\begin{align}
    \hat{\Phi}_{n} &= \sqrt{\frac{\hbar Z_{n}}{2}} \left(\hat{a}_n + \hat{a}_n^\dagger\right),\\  \hat{Q}_{n} &= \frac{1}{i}\sqrt{\frac{\hbar }{2Z_{n}}} \left(\hat{a}_n - \hat{a}_n^\dagger\right),\\
    \hat{\Phi}_{q_n} &= \sqrt{\frac{\hbar Z_{q_n}}{2}} \left(\hat{a}_{q_n} + \hat{a}_{q_n}^\dagger\right),\\
    \hat{Q}_{q_n} &= \frac{1}{i}\sqrt{\frac{\hbar }{2Z_{q_n}}} \left(\hat{a}_{q_n} - \hat{a}_{q_n}^\dagger\right),
\end{align}
\end{subequations}
where the effective impedance of the metamaterial resonator (qubit) at the $n$th unit cell is given by
\begin{equation}
    Z_{n} = \sqrt{\frac{(\boldsymbol{C}^{-1})_{n,n}}{(\boldsymbol{L}^{-1}_w)_{n,n}}}, \quad Z_{q_n} = \sqrt{\frac{(\boldsymbol{C}^{-1})_{q_n,q_n}}{\left({2\pi}/{\Phi_0}\right)^2 E_{J,n}}}.
\end{equation}
Using the above expressions, Eq.~\eqref{eq:numerical-model-Hamiltonian-diagonal} can be rewritten under the RWA as
\begin{align}
    \hat{H}_0/\hbar &= \sum_{n}\bigg[{\omega_{n}}\hat{a}_{n}^\dagger\hat{a}_{n} + {\omega_{q_n}}\hat{a}_{q_n}^\dagger\hat{a}_{q_n}
    + \frac{U_{q_n}}{2}(\hat{a}_{q_n}^\dagger)^2(\hat{a}_{q_n})^2\bigg],\label{eq:numerical-model-Hamiltonian-diagonal-second-quantized}
\end{align}
where the bare resonator frequency $\omega_{n}$, the transition frequency $\omega_{q_n}$ and the anharmonicity $U_{q_n}$ of the bare qubit at the $n$th unit cell are given by
\begin{subequations}
\begin{align}
\omega_n &= \sqrt{(\boldsymbol{C}^{-1})_{n,n}(\boldsymbol{L}^{-1}_w)_{n,n}}\\
\omega_{q_n} &= \frac{\sqrt{8E_{J,n}E_{C,n}}-E_{C,n}}{\hbar}\\
U_{q_n} &= -E_{C,n}.
\end{align}
\end{subequations}
Here, $E_{C,n}=e^2 (\boldsymbol{C}^{-1})_{q_n,q_n} / 2$ is the charging energy of the qubit at the $n$th unit cell where $e$ is the electron charge. The coupling terms in Eq.~\eqref{eq:numerical-model-Hamiltonian-offdiag} can also be written in the form of
\begin{align}
    \hat{V}/\hbar &= \sum_{n,n'}\bigg[t_{n,n'}\left(\hat{a}_{n}^\dagger \hat{a}_{n'} + \hat{a}_{n'}^\dagger \hat{a}_{n}\right) \nonumber \\
    &\qquad\quad + g_{n,q_{n'}}\left(\hat{a}_{n}^\dagger \hat{a}_{q_{n'}} + \hat{a}_{q_{n'}}^\dagger \hat{a}_{n}\right)\nonumber \\
    &\qquad\quad + J^{\prime}_{q_{n},q_{n'}}\left(\hat{a}_{q_{n}}^\dagger \hat{a}_{q_{n'}} + \hat{a}_{q_{n'}}^\dagger \hat{a}_{q_{n}}\right)\bigg].\label{eq:numerical-model-Hamiltonian-offdiag-second-quantized}
\end{align}
where $t_{n,n'}$ is the coupling between metamaterial resonators, $g_{n,q_{n'}}$ is the coupling between a qubit and a metamaterial resonator, and $J'_{q_n,q_{n'}}$ is the parasitic direct coupling between qubits, given by
\begin{subequations}
\begin{align}
    t_{n,n'} &= \frac{1}{2}\left[\frac{(\boldsymbol{C}^{-1})_{n,n'}}{\sqrt{Z_n Z_{n'}}} + {(\boldsymbol{L}^{-1}_w)_{n,n'}}{\sqrt{Z_n Z_{n'}}}\right], \\
    g_{n,q_{n'}}&= \frac{(\boldsymbol{C}^{-1})_{n,q_{n'}}}{2\sqrt{Z_n Z_{q_{n'}}}}, \\
    J'_{q_n,q_{n'}} &= \frac{(\boldsymbol{C}^{-1})_{q_{n},q_{n'}}}{2\sqrt{Z_{q_n} Z_{q_{n'}}}}.
\end{align}
\end{subequations}

\begin{figure*}[tbhp]
\begin{center}
\includegraphics[width=1\textwidth]{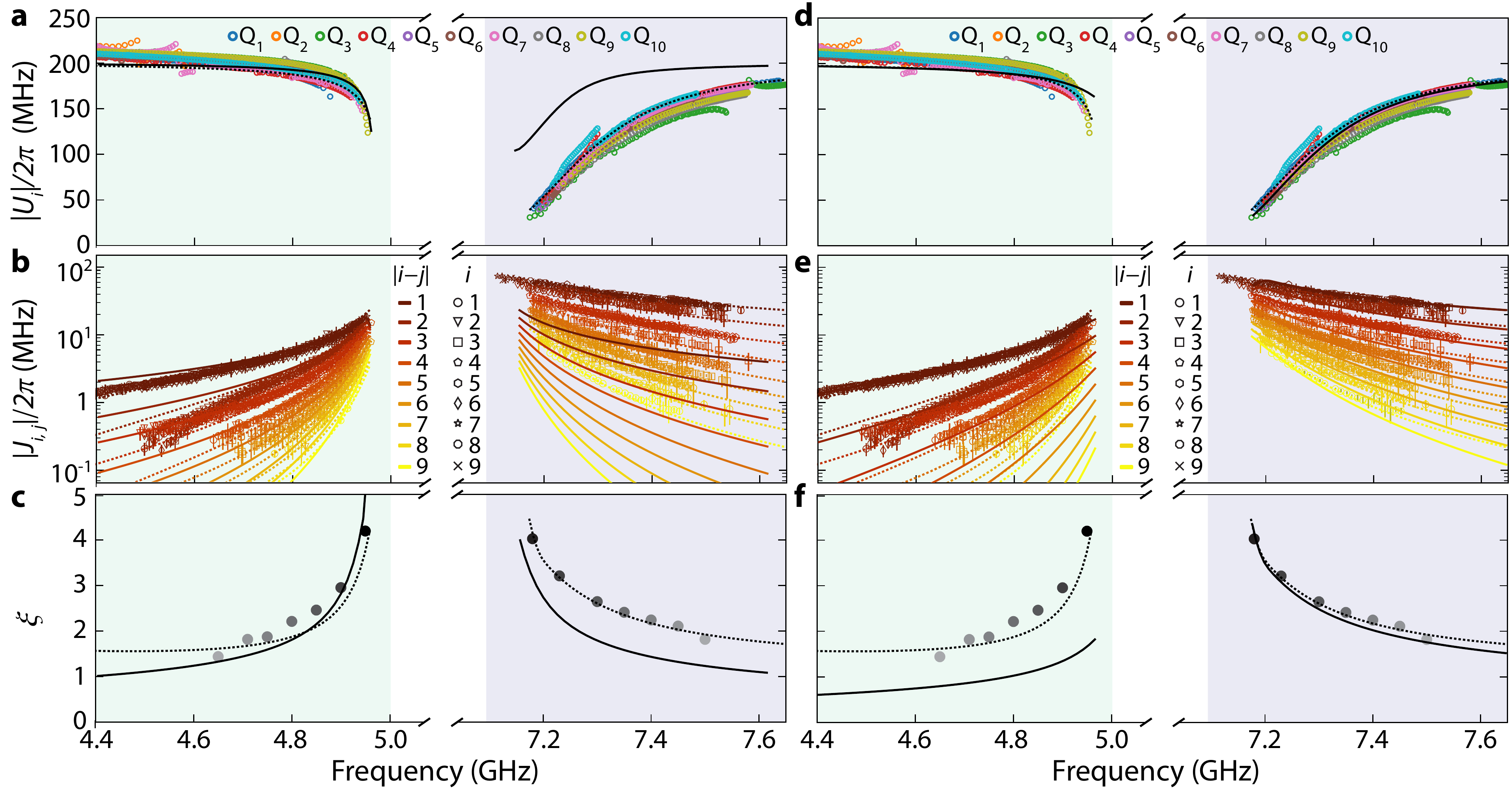}
\caption{\textbf{Comparison of numerical modeling with the experimental data.} 
Magnitude of on-site interaction $U_i$ (panels \textbf{a} and \textbf{d}), amplitude of hopping $J_{i,j}$ (panels \textbf{b} and \textbf{e}), and localization length $\xi$ (panels \textbf{c} and \textbf{f}) as a function of frequency with the solid curves in each panel showing the fitting based on numerical modeling. The fitting curves in panels \textbf{a}, \textbf{b}, and \textbf{c} are obtained from numerical optimization assuming the approximate tight-binding Hamiltonian in Eq.~\eqref{eq:quantized-Hamiltonian-real-space} and the panels \textbf{d}, \textbf{e}, and \textbf{f} assuming the nearest-neighbor-coupled circuit model with the form of Hamiltonian in Eqs.~\eqref{eq:numerical-model-Hamiltonian-diagonal-second-quantized} and \eqref{eq:numerical-model-Hamiltonian-offdiag-second-quantized}. The dotted curves are obtained from fitting assuming the capacitive and inductive coupling beyond nearest neighbor elements, the full model of which shown in Fig.~\ref{fig:mmschematic-numerical}. The experimental data (colored or grayscale markers) and the fit curves of the full model are identical to the ones in Fig.~\ref{fig:FigureN2} of the main text.
}
\refstepcounter{SIfig}\label{fig:model} 
\end{center}
\end{figure*}
\subsubsection{Fitting based on the numerical model}
\label{subsubsec:numerical}
With a set of electrical circuit parameters, we can perform numerical diagonalization of the exact Hamiltonian in Eqs.~\eqref{eq:numerical-model-Hamiltonian-diagonal-second-quantized} and \eqref{eq:numerical-model-Hamiltonian-offdiag-second-quantized}, enabling us to calculate various properties of our simulator and fit the experimental data.
Assuming a qubit couples to the middle unit cell of a 50-resonator metamaterial waveguide with an open boundary condition, we numerically diagonalize the Hamiltonian in the single- and two-excitation manifold to obtain the eigenfrequency $\omega_{01}$ ($\omega_{02}$) of the first (second) excited state $\ket{1}$ ($\ket{2}$) relative to the ground state $\ket{0}$ as a function of $E_J$. 
The on-site interaction can be calculated by $U = \omega_{02} - 2\omega_{01}$, which is shown in Fig.~\ref{fig:FigureN2}a of the main text.
The effective coupling $J_{i,j}$ between two transmon sites Q$_i$ and Q$_j$ is obtained by diagonalizing the Hamiltonian of two qubits coupled to two unit cells separated by a distance $|i-j|$ near the center of the 50-resonator metamaterial.
The two single-excitation eigenstates in the bandgaps correspond to the even and odd superposition of the two bound states with the eigenenergy difference (sum) of $2J_{i,j}$ ($2\omega_{01}$).
This allows us to numerically calculate $J_{i,j}$ as a function of $\omega_{01}$ shown in Fig.~\ref{fig:FigureN2}b of the main text. 
Fitting the exponential decay of $J_{i,j}$ as a function of distance $|i-j|$ at a fixed frequency $\omega_{01}$ gives the localization length $\xi$ plotted in Fig.~\ref{fig:FigureN2}c of the main text. 
The solid curves in Fig.~2a, b, and c are calculated from an identical set of circuit parameters, with $L_0=2.04\,$nH extracted from electromagnetic simulation (Sonnet\textsuperscript{\textregistered}), $C_{q\Sigma}=92.7\,$fF calculated from the average charging energy of $\overline{E_C}/h = e^2/(2hC_{q\Sigma})=220\,$MHz measured at the lower sweet spot (see Table.~\ref{tb:device-characterization}), and the free parameters obtained from numerical fitting of the experimental data (Nelder-Mead optimization) given by $C_0=242.19\,$fF, $C_{t,1}=60.17\,$fF, $C_{t,2}=0.542\,$fF, $M_1=-18.1\,$pH, $M_2=13.5\,$pH, $M_3=-1.09\,$pH, $M_4=0.438\,$pH, $C_{g,0}=9.19\,$fF, $C_{g,1}=0.368\,$fF, and $C_{qq,1}=10.1\,$aF. 
The long-range coupling capacitance assumes the form $C \propto 1/r^3$ where $r$ is the distance between two planar electrodes \cite{martinis2014calculation}, giving
\begin{align*}
C_{t, |x|}&=C_{t,2}(2/|x|)^3 \quad (|x|\ge 2)
\end{align*}
and the form of $C_{qq, |x|}$, $C_{g,|x|}$ ($|x|>1$) following the physical distance on the device.
In addition, the long-range mutual inductance between metamaterial resonators assumes $M_{|x|} \propto (-1)^x \ln{(1 + 1/|x|)}$ for $|x|\ge 4$, adapted from the form of mutual inductance between two parallel wires \cite{rosa1908self}.

The exact Hamiltonian in Eqs.~\eqref{eq:numerical-model-Hamiltonian-diagonal-second-quantized} and \eqref{eq:numerical-model-Hamiltonian-offdiag-second-quantized} taking into account both the long-range capacitance and mutual inductance was necessary to sufficiently explain the data in Fig.~2a, b, and c in both the LBG and the UBG. For example, assuming an approximate form of Hamiltonian in Eq.~\eqref{eq:quantized-Hamiltonian-real-space} with only nearest-neighbor coupling, we were not able to reproduce the wide tuning range of $U_i$ and large $|J_{i,j}|$ in the UBG (Fig.~\ref{fig:model}a, b, and c).
This indicates that the asymmetry between LBG and UBG partly originates from beyond-nearest-neighbor coupling in the Hamiltonian. Additionally, the capacitance ratio from the fitted parameters above is $C_{t,1}/(C_{0}+2C_{t,1}+C_{g,0}) = 0.162$, which suggests the invalidity of the small-coupling approximation in Eq.~\eqref{eq:Cinv_approx} in the derivation of this analytically solvable model.
Another numerical fitting using an exact Hamiltonian in Eqs.~\eqref{eq:numerical-model-Hamiltonian-diagonal-second-quantized} and \eqref{eq:numerical-model-Hamiltonian-offdiag-second-quantized} with zero capacitive and inductive coupling beyond nearest-neighbor, corresponding to the circuit diagram in Fig.~\ref{fig:mmschematic-analytic} with inductive coupling between adjacent resonators, also fails to fit the experimental data with a good agreement.
Although this model captures the behavior of $|U_i|$, $J_{i,j}$, and $\xi$ inside the UBG, as shown in Fig.~\ref{fig:model}d, e, and f, the long-range part in $J_{i,j}$ inside the LBG is underestimated, resulting in a smaller $\xi$ than the experimental results. Only when the full circuit model in Fig.~\ref{fig:mmschematic-numerical} is assumed can we recover the trend in the LBG, implying the importance of long-range capacitive and inductive coupling especially when the hopping strength $|J_{i,j}|$ is small. 

\section{Device characterization and experimental setup}
\label{sec:SII}
In this section, we summarize the details of the device and the experimental setup used in our work.

\renewcommand{\arraystretch}{1.3}
\begin{table*}[tbhp]
\begin{tabular}{P{4.0cm}P{2.3cm}|P{0.79cm}P{0.79cm}P{0.79cm}P{0.79cm}P{0.79cm}P{0.79cm}P{0.79cm}P{0.79cm}P{0.79cm}P{0.79cm}|P{0.79cm}P{0.79cm}} 
\hline\hline
\multicolumn{2}{c|}{Parameters} & Q$_1$ & Q$_2$ & Q$_3$ & Q$_4$ & Q$_5$ & Q$_6$ & Q$_7$ & Q$_8$ & Q$_9$ & Q$_{10}$ & Avg. & Stdev. \\ \hline 
\multicolumn{2}{c|}{Lower sweet spot frequency $\omega_{01,\mathrm{min}}/2\pi$\,(GHz)}
& 3.738 & 3.520 & 3.601 & 3.513 & 3.743 & 3.467 & 3.671 & 3.532 & 3.557 & 3.396 & 3.574 & 0.108 \\
\multicolumn{2}{c|}{Upper sweet spot frequency $\omega_{01,\mathrm{max}}/2\pi$\,(GHz)}
& 7.636 & 7.506 & 7.650 & 7.577 & 7.575 & 7.494 & 7.584 & 7.573 & 7.578 & 7.483 & 7.566 & 0.053 \\ \hline
Lifetime $T_1\,$($\upmu$s) & \multirow{4}{*}{\centering at $\omega_{01,\mathrm{min}}$}
& 7.56 & 75.4 & --   & 73.0 & 5.13 & 74.2 & 22.4 & 54.3 & 24.9 & 39.1 & 41.8 & 26.9 \\
Ramsey $T_2^*\,$($\upmu$s) & 
& 4.69 & 10.4 & --   & 0.81 & 1.48 & 13.6 & 6.97 & 16.7 & 4.80 & 16.2 & 8.4 & 5.7 \\
Hahn echo $T_{2E}$\,($\upmu$s) & 
& 10.8 & 70.1 & --   & 51.7 & 7.34 & 82.5 & 24.8 & 52.4 & 20.4 & 32.2 & 39.1 & 24.9 \\ 
Anharmonicity $U/2\pi$\,(MHz) &
& $-211$ & $-222$ &  --  & $-223$ & $-239$ & $-220$ & $-207$ & $-221$ & $-215$ & $-224$ & $-220$ & $9$ \\
\hline
Lifetime $T_1\,$($\upmu$s) & \multirow{4}{*}{\parbox{2.6cm}{\centering at $\omega_{01}/2\pi\approx 4.72\,$GHz}}
& 4.89 & 31.0 & 22.6 & 25.0 & 14.5 & 33.8 & 11.3 & 32.7 & 12.1 & 28.0 & 21.6 & 9.7 \\
Ramsey $T_2^*\,$($\upmu$s) & 
& 1.08 & 1.05 & 1.07 & 1.03 & 1.18 & 1.37 & 1.25 & 1.20 & 1.26 & 1.14 & 1.16 & 0.10 \\
Hahn echo $T_{2E}$\,($\upmu$s) & 
& 4.67 & 5.87 & 6.84 & 6.20 & 5.64 & 6.83 & 4.83 & 5.35 & 5.48 & 4.73 & 5.64 & 0.76 \\ 
Anharmonicity $U/2\pi$\,(MHz) &
& $-195$ & $-203$ & $-205$ & $-194$ & $-202$ & $-199$ & $-195$ & $-203$ & $-205$ & $-198$    & $-200$ & $4$ \\
\hline
Lifetime $T_1\,$($\upmu$s) & \multirow{4}{*}{\parbox{2.6cm}{\centering at $\omega_{01}/2\pi\approx 7.35\,$GHz}}
& 3.37 & 2.47 & 2.22 & 3.18 & 3.39  & 5.95 & 6.45 & 4.83 & 4.74 & 3.74 & 4.03 & 1.34\\
Ramsey $T_2^*\,$($\upmu$s) &
& 1.49 & 1.30 & 1.19 & 1.31 & 1.80 & 2.87 & 2.63 & 2.67 & 2.93 & 2.06 & 2.02 & 0.66 \\
Hahn echo $T_{2E}$\,($\upmu$s) &
& 3.18 & 1.89 & 1.31 & 1.73 & 3.28 & 4.53 & 6.26 & 4.83 & 5.56 & 3.28 & 3.59 & 1.59 \\ 
Anharmonicity $U/2\pi$\,(MHz) &
& $-134$ & $-126$ & $-112$ & $-131$ & $-132$ & $-125$ & $-129$ & $-118$ & $-123$ & $-134$ & $-126$ & $7$ \\ \hline
Lifetime $T_1\,$($\upmu$s) & \multirow{4}{*}{\parbox{2.6cm}{\centering at  $\omega_{01,\mathrm{max}}$}}
& 4.58 & 8.52 & 2.60 & 14.2 & 7.65  & 10.1 & 9.64 & 10.4 & 4.37 & 8.49 & 8.1 & 3.3 \\
Ramsey $T_2^*\,$($\upmu$s) &
& 5.53 & 5.78 & 3.02 & 12.0 & 7.34 & 5.67 & 9.54 & 10.6 & 6.09 & 10.5 & 7.6 & 2.7 \\
Hahn echo $T_{2E}$\,($\upmu$s) &
& 7.12 & 5.67 & 3.18 & 16.7 & 9.29 & 11.5 & 13.6 & 10.9 & 7.28 & 12.5 & 9.8 & 3.9 \\ 
Anharmonicity $U/2\pi$\,(MHz) &
& $-181$ & $-162$ & $-177$ & $-177$ & $-170$ & $-161$ & $-176$ & $-165$ & $-168$ & $-166$ & $-170$ & $7$ \\ \hline
\multicolumn{2}{c|}{Readout resonator frequency $\omega_r/2\pi$\,(GHz)}  &    5.833   &  6.084   &    6.328   &    5.574  &  6.008   &  5.907 & 5.622 & 6.236 & 6.169 & 5.741 & 5.950 & 0.245 \\
\multicolumn{2}{c|}{Readout resonator decay rate $\kappa_r/2\pi$\,(MHz)}  &    11.47   &  8.38   &    16.95  &    14.08  &  9.85   &  8.09 & 10.46 & 10.57 & 21.95 & 6.16 & 11.80 & 4.46 \\ \hline\hline
 \end{tabular}
\caption{\textbf{Basic characterization of qubits and readout resonators.} Various parameters of the qubits and the readout resonators used in this work are summarized. The last two columns of the table show the average and the standard deviation of each parameter over all qubits or readout resonators.}
\label{tb:device-characterization}
\end{table*}

\subsection{Device details}
\subsubsection{Qubit}
The ten transmon qubits in the device are designed to be nominally identical, with asymmetric Josephson junctions on superconducting quantum interference device (SQUID) loop to reduce sensitivity to flux noise while maintaining a tuning range wide enough to cover both the lower bandgap (LBG) and the upper bandgap (UBG). 
Each qubit is individually addressed by a charge drive line (XY control) with a designed capacitance of $C_{d}=80\,\mathrm{aF}$ and a flux bias line (Z control) with a mutual inductance of $M_{\Phi}\approx 1\,\Phi_0/\mathrm{mA}\approx 2\,\mathrm{pH}$ to the qubit's SQUID loop. 
The staggered qubit placement with respect to the metamaterial, i.e., Q$_{i}$ and Q$_{i + 1}$ located on the opposite sides of the metamaterial waveguide, is employed in order to minimize the parasitic near-field coupling between qubits. 
We perform basic characterization of each qubit in its entire tuning range inside the bandgaps by sweeping over DC current sent along its flux bias line while parking the remaining qubits inside the opposite bandgap. We measure the lifetime $T_1$, Ramsey coherence time $T_2^*$, Hahn echo coherence time $T_{2E}$, and anharmonicity $U$ as a function of qubit frequency, which are summarized in Table~\ref{tb:device-characterization} at a few different frequencies. 

\subsubsection{Readout resonator}
The compact readout resonators in our device consists of a meander inductor of $1\,\upmu$m pitch and a planar capacitor. They are designed to have frequencies near the center of the passband, enabling dispersive readout when the qubits are tuned inside the bandgaps. 
The resonant frequencies of readout resonators, controlled by the length of the meander inductor, are designed to have larger separation for physically adjacent readout resonators in order to avoid deleterious effects from parasitic near-field coupling.
The readout resonator frequency $\omega_r$ measured from waveguide transmission spectroscopy and the decay rate $\kappa_r$ extracted from ring-down measurement are shown in Table~\ref{tb:device-characterization}. The variations in the decay rates originate from the dispersion of the metamaterial waveguide. 
To achieve a high-fidelity readout, we design the coupling between a qubit and its readout resonator to be $g_{qr}/2\pi=250\,$MHz, giving the dispersive shift of $\chi/2\pi \approx 6\,$MHz when the qubit is parked at 4.5\,GHz or 7.5\,GHz.
\begin{figure*}[tbhp]
    \begin{center}
    \includegraphics[width=1.0\textwidth]{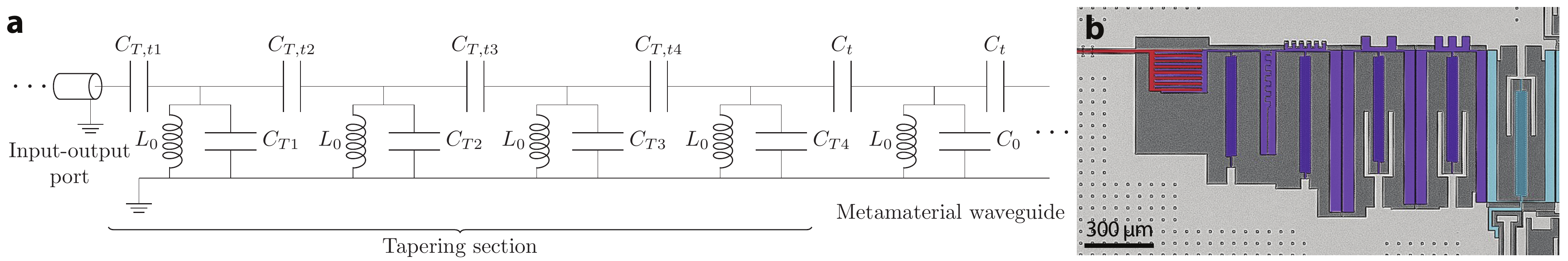}
    \caption{\textbf{Tapering section.} \textbf{a}, Schematic diagram of the tapering section consisting of four LC resonators coupled to an external input-output port on the left indicated with a transmission line symbol and the regular metamaterial waveguide section on the right. \textbf{b}, Optical micrograph (false colored) of the tapering section of the fabricated device. Here, the metamaterial waveguide, the tapering section, and the input-output port are colored blue, purple, and red, respectively.}
    \refstepcounter{SIfig}\label{fig:tapering}
    \end{center}
\end{figure*}
\subsubsection{Metamaterial waveguide}
As mentioned in the main text, the metamaterial waveguide consists of an array of nominally identical 42 compact resonators each formed by a meander inductor of 2$\,\upmu$m pitch and a planar capacitor. Each resonator of the ten inner unit cells of the metamaterial waveguide (labeled by $i=1$--10) is capacitively coupled to a qubit Q$_i$ and simultaneously to its readout resonator R$_i$. 
On each metamaterial resonator without a qubit, we keep the capacitors of a qubit and a readout resonator to maintain the total capacitance of the metamaterial resonator and minimize the discrepancy in resonator frequencies. 
The two metamaterial resonators close to the auxiliary qubits (colored yellow in Fig.~\ref{fig:FigureN1}e of the main text) have capacitors which are redesigned to compensate for the absence of qubit and readout resonator capacitors. The metamaterial waveguide is connected to external input-output ports via tapering sections.
\subsubsection{Tapering section}\label{sec:tapering-section}
The tapering section in our device is a network of inductors and capacitors designed to reduce impedance mismatch between the metamaterial waveguide and the external $50$-$\Omega$ input-output ports at the passband frequencies \cite{ferreira2021collapse, kim2021quantum}. This significantly decreases the level of ripples in the transmission spectrum near the center of the passband (shown in Fig.~\ref{fig:FigureN1}f of the main text), enabling the use of metamaterial waveguide as a resource-efficient feedline for qubit readout. We utilize a design illustrated in Fig.~\ref{fig:tapering}a where four LC resonators with an identical inductance $L_0$ are capacitively coupled to each other with gradually changing capacitance values given by $C_{T,t1}=222.8\,$fF, $C_{T1}=51.0\,$fF, $C_{T,t2}=77.3\,$fF, $C_{T2}=210.7\,$fF, $C_{T, t3}=53.8\,$fF, $C_{T3}=298.1\,$fF, $C_{T, t4}=65.3\,$fF, and $C_{T4}=293.1\,$fF. An optical micrograph of the tapering section realized in our device is shown in Fig.~\ref{fig:tapering}b. Note that our tapering design becomes less efficient at frequencies close to the band-edges, creating a dense spectrum of high-$Q$ modes. 
\subsection{Experimental setup}
\begin{figure}[tbhp]
\begin{center}
\includegraphics[width=0.5\textwidth]{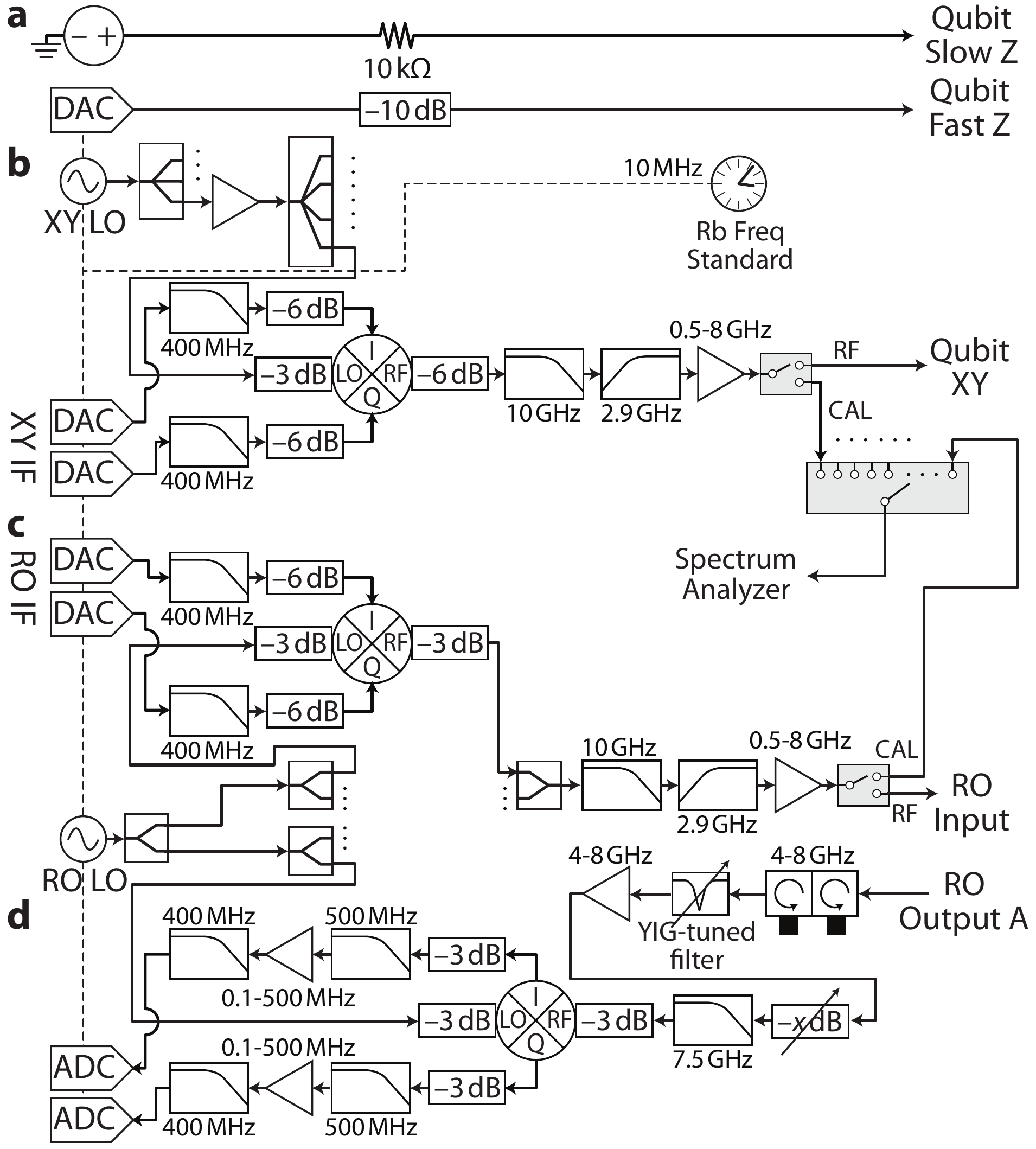}
\caption{\textbf{Schematic of the room-temperature electronic setup outside the dilution refrigerator.} The diagram for synthesizing the signals for qubit frequency control (slow and fast Z control, panel \textbf{a}), qubit drive (XY control, panel \textbf{b}), and readout input (panel \textbf{c}) are shown, together with the analog downconversion and the filtering procedures for the output readout signals (panel \textbf{d}). The manufacturers and the model numbers of the parts and the instruments used in the diagram are enumerated in Sec.~\ref{sec:room-temp-electronics}.}
\refstepcounter{SIfig}\label{fig:room-temp-electronics} 
\end{center}
\end{figure}

\subsubsection{Room-temperature electronics}\label{sec:room-temp-electronics}
The electronic setup at the room temperature for synthesis of qubit control signals and processing of qubit readout signals is illustrated in Fig.~\ref{fig:room-temp-electronics}.
Static component of qubit frequency control (slow Z) signal is generated by a stable DC voltage source (QDevil, QDAC) passed through a 10\,k$\Omega$ resistor. Dynamic qubit frequency control (fast Z) is achieved by employing arbitrary waveforms generated from a digital-to-analog converter (DAC) channel with an analog bandwidth of 400\,MHz at 1\,ns temporal resolution, sent through a 10\,dB attenuator before entering the dilution refrigerator. For microwave synthesis of drive signals on each qubit (XY), we prepare a pair of intermediate frequency (IF) signals from DAC channels passed through a low-pass filter with 400\,MHz cutoff. The pair of IF signals are attenuated and multiplied to a local oscillator (LO) signal generated by a microwave signal generator (Rohde \& Schwarz, SMB100A) by using a IQ mixer (Marki Microwave, MMIQ-0218L) for upconversion, enabling synthesis of signals in approximately 800\,MHz-wide frequency band about the LO frequency. 
This is followed by attenuation and filtering with a low-pass filter with 10\,GHz cutoff (Mini-Circuits, ZXLF-K14+) and a high-pass filter with 2.9\,GHz cutoff (Mini-Circuits, VXHF-292M+) and a subsequent low-noise amplification (Mini-Circuits, ZX60-83-LN-S+). Note that we distribute the LO signal generated from a single, common microwave signal generator to multiple IQ mixers used for synthesis of qubit drive signals by using a suitable combination of power splitters and amplifiers. Together with synchronized DAC channels used on all qubits, this ensures that the phase relation between qubit XY drive signals sent to different qubits is deterministic and constant over the repetition of experiments. The readout input signals are synthesized in a way similar to the XY signals by IQ mixing with a separate LO dedicated for readout. Each qubit XY and readout input signals are optionally directed to a spectrum analyzer by digitally controlled microwave switches, enabling calibration of each IQ mixers to suppress their spurious LO and image leakage tones. The readout output signals from the refrigerator are passed through a room-temperature dual-junction circulator (Fairview Microwave, FMCR1019) and a yttrium iron garnet (YIG)-tuned band-reject filter (Micro Lambda Wireless, MLBFR-0212) to suppress leakage tones at the JTWPA pump frequency. This signal is then amplified with a high-gain low-noise amplifier (L3Harris Narda-MITEQ, LNA-40-04000800-07-10P), a tunable attenuator (Vaunix, Lab Brick LDA-133), and a low-pass filter with 7.5\,GHz cutoff (Marki Microwave, FLP-0750). Then, the signals are downconverted to IF band by using IQ mixers pumped by the same readout LO used for upconversion of readout input signals and suitably filtered (Mini-Circuits, VLFX-400+ and VLFX-500+) and amplified (Mini-Circuits, ZFL-500LN+) before digitization at analog-to-digital converter (ADC) channels. The DAC and ADC channels used in this work are analog output and analog input channels, respectively, of Quantum Machines OPX+. All the microwave instruments used in our work are referenced to an external 10\,MHz reference from a Rubidium frequency standard (Stanford Research Systems, FS725).

\begin{figure}[t!]
\begin{center}
\includegraphics[width=0.5\textwidth]{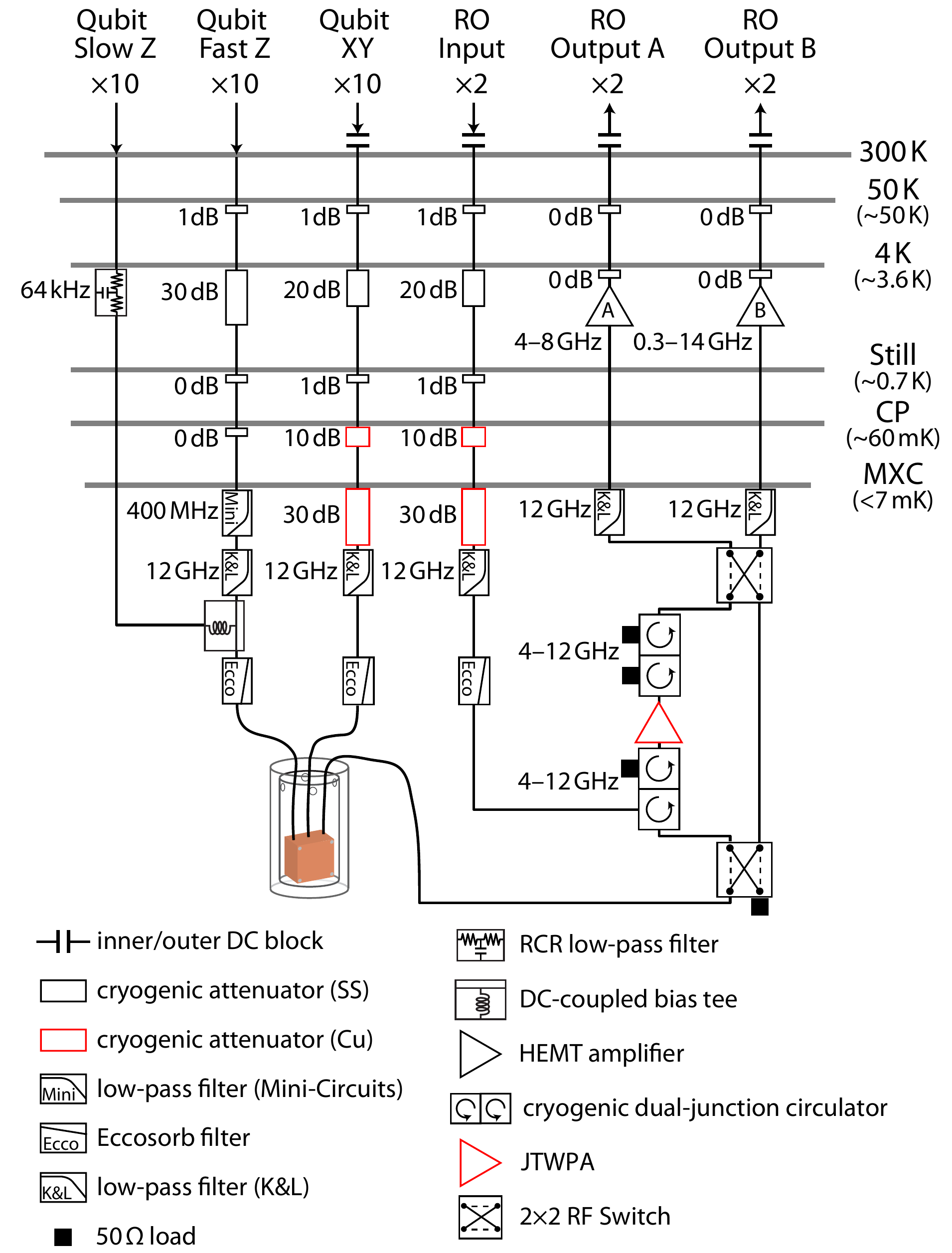}
\caption{\textbf{Schematic of the experimental setup inside the dilution refrigerator.} The cryogenic setup consists of slow and fast Z lines for qubit frequency control, XY lines for qubit drive, and RO input/output lines for qubit readout. 
The meaning of the symbols used in the diagram is enumerated at the bottom. The attenuation values of cryogenic attenuators, cutoff frequencies of low-pass filters, and bandwidths of circulators and amplifiers are shown next to the components. The input pump line for JTWPA's are not shown in the diagram for brevity.  The manufacturers and the model numbers of the parts used in the diagram are enumerated in Sec.~\ref{sec:cryogenic-setup}.}
\refstepcounter{SIfig}\label{fig:fridge} 
\end{center}
\end{figure}

\subsubsection{Cryogenic setup}\label{sec:cryogenic-setup}
The experimental setup inside the cryogen-free dilution refrigerator (Bluefors, BF-LD250) used in our work is illustrated in Fig.~\ref{fig:fridge}. The refrigerator consists of multiple temperature stages named 50\:K flange, 4\:K flange, still flange, cold plate (CP), and mixing chamber (MXC) flange with typical temperatures of 50\,K, 3.6\,K, 0.7\,K, 60\,mK, and under 7\,mK, respectively, during normal operations. The frequency of each qubit is controlled statically by a DC bias (slow Z) passing through a resistor-capacitor-resistor (RCR) low-pass filter (Aivon Ltd., Therma-uD25-G) with 64$\,$kHz cutoff thermalized to the 4\,K plate and dynamically by RF pulses (fast Z) passing through a series of XMA cryogenic attenuators (2082-6418-$\square\square$-CRYO) with stainless steel enclosure whose attenuation values are given by 1\,dB, 30\,dB, 0\,dB and 0\,dB at 50\,K, 4\,K, Still and CP stages, respectively. This is followed by reflective low-pass filtering with a $400\,$MHz filter (Mini-Circuits, VLFX-400+) and a 12\,GHz ``clean-up'' filter (K\&L Microwave, 6L250-12000/T26000-OP/O) at the mixing chamber plate. 
The DC bias and the RF pulses for Z control are then combined with a DC-coupled bias tee obtained by shorting the capacitor of a Mini-Circuits ZFBT-4R2GW+ bias tee, followed by infrared filtering with an Eccosorb filter at the mixing chamber before connecting to the flux bias line on the device. 
The microwave signals for XY drive of each qubit is attenuated by 1\,dB, 20\,dB, and 1\,dB at the 50\,K, 4\,K, and Still stages, respectively, using the XMA cryogenic attenuators. Also, 10\,dB and 30\,dB cryogenic attenuators made with oxygen-free high-conductivity (OFHC) copper enclosure (Quantum Microwave, QMC-CRYOATT-$\square\square$) are placed at the CP and MXC stages, respectively, in order to achieve good thermalization at temperatures below 100\,mK (similar to the attenuators in Ref.~\cite{yeh2017microwave}). This is followed by infrared filtering by a 12$\,$GHz K\&L low-pass filter and an Eccosorb filter before entering the device. 
A pair of readout input lines go through the same attenuation and filtering as the XY drive lines and connect to wideband dual-junction circulators (Low-Noise Factory, LNF-CICIC4\_12A).
Each circulator connects to an input-output port of the metamaterial waveguide via a 2$\times$2 RF switch (Radiall, R577432000), playing the role of both sending waveguide input signal into the device and directing waveguide output signal to amplification chains of readout output lines. 
The output signal is amplified by a Josephson traveling-wave parametric amplifier (JTWPA) from MIT Lincoln Laboratory sandwiched by two sets of dual-junction circulators at the mixing chamber and then a high-electron-mobility transistor (HEMT) amplifier (Low-Noise Factory, LNF-LNC4\_8C and LNF-LNC0.3\_14A for amplification in the 4--8\,GHz and 0.3--14\,GHz bands, respectively) at the 4K flange before further amplification at room temperature. The two external ports (1 and 2) of the metamaterial waveguide connect to two individual sets of input-output line and amplification chain. With four $2\times 2$ RF switches in total, we can configure the setup to measure both transmission (S$_{21}$ and S$_{12}$) and reflection (S$_{11}$ and S$_{22}$) from both waveguide ports with an option to bypass JTWPA in transmission measurements. Also, with individual amplification chains, the transmitted and the reflected signals can be measured simultaneously (e.g., S$_{21}$ and S$_{11}$). We place inner/outer DC blocks (Inmet, 8039 or Centric RF, CD9519) along all the cryogenic coaxial lines before connection to room-temperature electronics to break ground loops, except for the qubit Z control lines where a full bandwidth from DC to 400\,MHz is necessary in order to  minimize pulse distortion.

\section{Details of the metamaterial Purcell filter}
\label{sec:SIII}
\begin{figure*}[t!]
\begin{center}
\includegraphics[width=\textwidth]{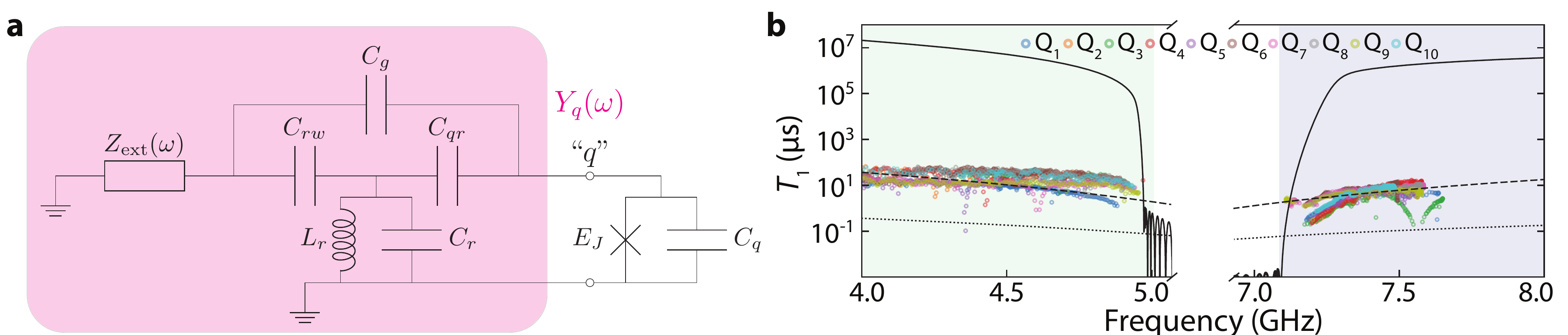}
\caption{\textbf{Metamaterial Purcell filter.} 
\textbf{a}, General circuit model for dispersive qubit readout. A qubit, represented as a parallel circuit of a Josephson junction (Josephson energy $E_J$) and a capacitor (capacitance $C_q$), couples to an external port with impedance $Z_\mathrm{ext}(\omega)$ via a readout resonator (inductance $L_r$ and capacitance $C_r$). The circuit elements in the shaded region is taken into account when calculating the admittance $Y_q(\omega)$ seen from the qubit node ``$q$". \textbf{b}, Qubit lifetime $T_1$ plotted against frequency. Calculated Purcell-limited lifetime is shown (with readout resonator frequency $\omega_\mathrm{R_5}/2\pi=6.01\,$GHz, decay rate $\kappa_\mathrm{R_5}/2\pi=9.85\,$MHz, and qubit-resonator coupling of 250$\,$MHz) when the readout resonator couples to an external port directly (dotted curve), via a single-pole Purcell filter with $Q = 15$ (dashed curve), and via the metamaterial Purcell filter in this device (solid curve). The measured $T_1$ of ten qubits are shown in colored circles. The lower and the upper bandgaps of the metamaterial waveguide are shaded green and purple, respectively.}
\refstepcounter{SIfig}\label{fig:Purcell} 
\end{center}
\end{figure*}

Besides providing a frequency band for readout, the metamaterial waveguide also acts as a hardware-efficient Purcell filter. 
The dispersion of the metamaterial and the large number of unit cells in this realization effectively prevent qubits at bandgap frequencies from accessing the external ports, as evidenced by the large extinction ratio and sharp transition at the band-edges illustrated in Fig.~\ref{fig:FigureN1}f of the main text.
Here, both the direct decay of qubits into the metamaterial and the Purcell decay via readout resonators are strongly suppressed. To show this, we compare in this section the performance of our metamaterial Purcell filter with traditional circuit QED settings \cite{Krantz:2019, Blais:2021} where readout resonators are coupled to external ports directly or via a single-pole Purcell filter. 

\subsection{Purcell decay}
A general scheme for qubit readout in circuit QED involves a qubit coupled to a readout resonator which could be accessed from an external port, an example electrical circuit of which is illustrated in Fig.~\ref{fig:Purcell}a. In the dispersive regime where the qubit-resonator coupling $g_{qr}$ is small compared to the magnitude of their detuning $\Delta_{qr} \equiv \omega_{01} - \omega_r$, the frequency of the readout resonator depends on the state of the qubit, enabling quantum non-demolition readout \cite{blais2004cavity, Schuster:2005}. However, this coupling simultaneously introduces an unwanted qubit decay channel into the external port via the readout resonator, known as the Purcell decay, which could be mitigated by engineering a Purcell filter \cite{houck2008controlling, reed2010fast, Jeffrey:2014} that suppresses the density of states at the qubit frequency. The rate of Purcell decay when the readout resonator is coupled to an external port with impedance $Z_\mathrm{ext}(\omega)$ is given by \cite{cleland2019mechanical}
\begin{equation}
\label{eq:gamma}
    \Gamma_{1}^\mathrm{Purcell} = \frac{g_{qr}^2}{\Delta_{qr}^2}\frac{\mathrm{Re}[Z_\mathrm{ext}(\omega_{01})]}{\mathrm{Re}[Z_\mathrm{ext}(\omega_r)]}\kappa_r,
\end{equation}
where $\kappa_r$ is the decay rate of the readout resonator. 


\subsection{Modeling of the metamaterial Purcell filter}
In the case of a metamaterial waveguide, the external impedance $Z_\mathrm{ext}(\omega)$ is highly frequency-dependent with an impedance close to $Z_0=50\,\Omega$ inside the passband after tapering and in principle a purely imaginary Bloch impedance inside the bandgaps \cite{Pozar:2012}. 
As a result, the Purcell decay rate $\Gamma_{1}^\mathrm{Purcell}$ in Eq.~\eqref{eq:gamma} is strongly suppressed by the ratio of the real part of the external impedance at the qubit transition frequency $\omega_{01}$ to that at the readout resonator frequency $\omega_r$.

In our circuit construction, the real part $\mathrm{Re}[Z_\mathrm{ext}(\omega_{01})]$ of external impedance at the qubit transition frequency is not zero due to finite number of unit cells. 
Taking into account the direct coupling capacitance $C_g$ between the qubit and the metamaterial waveguide (see Fig.~\ref{fig:Purcell}a), we calculate the qubit decay rate into the external port by using the relation \cite{houck2008controlling, reed2010fast}
\begin{equation}
\label{eq:gamma1}
    \Gamma_{1}^\mathrm{Purcell} = \frac{\mathrm{Re}[Y_{q}(\omega_{01})]}{C_{q\Sigma}},
\end{equation}
where $Y_{q}(\omega)$ is the admittance seen from the qubit node ``$q$'' illustrated in Fig.~\ref{fig:Purcell}a. 
To numerically evaluate the expression, we use the readout resonator R$_5$ as an example with parameters $C_{r}=130.5\,$fF and $L_{r}=4.518\,$nH,  together with metamaterial parameters from the numerical fit results discussed in Sec.~\ref{subsubsec:numerical} and the designed coupling capacitance values $C_{qr}=10.3\,$fF, $C_{rw}=6.8\,$fF. 
Note that the admittance $Y_{q}(\omega)$ also depends on the details of the tapering section, design parameters of which are enumerated in Sec.~\ref{sec:tapering-section}. 
With the set of circuit parameters listed above, we utilize AWR Microwave Office\textsuperscript{\textregistered} to calculate the Purcell-limited lifetime $T_{1}^\mathrm{Purcell}=1/\Gamma_{1}^\mathrm{Purcell}$ based on Eq.~\eqref{eq:gamma1}, indicated by the black solid curve in Fig.~\ref{fig:Purcell}b.

\subsection{Comparison to traditional qubit readout settings}
For comparison, we also consider two conventional qubit readout scenarios in circuit QED where the readout resonator is coupled to an external port directly or via a single-pole Purcell filter ($C_g=0$ is assumed in these cases). 
\subsubsection{Direct coupling}
In the case of direct coupling of a readout resonator to an external port, the external impedance is simply given by $Z_\mathrm{ext}(\omega)=Z_0=50\,\Omega$ with no frequency dependence and hence Eq.~\eqref{eq:gamma} recovers the basic form of the Purcell decay \cite{blais2004cavity} at a rate of 
\begin{equation}
    \Gamma_{1,\mathrm{direct}}^\mathrm{Purcell}=\left(\frac{g_{qr}}{\Delta_{qr}}\right)^2\kappa_r,
\end{equation}
which is utilized to obtain the dotted curve in Fig.~\ref{fig:Purcell}b assuming the capacitance values listed above. 
\subsubsection{Single-pole Purcell filter}
In the case of a single-pole Purcell filter \cite{Jeffrey:2014, sete2015quantum} consisting of a resonator with decay rate $\kappa_r$ at frequency $\omega_r$, it can be shown that the external impedance is given by the form \cite{Pozar:2012}
\begin{equation}
Z_\mathrm{ext}(\omega) = \frac{Z_0}{1 + 2j(\omega-\omega_f)/\kappa_f}. \label{eq:Zext-singlepolePurcell}
\end{equation}
Substituting Eq.~\eqref{eq:Zext-singlepolePurcell} into Eq.~\eqref{eq:gamma} gives the Purcell decay rate of
\begin{equation}
    \Gamma_{1,\text{single-pole}}^\mathrm{Purcell} = \frac{g_{qr}^2}{\Delta_{qr}^2}\frac{1}{1 + (2\Delta_{qf} / \kappa_f)^2}\tilde{\kappa}_r,
\end{equation}
where $\Delta_{qf}=\omega_{01} - \omega_f$ is the detuning of the qubit from the filter resonator and $\tilde{\kappa}_r=\kappa_r \{1 + [2(\omega_r - \omega_f) / \kappa_f]^2\}$ is the decay rate of the readout resonator modified by the presence of the filter resonator. 
Here, we need to choose a bandwidth of the filter wide enough to accommodate ten readout resonators, playing a role similar to our metamaterial waveguide. 
As an example, we assume a filter quality factor of $Q_f=\omega_f/\kappa_f=15$ (half the value used in Ref.~\cite{Jeffrey:2014}), corresponding to a full-width half-maximum linewidth of $\kappa_f/2\pi\approx400$\,MHz, with the Purcell filter resonant to the readout resonator at $\omega_{\mathrm{R}_5}/2\pi=6.01\,$GHz. The expected Purcell-limited lifetime using this set of parameters for the single-pole Purcell filter is indicated by the dashed curve in Fig.~\ref{fig:Purcell}b.


\subsection{Discussion}

In Fig.~\ref{fig:Purcell}b, the Purcell-limited qubit lifetime is shown to be below 1$\,\upmu$s in the case of direct coupling and approximately 10$\,\upmu$s in the case of a single-pole Purcell filter \cite{Jeffrey:2014} with $Q = 15$, suggesting that such strategies for mitigating qubit decay are incompatible with the desired rapid qubit readout.  However, with the metamaterial waveguide acting as a Purcell filter, the expected decay rate of a qubit into the external ports is suppressed by more than five orders of magnitude, lifting the Purcell limitation on qubit lifetime.  The measured lifetime $T_1$ of qubits in our device lies in the range of 10.3--47.2$\,\upmu$s (3.6--9.8$\,\upmu$s) at 4.5$\,$GHz (7.45$\,$GHz) in the LBG (UBG), which surpasses the Purcell limit of the aforementioned traditional settings but remains far below the ideal prediction due to other material-related decay channels.

\section{Qubit readout methods and characterization} \label{sec:RO}
In this section, we provide details of the qubit readout procedures used in this work and additional readout characterization results.

\subsection{Qubit state discrimination}\label{sec:qubit-state-discrimination}
As discussed in the main text, we utilize both the reflected and the transmitted fields from readout resonators to perform qubit state discrimination in order to achieve the best quantum efficiency available in our setup. In the simplest case of a readout resonator symmetrically coupled to a readout feedline \cite{barends2013coherent}, qubit state information from the readout resonator is equally divided into the forward (transmitted) and the backward (reflected) directions. Therefore, collection of both the reflected and the transmitted fields is expected to give readout signal-to-noise ratio (SNR) about 3\,dB larger than what could be obtained by using only one of the two fields. In our device, however, we observe frequency-dependent asymmetric coupling of readout resonators to the two external input-output ports associated with the dispersion of the metamaterial waveguide. Here, the transfer functions from a qubit to the external ports becomes asymmetric away from the eigenfrequencies of the metamaterial. While the tapering sections help mitigate this effect by reducing the sharpness of the transmission response, the ratio of forward and backward decay rates of readout resonators is expected to be as large as ten at certain frequencies under numerical modeling. For these reasons, it is essential to collect the readout signals from both input-output ports of the device for qubit state discrimination, which gives a smooth experimental workflow with the highest readout SNR without concerns about this undesired asymmetry.

In order to achieve the highest readout SNR from the addition of the two readout fields, we pass the two fields through independent analog signal processing branches consisting of near-quantum-limited amplification, filtering, and downconversion, which gives a pair of complex-valued IF signals $\boldsymbol{s}_p(t)= \boldsymbol{a}_{p}(t) e^{i\omega_\mathrm{IF} t}$ from the ports $p=1,2$. Here, $\boldsymbol{a}_{p}(t)$ denotes the complex-valued baseband waveform and $\omega_\mathrm{IF}$ is the IF frequency of the readout. The pair of IF signals are independently digitized into discrete-time waveforms $\boldsymbol{s}_p(t_m)$ ($m=0,1,\cdots,M-1$), which undergo digital demodulation
\begin{align}
\boldsymbol{x}_p = \sum_{m=0}^{M-1} \boldsymbol{w}_p^*(t_m)\boldsymbol{s}_p(t_m) e^{-i\omega_\mathrm{IF}t_m} = \sum_{m=0}^{M-1} \boldsymbol{w}_p^*(t_m)\boldsymbol{a}_p(t_m)\label{eq:digital-demodulation}
\end{align}
weighted by complex-valued integration weights $\boldsymbol{w}_p(t_m)$. Here, the integration weights $\boldsymbol{w}_p(t_m)$ can be chosen to be either a constant, e.g., $\boldsymbol{w}_p(t_m)=1$, or samples optimized for maximum separability of qubit-state-dependent readout signals \cite{ryan2015tomography, magesan2015machine}. The demodulated complex scalar variables $\boldsymbol{x}_p$ are then summed with weights $\boldsymbol{v}_p$ according to
$\boldsymbol{x} = \sum_{p} \boldsymbol{v}_p^* \boldsymbol{x}_p,
$
followed by thresholding to discriminate the state of a qubit, which is based on a pre-calibrated distribution of $\boldsymbol{x}$ obtained with initialization of the qubit in its standard basis states. The optimal values of the weights $\boldsymbol{v}_p$ maximizing the separability in $\boldsymbol{x}$ can be obtained from the distribution of demodulated scalars $\boldsymbol{x}_p$ of each port by performing linear discriminant analysis (LDA) \cite{fisher1936, scikit-learn}. Our method is superior to the approach used in Ref.~\cite{wang2021improved} where the placement of a power combiner before the first amplification stage led to reduction in SNR by at least 3\,dB, counteracting the 3\,dB gain expected from the addition. Also,  our method of applying digital signal processing to combine the two signals takes into account the phase and gain imbalance of the independent branches, allowing for noise-matched and phase-coherent addition of the readout signals, with the effective SNR being the sum of SNR available from each branch.

Utilizing a FPGA-based control architecture (Quantum Machines OPX+) to synthesize and analyze readout signals, we carry out the digital signal processing procedures mentioned above in real time, including weighted demodulation of digitized IF waveforms, weighted sum of demodulated quadrature variables, and thresholding. This allows us to perform low-latency ($<1\,\upmu$s) feedback control over qubits conditioned on measurement outcomes, a demonstration of which in the form of active qubit reset is described in Sec.~\ref{sec:qubit-reset}.

\subsection{Readout control parameter tune-up}\label{sec:readout-tuneup}
An important prerequisite to high-fidelity single-shot readout of a qubit is to find a set of parameters that determine pulses sent to a readout resonator. In a regular setting \cite{Krantz:2019}, a frequency between the readout resonator frequencies $\omega_r^{\ket{0}}=\omega_r + \chi$ and $\omega_r^{\ket{1}}=\omega_r - \chi$ under qubit preparation in states $\ket{0}$ and $\ket{1}$, respectively, is chosen to maximize the $\ket{0}$-$\ket{1}$ signal separation at a fixed readout power. This is followed by a readout power sweep at this frequency to find an optimal power that maximizes the readout fidelity while being unaffected by parasitic state transitions associated with the measurement \cite{slichter2012measurement, sank2016measurement}. In our experiment, however, the dispersion of the metamaterial waveguide affects the validity of this method in two ways. First, the colorful transmission response of the metamaterial waveguide on the order of a few dB prohibits us from making a frequency sweep with a fixed power arriving at the resonator, making it challenging to find the optimal readout frequency. Also, as noted in Sec.~\ref{sec:qubit-state-discrimination}, the dispersion of the metamaterial waveguide causes the decay rate of readout resonator to be highly frequency-dependent in cases with a large dispersive shift $\chi$. In this case, the intra-resonator photon occupation $\overline{n}_r \propto 1/[\Delta_r^2 + (\kappa_r/2)^2]$ during the readout \cite{gambetta2006qubit} becomes highly asymmetric between the qubit states $\ket{0}$ and $\ket{1}$, making the behavior associated with readout-induced state transition hard to predict.

Due to the challenges mentioned above, we devise a more general readout tune-up procedure based on a fast real-time sweep over both frequency and power of the readout. Given a readout pulse envelope, we sweep over frequency and power of readout to find a combination that maximizes the readout fidelity without parasitic qubit state transitions. This is achieved by repeating the qubit readout under initialization of qubit in each standard basis state $\{\ket{0},\ket{1}\}$ for $n_\mathrm{rep}$ times. From the resulting distribution of demodulated quadrature variables $\boldsymbol{x}_p$ in Eq.~\eqref{eq:digital-demodulation}, we perform LDA \cite{scikit-learn} to extract the assignment fidelity and outlier counting for estimating the probability of parasitic state transitions, similar to the method discussed in Ref.~\cite{chen2018metrology}, at each combination of readout frequency and power. Specifically, we choose the first $n_\mathrm{rep}/2$ samples in the dataset to train the LDA discriminator and the remainder are used to determine the readout fidelity with out-of-sample validation at a low statistical error \cite{abu2012learning}. For outlier counting, we draw a $3\sigma$-radius circle  with respect to the center of distribution calculated from the LDA and compute the portion of datapoints lying outside the regions enclosed by the circles. We choose a frequency and power combination that realizes the maximum readout fidelity, simultaneously having an outlier probability below a certain user-defined threshold.


\begin{table*}[tbhp]
\begin{tabular}{P{5.0cm}|P{0.79cm}P{0.79cm}P{0.79cm}P{0.79cm}P{0.79cm}P{0.79cm}P{0.79cm}P{0.79cm}P{0.79cm}P{0.79cm}|P{0.79cm}P{0.79cm}} 
\hline\hline
{Parameters} & Q$_1$ & Q$_2$ & Q$_3$ & Q$_4$ & Q$_5$ & Q$_6$ & Q$_7$ & Q$_8$ & Q$_9$ & Q$_{10}$ & Avg. & Stdev.\\ \hline
Qubit frequency $\omega_{01}/2\pi$\,(GHz)         & 4.02  & 4.78  & 4.44  & 4.20  & 4.42  & 4.48  & 3.80  & -- & 4.78  & 4.50 \\ 
Readout frequency $\omega_\text{RO}/2\pi$\,(GHz)  & 5.845 & 6.122 & 6.339 & 5.615 & 6.047 & 5.935 & 5.687 & -- & 6.196 & 5.788 \\
Readout fidelity $\mathcal{F}_\mathrm{1Q}$        & 0.983 & 0.980 & 0.974 & 0.991 & 0.983 & 0.985 & 0.984 & -- & 0.968 & 0.989 & 0.982 & 0.007 \\
 \hline\hline
 \end{tabular}
\caption{\textbf{Characterization of rapid single-qubit readout.} The highest readout fidelity achieved for each qubit, utilizing a rapid readout method described in Sec.~\ref{sec:rapid_readout}, is summarized with corresponding qubit and readout frequencies used for the characterization. }
\label{tb:single-qubit-readout}
\end{table*}

This sweep is first coarsely performed with a small number of repetitions $n_\mathrm{rep}\sim 10^{2}$, where reset is performed by waiting about 400\,$\upmu$s for the qubit to relax between pulse sequences. The next round of the sweep is performed with a larger number of repetition $n_\mathrm{rep}\sim 10^{4}$ at a much higher repetition rate  to reduce statistical fluctuations, efficiently implemented by employing active reset (see Sec.~\ref{sec:qubit-reset}) based on the single-shot readout calibrated from the first round of the sweep. Optionally, an additional round of a finer version of the sweep is performed to find a condition that achieves the highest readout fidelity. The total time spent on this tune-up procedure is about 2--4 minutes, limited by communication latency associated with data retrieval from the stream processor of OPX$+$.

For multiplexed readout, we follow a similar tune-up procedure involving simultaneous real-time sweep over the frequency and power sent into all ten readout resonators, whose sweep ranges are centered at the optimal condition determined from the single-qubit readout calibration described above. Such multiplexed readout calibration helps avoid spurious processes associated with parasitic near-field coupling between readout resonators and qubits as well as mitigate the decrease in readout fidelity due to the non-linearity of the measurement chain.

\subsection{Readout characterization}
Here, we summarize the characterization results of single-shot readout  achieved in our system.

\subsubsection{Rapid single-qubit readout}\label{sec:rapid_readout}
We quantify the performance of single-shot readout of each qubit with its assignment fidelity $\mathcal{F}_\mathrm{1Q}$, which is obtained by preparing the qubit in its standard basis states $\{\ket{0}, \ket{1}\}$ followed by measurement, repeated over $n_\mathrm{rep}\sim 10^{5}$ experimental counts. 
A rapid readout is achieved by employing a 148\,ns-long square pulse with an initial 20\:ns kick with two times large amplitude for fast ring-up \cite{mcclure2016rapid}, convolved with a Gaussian envelope of $10$\,ns standard deviation.
The resulting output readout signals are demodulated and integrated with optimal weights \cite{ryan2015tomography, magesan2015machine}. 
Here, we use a 100\,ns-longer integration window to collect transient readout signals associated with ring-up and ring-down of the readout resonator, timescales of which are about $1/\overline{\kappa_{\text{R}_i}}\approx 13.5\,$ns (see Table~\ref{tb:device-characterization}).

The characterization results of the highest readout fidelity achieved on each qubit, on par with the state-of-the-art performance achieved in superconducting quantum circuits \cite{Walter:2017, Heinsoo:2018}, are summarized in Table~\ref{tb:single-qubit-readout}.

\begin{figure*}[t!]
\begin{center}
\includegraphics[width=1\textwidth]{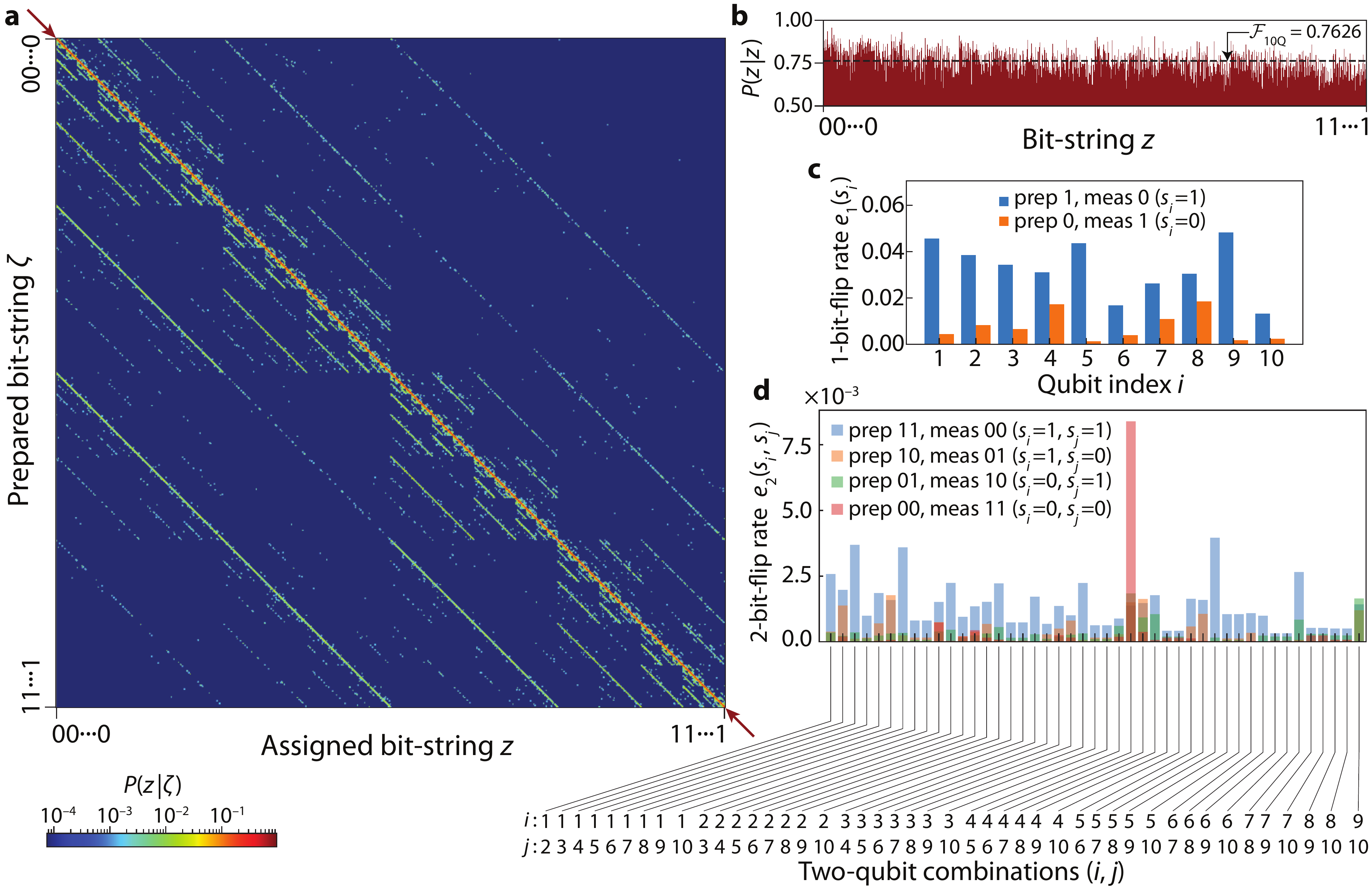}
\caption{\textbf{Characterization of multiplexed readout of qubits.} 
\textbf{a}, Assignment probability matrix $P(z|\zeta)$ of prepared bit-strings $\zeta$ and assigned bit-strings $z$ arranged in the ascending order. \textbf{b}, The diagonal elements $P(z|z)$ of the assignment probability matrix, indicated by the dark red arrows in panel \textbf{a}, is plotted as a function of 10-bit strings $z$ in the ascending order. The dashed line indicates the average of diagonal elements over all bit-strings, corresponding to the 10-qubit readout fidelity $\mathcal{F}_\text{10Q}=0.7626$. \textbf{c}, Single-qubit bit-flip error rate $e_1(s_i)$ during the readout, given by the average of assignment probability matrix elements corresponding to preparation of a state $s_i$ and assignment of the flipped state $\neg{s_i}$ on qubit Q$_i$ while the states of the remaining qubits are fixed. \textbf{d}, Two-qubit bit-flip error rate $e_2(s_i, s_j)$ during the readout, corresponding to the average of assignment probability matrix elements corresponding to preparation of states $(s_i, s_j)$ and assignment of the flipped states $(\neg{s_i}, \neg{s_j})$ on qubits (Q$_i$,Q$_j$) while the states of the remaining qubits are fixed.}
\refstepcounter{SIfig}\label{fig:multiRO} 
\end{center}
\end{figure*}

\subsubsection{Multiplexed readout}
We characterize multiplexed readout when the qubits are parked at their idle frequencies shown in Fig.~\ref{fig:Sequence}. Here, we use 400\,ns-long square pulses, convolved with a Gaussian envelope of 10\,ns standard deviation, sent to all readout resonators by frequency multiplexing (calibrated using the tune-up procedures discussed in Sec.~\ref{sec:readout-tuneup}). The output signals from readout resonators are demodulated and integrated with constant weights in 100\,ns-longer integration window. The characterization is performed by initializing the system in a state of random 10-bit-string by a choice of local gates ($I$ or $X$) on all qubits sent in parallel followed by the multiplexed readout,
repeated over $n_\mathrm{rep}>10^{6}$ experimental realizations.

The performance of the multiplexed readout is quantified by the probability $P(z|\zeta)$ of assigning a bit-string $z$ when the system is prepared in a bit-string $\zeta$, known as the assignment probability matrix, which is illustrated in Fig.~\ref{fig:multiRO}a. The average probability of correct assignment, i.e., mean of diagonal elements $P(z|z)$ of the assignment probability matrix, gives the fidelity of multiplexed readout for discriminating $2^{10}$ standard basis states of 10 qubits, which is calculated to be $\mathcal{F}_\text{10Q}=0.7626$ (see Fig.~\ref{fig:multiRO}b). In the absence of correlated multi-qubit readout errors, this corresponds to an average single-qubit readout fidelity of $(\mathcal{F}_\text{10Q})^{1/10}\approx 0.9733$, which is slightly lower than the average of best single-qubit rapid readout fidelity shown in Table~\ref{tb:single-qubit-readout}.

The sources of infidelity can be qualitatively understood by noting the non-zero off-diagonal elements of the assignment probability matrix. For example, the dominant infidelity components of the assignment matrix in Fig.~\ref{fig:multiRO}a, on the order of $10^{-2}$, form a fractal pattern situated in the lower triangular part of the matrix. This corresponds to a single-qubit decay error, where a bit of a bit-string prepared in state 1 is assigned to be in state 0 while the remaining bits of the assigned bit-string are identical to the originally prepared ones. In order to perform a quantitative analysis of the errors during the multiplexed readout, we define bit-flip error rates calculated from the infidelity components of the assignment probability matrix (see Fig.~\ref{fig:multiRO}c and d). The single-qubit bit-flip error rate is written as
\begin{equation}
    e_1(s_i)\equiv \overline{{P(\cdots\neg{s_i}\cdots|\cdots{s_i}\cdots)}},
\end{equation}
which is the average of assignment probability matrix elements corresponding to preparation of a state $s_i$ and assignment of the flipped state $\neg{s_i}$ on qubit Q$_i$ while the assigned bits of the remaining qubits are identical to the prepared ones. It is observed in Fig.~\ref{fig:multiRO}c that the single-qubit decay $\ket{1}\rightarrow\ket{0}$ (few percent on average) is the dominant contributor to the infidelity of the readout, which is believed to be associated with the limited lifetimes $T_1$ of qubits resulting in non-negligible decay during the on-time of readout pulses. Similarly, the two-qubit bit-flip error rate is defined as
\begin{equation}
    e_2(s_i, s_j)\equiv \overline{{P(\cdots\neg{s_i}\cdots\neg{s_j}\cdots|\cdots{s_i}\cdots{s_j}\cdots)}},
\end{equation}
which is the average of assignment probability matrix elements corresponding to preparation of states $(s_i, s_j)$ and assignment of the flipped states $(\neg{s_i}, \neg{s_j})$ on qubits (Q$_i$,Q$_j$) while the other bits remain identical between the preparation and the assignment. We find that the two-qubit error rates shown in Fig.~\ref{fig:multiRO}d are dominated by two-qubit decay process $\ket{11}\rightarrow\ket{00}$ in most cases with error rates an order of magnitude smaller than the single-qubit error rates in Fig.~\ref{fig:multiRO}c. Such behavior is expected in a system with independent single-qubit error sources. We also note the prevalence of non-trivial two-qubit error processes in few combinations of qubit pairs, e.g., two-qubit excitation $\ket{00}\rightarrow\ket{11}$ on (Q$_4$,Q$_6$) and population swap $\ket{01}\rightarrow\ket{10}$ on (Q$_9$,Q$_{10}$). 
We attribute these processes to state preparation errors associated with always-on qubit-qubit couplings and frequency collision in our system as well as parasitic readout-induced state transitions, which are under active investigation. We observe that the higher-order error processes make negligible contributions to the 10-qubit readout fidelity $\mathcal{F}_\text{10Q}$.


\section{Qubit control methods}
\label{sec:SV}
In this section, we provide details about single- and multi-qubit control methods, including qubit XY and Z control methods, qubit reset procedures, and experimental pulse sequences described in the main text.

\subsection{Qubit XY control}
The XY control of a qubit is realized by sending a microwave pulse on the qubit's charge drive line at the transition frequency. Throughout the experiments described in this work, we use 40-ns-long Gaussian pulses with a standard deviation of 10\,ns, corrected with the derivative removal by adiabatic gate (DRAG) method \cite{motzoi2009simple}.
The pulses are calibrated following the procedures outlined in Ref.~\cite{gao2021practical}.

\subsection{Qubit Z control}
The transition frequency of a qubit is controlled by current sent into its flux-bias line, which generates magnetic field threading the qubit's SQUID loop. 
The DC flux bias (slow Z) sets the static frequency while the flux bias pulses (fast Z) dynamically tune the qubit to, e.g., the interaction frequency during an experimental sequence.
Here, we provide details on the corrections applied on the flux bias of the qubits.

\begin{figure*}[t!]
\begin{center}
\includegraphics[width=1\textwidth]{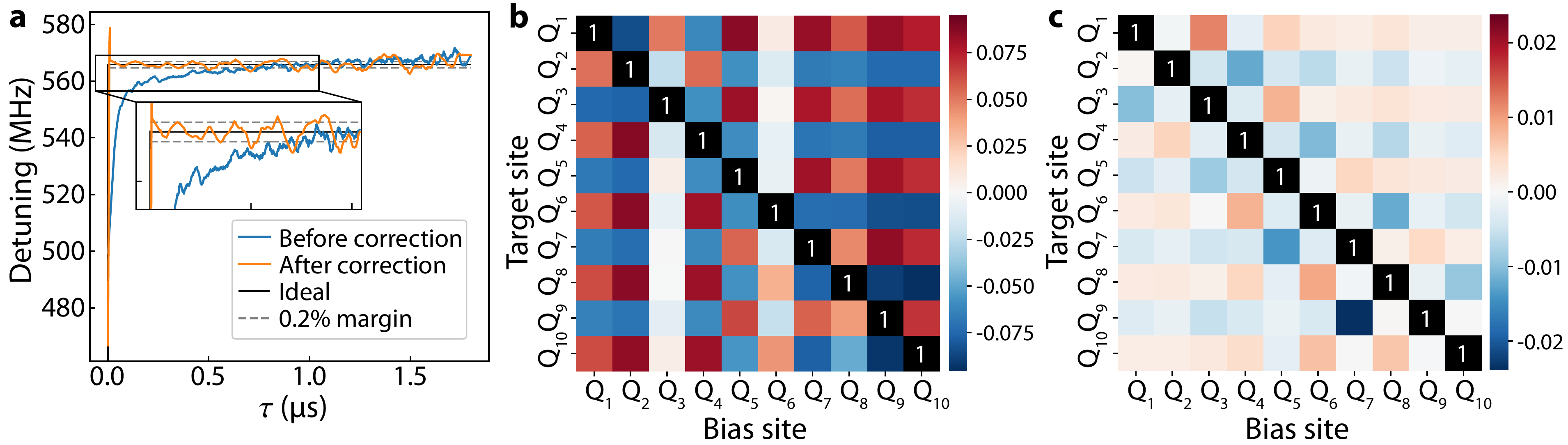}
\caption{\textbf{Flux bias corrections.} 
\textbf{a}, Qubit detuning by a flux pulse as a function of pulse duration $\tau$. The detuning is plotted for the flux pulse before distortion correction (blue), after distortion correction (orange), and for the ideal pulse at the qubit (black). The margin of $\pm0.2\%$ around the ideal pulse is shown in dashed gray lines. The inset gives a magnified view of the initial part of the pulses.
\textbf{b} (\textbf{c}),  The crosstalk matrix for the slow (fast) flux bias with the diagonal elements equal to unity and values of off-diagonal elements indicated by colors.}
\refstepcounter{SIfig}\label{fig:Zpulse} 
\end{center}
\end{figure*}
\subsubsection{Pulse distortion}
A flux pulse is distorted after passing through microwave components along a flux bias line, which introduces a significant error in the frequency control of qubits.
Here, we characterize the pulse distortion utilizing an in situ method involving a Ramsey-like sequence \cite{rol2020time} and perform correction by using pre-distorted waveforms.
To show the effectiveness of the correction, we measure the frequency detuning of a qubit during the on-time of a square flux pulse with and without correction, which are illustrated in Fig.~\ref{fig:Zpulse}a.
We observe a low-pass response in the uncorrected pulse and a response close to the ideal step function after the correction. 
The rise time of the remaining low-pass response of the corrected pulse is 6\,ns with approximately 2\,ns-long overshoot originating from the finite bandwidth of the waveform generator to faithfully implement the pre-distortion.
The deviation of the corrected pulse remains mostly within the $\pm0.2\%$ margin of the ideal response for short pulses, while a larger fluctuation is observed at pulse durations greater than 1$\,\upmu$s limited by the coherence of the qubit during the characterization.
\subsubsection{Flux crosstalk}
For a multi-qubit device, the flux crosstalk, i.e., the flux bias on one qubit affecting the frequency of another qubit, poses challenges on precise simultaneous control over the frequency of multiple qubits.
To compensate for the flux crosstalk, we characterize a crosstalk matrix $\boldsymbol{C}_V$ defined by $\boldsymbol{V}_\mathrm{eff}=\boldsymbol{C}_{V}\boldsymbol{V}_\mathrm{app}$, where $\boldsymbol{V}_\mathrm{eff}$ $(\boldsymbol{V}_\mathrm{app})$ is a vector of effective (applied) flux bias voltages on all the qubits.
Assuming that the diagonal components of the crosstalk matrix are close to the unity, i.e., $[\boldsymbol{C}_{V}]_{i,i}=\Delta V_{\mathrm{eff},i}/\Delta V_{\mathrm{app},i}\approx 1$, we characterize each off-diagonal crosstalk element $[\boldsymbol{C}_{V}]_{i,j}$ ($i\neq j$) by measuring the ratio of flux tuning rates of a target site Q$_i$ from biasing the sites Q$_j$ and Q$_i$, giving
\begin{equation}
    [\boldsymbol{C}_{V}]_{i,j} \equiv \frac{\Delta V_{\mathrm{eff}, i}}{\Delta V_{\mathrm{app}, j}} \approx \frac{\Delta V_{\mathrm{app},i}}{\Delta V_{\mathrm{app}, j}} = \frac{\Delta \omega_{01,i}/\Delta V_{\mathrm{app},j}}{\Delta \omega_{01,i}/\Delta V_{\mathrm{app}, i}}.
\end{equation}
The measurement of flux tuning rate $\Delta \omega_{01,i}/\Delta V_{\mathrm{app},j}$ is performed at biases where the frequency of a qubit is nearly linear in the flux bias.
During the characterization of $[\boldsymbol{C}_{V}]_{i,j}$, sites other than the target site Q$_i$ is detuned by at least 3\,GHz to minimize the influence of coupling between sites.
In the DC flux crosstalk characterization, the target qubit frequency tuning $\Delta \omega_{01,i}$ is calibrated from Ramsey fringes experiments, yielding an average magnitude of crosstalk elements of 0.059 (see Fig.~\ref{fig:Zpulse}b). 
For the fast flux pulse crosstalk, $\Delta \omega_{01,i}$ is measured from a Ramsey sequence with a flux pulse of increasing duration (up to 1$\,\upmu$s) between two $\pi/2$ pulses. 
The resulting flux pulse crosstalk matrix is shown in Fig.~\ref{fig:Zpulse}c with the average magnitude of crosstalk elements calculated as 0.004, more than an order of magnitude smaller than its DC flux counterpart.
The crosstalk matrices can then be used for calculating the bias voltage to apply in order to achieve the effective ideal bias by $\boldsymbol{V}_\mathrm{app}=\boldsymbol{C}_{V}^{-1}\boldsymbol{V}_\mathrm{eff}$. 
We iterate the crosstalk calibrations based on existing corrections until the remainder of the crosstalk element exhibits no further decrease in magnitude, giving the crosstalk level below $1\times 10^{-4}$ for the majority of elements.

\subsection{Qubit reset}\label{sec:qubit-reset}
Before the start of an experimental sequence, the qubits are required to be in the ground state, which is achieved by qubit reset.
The traditional way to reset a qubit is to wait for a time much longer than its lifetime to ensure the population decay.
For the qubits on this device, wait time around 1\,ms is necessary, limiting the repetition rate of the experiment below $10^{3}$ counts per second.
In most experiments, we utilize the ability of real-time feedback operations with an FPGA-based controller (Quantum Machines OPX$+$) to perform active reset of the qubits.
The single-qubit reset is achieved by measuring the state of the qubit and applying a subsequent $\pi$ pulse if it is detected to be in state $|1\rangle$ \cite{riste2012feedback}. 
Similarly, the active reset of multiple qubits is implemented by performing multiplexed readout and applying subsequent $\pi$ pulses on the qubits measured to be in state $|1\rangle$. 
The procedure is deemed successful once we measure the single qubit or all qubits to be in the ground state for $N_r$ consecutive rounds of repeated readout and reset. 
Typically, we use $N_r=2$ or 6 for single-qubit or 10-qubit experiments, respectively. 
The duration of a round of active reset includes the readout time (about 500\,ns) and the feedback latency (below 1$\,\upmu$s).
The active reset significantly reduces the reset time and enables the fast repetition rate of our experiments, which is essential for efficiently analyzing the bit-string statistics with low statistical fluctuations.

\subsection{Experimental pulse sequence}

\begin{figure}[t!]
\begin{center}
\includegraphics[width=0.48\textwidth]{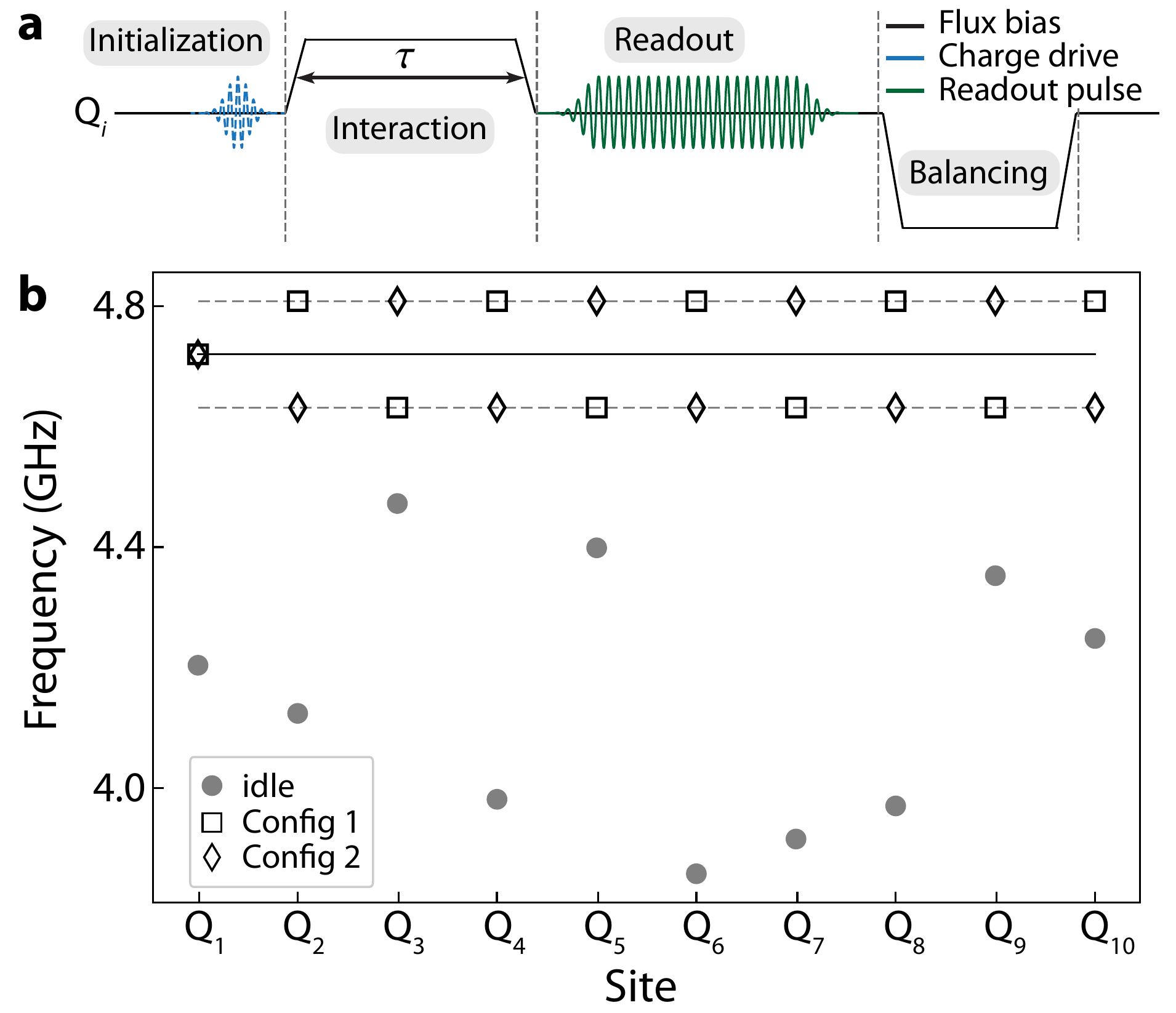}
\caption{\textbf{Experimental pulse sequence.} 
\textbf{a}, Detailed illustration of pulses on site Q$_i$ for the sequence shown in Fig.~\ref{fig:FigureN2}e of the main text. The flux bias pulse, the charge drive pulse, and the readout pulse are shown in black, blue, and green, respectively. The dashed lineshape during the initialization represents either a $\pi$-pulse or a wait applied to the qubit. 
\textbf{b}, Frequencies of sites during the pulse sequence and the frequency alignment calibration. The idle frequencies of qubits and the interaction frequency $\omega_\mathrm{int}/2\pi=4.72\,$GHz for the 10-qubit experiments in the main text are indicated by circles and a solid line, respectively. The frequencies used during the frequency alignment calibration are shown in dashed gray lines together with two frequency configurations to align Q$_1$ to interaction frequency, indicated by square and diamond markers.}
\refstepcounter{SIfig}\label{fig:Sequence} 
\end{center}
\end{figure}

The experimental pulse sequence shown in Fig.~\ref{fig:FigureN2}e of the main text is detailed in Fig.~\ref{fig:Sequence}a for a single qubit. 
We park the qubit at its idle frequency during the initialization stage where a local XY gate is utilized to prepare the qubit in its standard basis states $\ket{0}$ or $\ket{1}$, realized by waiting or by applying a microwave $\pi$-pulse, respectively. 
A square flux pulse is then used to dynamically tune the qubit to the interaction frequency for time $\tau$ followed by the readout after the qubit returns to its idle frequency.
At the end of the sequence, we apply a flux balancing pulse to achieve zero-average flux on the bias line during the whole sequence. 

The idle frequencies are chosen to achieve high-fidelity multiplexed readout (see Sec.~\ref{sec:RO}) and decrease frequency collision of both $\omega_{01}$ and $\omega_{12}$ between sites with strong coupling.
The set of idle frequencies used in Figs.~\ref{fig:FigureN3} and \ref{fig:FigureN4} of the main text is shown in Fig.~\ref{fig:Sequence}b. 
During the interaction stage, all ten sites are aligned to the same frequency by flux pulses whose amplitudes are calibrated using the following procedures.
First, we perform a coarse tuning using the crosstalk matrices and the pre-calibrated tuning curve (site frequency vs.~bias voltage) for individual sites. 
To obtain the precise flux pulse amplitude for tuning a qubit to the interaction frequency $\omega_\mathrm{int}$ and account for remaining crosstalk, we perform a fine tuning by pulsing a target site Q$_i$ to $\omega_\mathrm{int}$ and other sites to $\omega_\mathrm{int} \pm 2\pi \times 88\,$MHz (calibration frequency), measuring the frequency of Q$_i$ by Ramsey sequence, and adjusting the pulse amplitude on Q$_i$ to compensate for deviations.
We repeat the above procedures until the measured frequency $\omega_{01,i}$ of the target qubit is sufficiently close to the desired interaction frequency $\omega_\mathrm{int}$.
To minimize the frequency shift induced by the coupling between Q$_i$ and other sites, we have the calibration frequencies staggered in two configurations illustrated in Fig.~\ref{fig:Sequence}b.
Then we use the average of the pulse amplitudes on Q$_i$ achieving $\omega_{01,i}=\omega_\mathrm{int}$ in the two configurations for the multi-qubit interaction stage, which effectively cancels the influence from possible residual flux crosstalk. 

\section{Hamiltonian learning using a many-body fidelity estimator}
\label{sec:SVI}
The goal of Hamiltonian learning is to find the set of parameters of an estimated Hamiltonian $\hat{H}'$ that is closest to the many-body Hamiltonian $\hat{H}$ realized in an experiment \cite{wiebe2014hamiltonian}. Here, we learn the Hamiltonian by comparing the measurement outcomes from evolving the experimental system under its Hamiltonian and the numerical results from simulating the evolution of an estimated Hamiltonian.
We experimentally prepare an initial state $\ket{\Psi_0}$ and evolve it for time $\tau$ under a Hamiltonian $\hat{H}$ realized in the experiments, which results in a state represented by a density matrix $\hat{\rho}(\tau)$. 
Assume we can classically simulate the same process for a guessed Hamiltonian $\hat{H}'$ and obtain the time-evolved state $e^{-i\hat{H}' \tau/\hbar}\ket{\Psi_0}$. 
The many-body fidelity is the overlap between the experimental and guessed states:
\begin{equation}
    F(\tau,\hat{H}') \equiv \langle \Psi_0|e^{i\hat{H}'\tau/\hbar} \hat{\rho}(\tau) e^{-i\hat{H}'\tau/\hbar}|\Psi_0\rangle. \label{eq:fidelity}
\end{equation}
If the fidelity can be efficiently estimated, the Hamiltonian can then be learned by maximizing $F(\tau, \hat{H}')$ over a family of trial Hamiltonians $\{\hat{H}'\}$. 
It is numerically found \cite{mark2022inprep} that this approach remains robust even in the presence of small errors from state preparation and measurement (SPAM) and decoherence processes.
In such scenarios, the optimized many-body fidelity $F$ is smaller than one, but is maximized at approximately the correct Hamiltonian parameter values.

\subsection{Many-body fidelity estimator $F_d$}
\begin{figure}[t!]
\begin{center}
\includegraphics[width=0.48\textwidth]{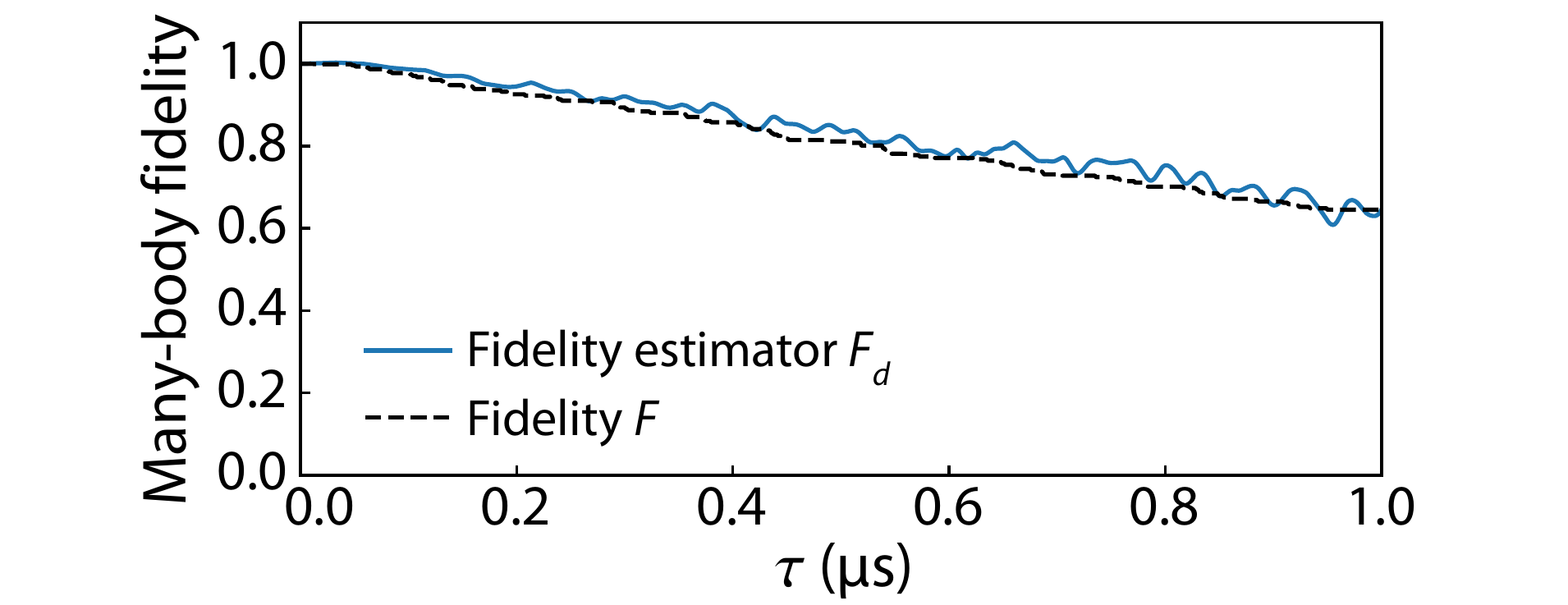}
\caption{\textbf{Numerical simulation of many-body fidelity $F$ and its estimator $F_d$ under dephasing errors.} 
The many-body fidelity $F$ is calculated using Eq.~\eqref{eq:fidelity} for an evolution under the system Hamiltonian described in Eq.~\eqref{eq:Ham} of the main text, with injected phase-flip errors. The initial state is chosen to be $z_\mathrm{init}=1100100101$ at random. The corresponding fidelity estimator $F_d$ is shown for comparison, which is obtained from Eq.~\eqref{eq:Fd} with the bit-string probabilities $p_z$ and $p_z^{\mathrm{(T)}}$ obtained from the numerical simulations of evolution with and without errors, respectively.}
\refstepcounter{SIfig}\label{fig:decoheredFd} 
\end{center}
\end{figure}

In practice, it is cumbersome to obtain the many-body fidelity $F$ by  either directly measuring non-local observables in the expression of $F$ \cite{buhrman2001quantum, flammia2011direct, cincio2018learning} or characterizing the experimental density matrix $\hat{\rho}(\tau)$ via quantum state tomography \cite{steffen2006measurement}.
Recently, Ref.~\cite{mark2022inprep} introduced the estimator $F_d$ to approximate the many-body fidelity between simulated and experimental time evolution in Eq.~\eqref{eq:fidelity}. 
Crucially, this estimator only requires measurements in a fixed basis---here the computational $z$-basis---and requires a relatively small number of measurement samples. 
Based on universal statistical fluctuations and operator spreading, this estimator works in a variety of quantum devices, including bosonic and fermionic itinerant particles on optical lattices, trapped ions, and arrays of Rydberg atoms.
Through $F_d$, the fidelity $F$ can be estimated with a small number of samples for different parameter sets in $\{\hat{H}'\}$, yielding parameter values that are most likely to generate the observed data. 
To illustrate the effectiveness of this estimator in our system,  we show in Fig.~\ref{fig:decoheredFd} that the $F_d$ closely tracks the many-body fidelity $F$ in a numerical simulation of our system Hamiltonian shown in Eq.~\eqref{eq:Ham} of the main text in the presence of random phase-flip errors. 

\subsection{$F_d$ calculation with experimental data}
\label{sec:Fd_experiment}
We describe the procedure used to calculate $F_d$ and obtain the optimized Hamiltonian parameters shown in Fig.~\ref{fig:FigureN2} of the main text. This procedure is adapted from the protocol detailed in Ref.~\cite{mark2022inprep}. 

On the experimental side, following the pulse sequence illustrated in Fig.~\ref{fig:Sequence}a, we prepare a randomly chosen five-excitation initial state, e.g., $z_\mathrm{init}=1001101010$, evolve the system for time $\tau$, and read out the states of ten sites to get a bit-string $z$.
Repeating the sequence $M=4000$ times for the same $z_\mathrm{init}$ gives the probability of measuring each bit-string.
We use the assignment probability matrix obtained from Sec.~\ref{sec:RO} to mitigate the effect of readout error, and only keep the bit-strings with five excitations considering the excitation-number-conserving Hamiltonian. This gives an estimate of the bitstring probability $p_z(\tau)$. 

In the numerical simulation, we compute the theoretical probability $p_z^{\mathrm{(T)}}(\tau,\hat{H}')$ of measuring the bit-string $z$ after evolution under $\hat{H}'$ for a time $\tau$
\begin{equation}
    p_z^{\mathrm{(T)}}(\tau,\hat{H}') \equiv |\langle z|\mathrm{exp}(-i\hat{H}'\tau/\hbar)|\Psi_0\rangle|^2~,
\end{equation}
and its time-average $\overline{p}_{z}^{\mathrm{(T)}}(\hat{H}')$
\begin{equation}
\label{eq:time_avg}
    \overline{p}_{z}^{\mathrm{(T)}}(\hat{H}') \equiv \lim_{\mathcal{T}\rightarrow \infty}\frac{1}{\mathcal{T}}\int_0^\mathcal{T} p_z^{\mathrm{(T)}}(\tau, \hat{H}') \mathrm{d}\tau
\end{equation}
where $|\Psi_0\rangle = |z_\mathrm{init}\rangle$.

Combining the results from experiments, the many-body fidelity estimator is given by
\begin{equation}
\label{eq:Fd}
    F_d(\tau,\hat{H}') = 2 \frac{\sum_z p_z(\tau)p_z^{\mathrm{(T)}}(\tau,\hat{H}')/ \overline{p}_{z}^{\mathrm{(T)}}(\hat{H}')}{\sum_zp_z^{\mathrm{(T)}}(\tau,\hat{H}')^2/ \overline{p}_{z}^{\mathrm{(T)}}(\hat{H}')} - 1.
\end{equation}
To reduce possible systematic errors, we repeat the above procedure for $M_\mathrm{init}=40$ different initial states.
In Fig.~\ref{fig:FigureN2}f of the main text, we plot the average and standard deviation over different initial states for $F_d(\tau,\hat{H}')$ with different parameter values of $\{\hat{H}'\}$.

\subsection{Parameter optimization}
\label{Sec:ParaOpti}
To address the complexity of optimizing over a large parameter space, a greedy algorithm was introduced in Ref.~\cite{mark2022inprep} for multi-parameter estimation. 
We follow the same procedure to obtain the optimized values for hopping terms $\{J_{i,j}\}$, including the nearest-neighbor hopping $J_{i, i+1}$ with $i=1, \cdots, 9$ and the longer-range hopping $\overline{J_{i, i+k}}$ with $k=2, \cdots, 9$, averaged over qubits $i$. 
We compute $F_d(\tau, \hat{H}')$ at a fixed evolution time $\tau$, for the family of Hamiltonians $\hat{H}'(\{J_{i,j}\})$ in Eq.~\eqref{eq:Ham} of the main text.
Here, we use the measured on-site interactions $U_i$ (Fig.~\ref{fig:FigureN2}a of the main text) and assume a constant site energy $\epsilon_i/2\pi=4.72\,$GHz. 
To avoid systematic errors, we use the averaged estimated fidelity $\overline{F_d}(\hat{H}')$ over several times $\tau \in \{76, 148, 260, 420, 600, 780\}\,$ns.
For each guess of $\{J_l\}$, we randomly choose one element $l\in\{1,\cdots,17\}$, then
maximize $\overline{F_d}(\hat{H}')$ over a 
single $J_l$ 
while keeping other parameters fixed. 
After optimizing for all 17 parameters, we repeat this process multiple times over distinct random permutations of $\{1,\cdots,17\}$. 
After 11 rounds, both $\overline{F_d}$ and the optimized $\{J_{i,j}\}$ converge and are displayed in Fig.~2f and g in the main text.

\subsection{Influence of decoherence}
\label{sec:Fd_decohere}
A unitary time evolution of a quantum state under a Hamiltonian cannot capture the effect of decoherence present in experimental systems. In the context of Hamiltonian learning, the dominant decoherence channels in our system---population decay and dephasing---can affect the processes discussed in the previous sections by lowering the estimated many-body fidelity value from its decoherence-free counterpart. 
Here, we explain possible mitigation strategies in the processing of experimental data and discuss how these effects can be taken into account in numerical simulations.

The population decay lowers the excitation number in the system, leading to a finite lifetime $T_1$.
Given the experimental Hamiltonian that conserves excitation numbers, we post-select the measured bit-strings $z$ with the excitation numbers the same as the initial bit-string $z_\mathrm{init}$.
The purpose of this post-selection is to analyze the outcomes without quantum jump events \cite{dalibard1992wave, plenio1998quantum}, thus mitigating the influence of population decay processes.
However, if the lifetimes of different qubits exhibit a large variance, the post-selection will favor states with excitations on long-lived qubits at evolution times longer than the qubit lifetimes, failing to mitigate the population decay.

On the other hand, the dephasing errors can destroy the phase coherence within the same excitation number sector and cannot be corrected by post-selection, which results in a lower many-body fidelity $F$. This can be numerically simulated by injecting random phase-flip errors during the state evolution under the system Hamiltonian, which is shown in Fig.~\ref{fig:decoheredFd}.
Here, the rate of phase-flip errors is chosen such that the resulting single-qubit coherence time is $T_2^*=1.14\,\upmu$s, close to the measured coherence time at $\omega_{01}/2\pi = 4.72\,$GHz shown in Table~\ref{tb:device-characterization}.
We note that the optimized $F_d$ in Fig.~\ref{fig:FigureN2}f of the main text gives numbers smaller than this numerical simulation accounting for dephasing errors.
This discrepancy could be due to experimental imperfections such as SPAM errors and imperfect flux pulses used to align all qubits at the interaction frequency.

\section{Numerical simulation of quantum walk at different interaction frequencies}
\label{sec:SVII}
\begin{figure*}[t!]
\begin{center}
\includegraphics[width=\textwidth]{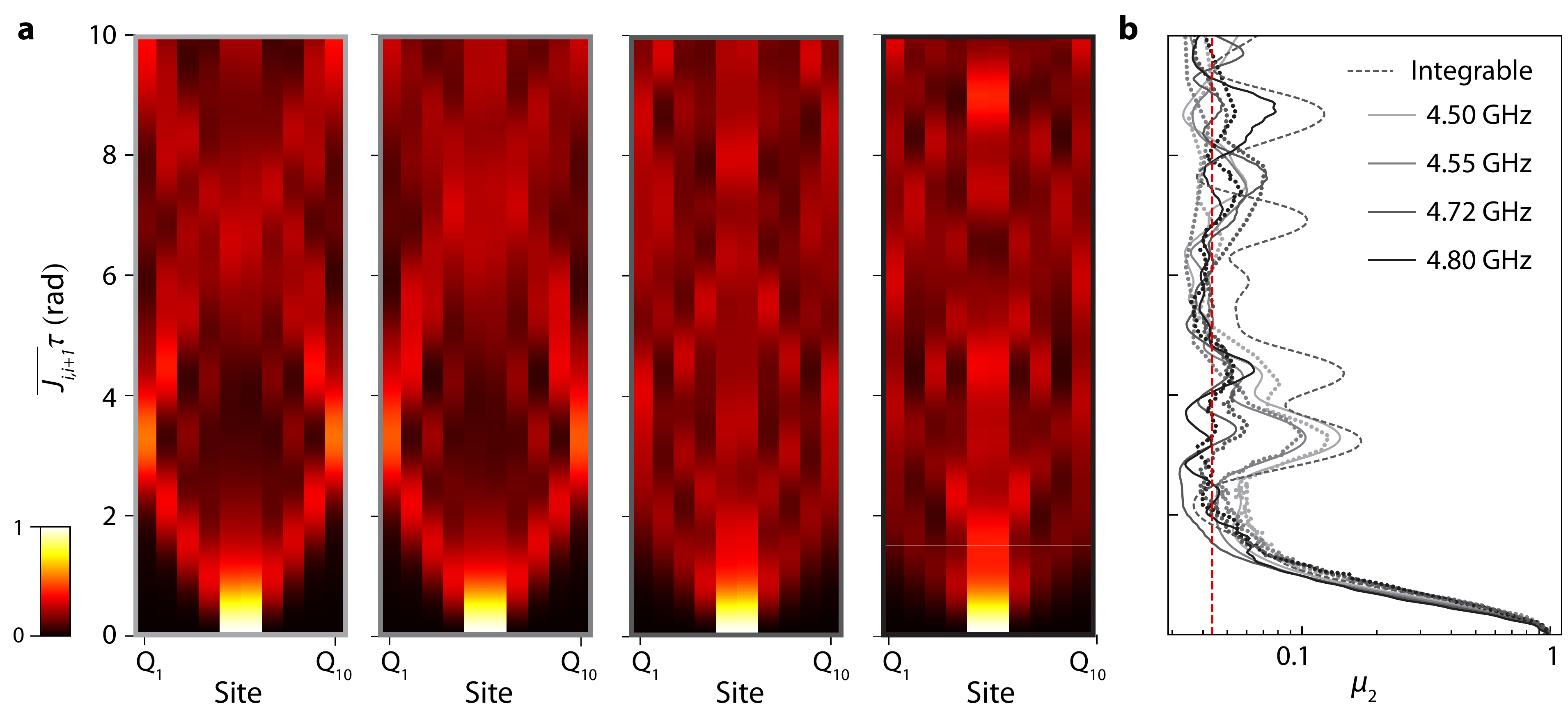}
\caption{\textbf{Numerical simulation of two-particle quantum walk with increasing hopping range.} 
\textbf{a}, Evolution of the population $\langle \hat{n}_i\rangle$ on sites Q$_1$--Q$_{10}$ as a function of normalized evolution time $\overline{J_{i,i+1}}\tau$. The system is initialized in $z_\textrm{init}=0000110000$ and the evolution occurs at $\omega_{01}/2\pi=4.50$\,GHz, 4.55\,GHz, 4.72\,GHz, and 4.80\,GHz from left to right. 
\textbf{b}, The second moment $\mu_2$ as a function of normalized evolution time $\overline{J_{i,i+1}}\tau$. Results calculated from the numerical simulation in panel \textbf{a} and the corresponding measurement in Fig.~\ref{fig:FigureN3}a of the main text are shown in gray-scale solid and dotted curves, respectively. The result from numerical simulation of the integrable Hamiltonian is shown as the gray dashed curve. The expected final value of the second moment $\mu_2^\mathrm{e}$ for a generic ergodic system is indicated by the red dashed line.}
\refstepcounter{SIfig}\label{fig:simuFig4} 
\end{center}
\end{figure*}

To corroborate the observations in Fig.~\ref{fig:FigureN3} of the main text, we numerically simulate the same set of evolutions using QuTiP \cite{johansson2012qutip}. 
We obtain the on-site interaction $U_i$ from Fig.~\ref{fig:FigureN2}a of the main text, assume the evolution at the interaction frequency of $\omega_{01}/2\pi=4.50$\,GHz, 4.55\,GHz, 4.72\,GHz, and 4.80\,GHz, and use the hopping terms $J_{i,j}$ from the parameter optimization procedure described in Sec.~\ref{Sec:ParaOpti} to simulate the quantum walk dynamics (Fig.~\ref{fig:simuFig4}a). 
The probability of measuring the bit-string $z$ with excitations on sites Q$_i$ and Q$_j$ ($i\neq j$) is $p_z = \langle \hat{n}_i \hat{n}_j\rangle$, which is used to calculate the second moment $\mu_2 = \sum_zp_z^2$ shown in Fig.~\ref{fig:simuFig4}b.
The simulated results exhibit good agreement with the experimental results in Fig.~\ref{fig:FigureN3} of the main text, confirming that Hamiltonians with longer hopping ranges will converge to the ergodic limit $\mu_2^\mathrm{e} = 2/(D+1)$ at earlier times. 

\section{Probing ergodic dynamics from global bit-string statistics}
\label{sec:SVIII}

\begin{figure*}[t!]
\begin{center}
\includegraphics[width=1\textwidth]{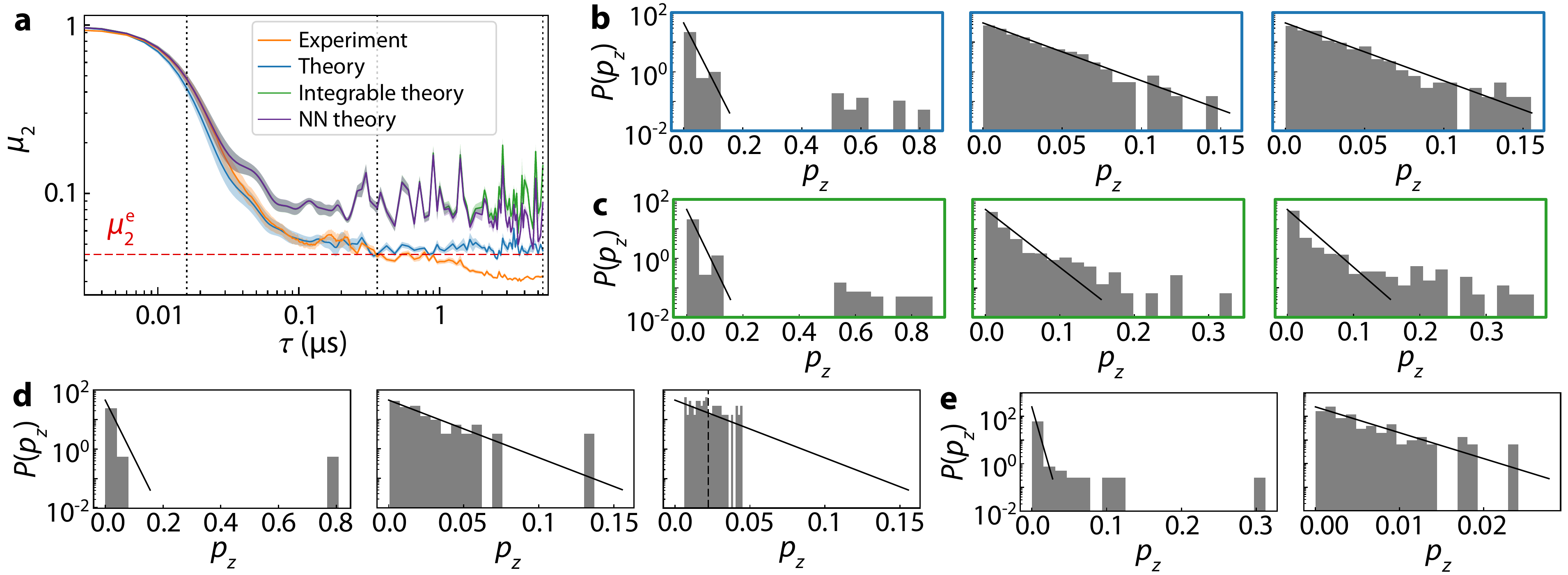}
\caption{\textbf{Additional results of bit-string statistics at 4.72$\,$GHz.} 
\textbf{a}, Second moment $\mu_2$ as a function of evolution time $\tau$ in our system from the experiment (orange) and the theory with the optimized parameter set in Fig.~\ref{fig:FigureN2}g of the main text (blue), compared to theoretical predictions of the integrable model (green) and the Bose-Hubbard model with NN hopping (purple). The shading on each curve corresponds to a standard deviation of the mean second moment for 20 randomly chosen initial bit-strings $z_\textrm{init}$ in the two-excitation sector, and the red dashed line represents the ergodic value $\mu_2^\mathrm{e}$.
\textbf{b} (\textbf{c}), Density histogram $P(p_z)$ of the distribution of theory (integrable theory) bit-string probabilities $\{p_z\}$ with the 20 initializations $z_\textrm{init}$'s at evolution times $\tau=16$\,ns, 360\,ns, and 5.4\,$\upmu$s from left to right (indicated by the dotted lines in panel \textbf{a}). The solid lines show the PT distribution.
\textbf{d}, Density histogram $P(p_z)$ of the distribution of experimental bit-string probabilities $\{p_z\}$ with a two-excitation initial state $z_\textrm{init}=0000110000$ at evolution times $\tau=16$\,ns, 360\,ns, and 5.4\,$\upmu$s from left to right.
\textbf{e}, Density histogram $P(p_z)$ of the distribution of experimental bit-string probabilities $\{p_z\}$ with a five-excitation initial state $z_\textrm{init}=0100111010$ at evolution times $\tau=16$\,ns and 360\,ns in the left and right panels, respectively.}
\refstepcounter{SIfig}\label{fig:ergodic} 
\end{center}
\end{figure*}

\subsection{Integrability of 1D Bose-Hubbard models}
In this section, we discuss the integrability of 1D Bose-Hubbard models in different parameter regimes.
The standard 1D Bose-Hubbard model with only nearest-neighbor (NN) hopping corresponds to the spin-1/2 XY model in the limit of diverging on-site interaction $|U| \rightarrow \infty$, where the Hilbert space of a site is truncated to only $|0\rangle$ and $|1\rangle$, i.e., the hard-core limit.
This model is known to be integrable and can be solved exactly with the Bethe ansatz \cite{nagaosa1999quantum}.
The integrability of the hard-core Bose-Hubbard model can be broken by either adding long-range hopping terms or taking a finite on-site interaction strength. 

Here, we emphasize that, in our system, finite on-site interaction $|U|$ is not the major factor that induces ergodicity.
This is because the experiments illustrated in Figs.~\ref{fig:FigureN3} and \ref{fig:FigureN4} of the main text are performed at $\omega_{01}/2\pi \le 4.80$\,GHz, featuring a finite $|U/J| > 36$, much greater than unity.
To further illustrate the effect of finite $|U/J|$, we simulate the standard Bose-Hubbard model with finite $|U/J|$ and without long-range hopping, where the Hamiltonian is obtained by removing the long-range hopping terms in Eq.~\eqref{eq:Ham} of the main text while using the on-site interaction $U_i$ from Fig.~\ref{fig:FigureN2}a of the main text. 
In this case, while the simulated model is ergodic in a strict sense, the resulting second moment $\mu_2$ is the same as the numerical simulation of the integrable model at evolution times $\tau < 2\,\upmu$s, gradually approaching the ergodic limit at long evolution times $\tau > 5\,\upmu$s (see the purple curve of Fig.~\ref{fig:ergodic}).
Therefore, we conclude that the experimentally observed ergodic behavior, which emerges at an intermediate evolution time of $\tau\approx360\,$ns, originates from the long-range hopping terms.

\subsection{Porter-Thomas distribution}
In this section, we first introduce the Porter-Thomas (PT) distribution \cite{porter1956fluctuations, boixo2018characterizing} and then explain why we expect the PT distribution in our system.
The PT distribution can be obtained from the distribution of overlap probabilities $p_z = \vert \langle z | \Phi\rangle \vert^2$ between a particular measurement outcome $\ket{z}$ and a state $\ket{\Phi}$ drawn from the Haar ensemble, the distribution of states on a Hilbert space that is invariant under any unitary operations. 
Specifically, in a Hilbert space of dimension $D$, the distribution of $p_z$ takes the form \cite{porter1956fluctuations, choi2021emergent}
\begin{equation}
    P(p_z)\:\mathrm{d}p_z = (D - 1)(1 - p_z)^{D - 2}\:\mathrm{d}p_z,
\end{equation}
which converges to the PT distribution in the limit of large $D$
\begin{equation}
    P(p_z) \xrightarrow{D\rightarrow\infty}  D\exp{(-Dp_z)}.
\end{equation}
Note that the second moment of $\{p_z\}$ from this distribution is given by
$$\mu_2 \equiv \sum_z p_z^2 = 
D\int_{0}^{1}\mathrm{d}p_z\: p_z^2P(p_z) = \frac{2}{D+1},$$ which is identical to the ergodic value $\mu_2^\mathrm{e}$ described in the main text. 

The PT distribution reflects the randomness of the measured wavefunction and has been shown to occur in Bose-Hubbard model with time-dependent random parameters \cite{neill2018blueprint} and deep random unitary circuits \cite{boixo2018characterizing, arute2019quantum}. 
Additionally, a large class of time-independent Hamiltonian is also shown to exhibit the normalized probability distribution $\tilde{p}_z \equiv p_z / \overline{p}_z$ following the PT distribution~\cite{mark2022inprep}, where $\overline{p}_z$ is the time-averaged probability defined in Eq.~\eqref{eq:time_avg}. 
This includes the ergodic Bose-Hubbard model with long-range hopping realized in our system.
Although our Hamiltonian in Eq.~\eqref{eq:Ham} of the main text conserves excitation number, the dynamics exhibits this universal randomness within the two-excitation, hard-core sector\footnote{We verify that the doublon states have a population of $\lesssim 1\%$ using numerical simulations, confirming that we are in the hard-core limit.} with a Hilbert space dimension of $D=45$.
Furthermore, we claim that the time-averaged probability $\overline{p}_z$ is approximately constant due to the effective infinite temperature\footnote{In detail, the effective temperature is infinite because in the rotating frame of $\omega = \omega_{01}$, the computational $z$-basis states have zero averaged energy $\langle \hat{H} \rangle_z = 0$ [see Eq.~\eqref{eq:Ham} of the main text].
It is the same as the averaged energy of a thermal state $\hat{\rho}_\beta$ at infinite temperature $\mathrm{Tr}(\hat{H}\hat{\rho}_{\beta=0})=\mathrm{Tr}(\hat{H}\mathbb{I})/D=0$, where $\beta=1/T^*$ is the inverse effective temperature and $\hat{\rho}_{\beta = 0} = \mathbb{I}/D$.
} of the initial and measurement states, and thus the unnormalized $\{p_z\}$ follows the PT distribution at intermediate evolution times, following the arguments below.

The notion of the effective temperature of a state is conventionally used to understand its \textit{local} properties \cite{gogolin2016equilibration}: it is believed that an initial state, when quench-evolved under an ergodic Hamiltonian, will quickly thermalize such that its expectation value of a local observable $\hat{A}$ is very close to the value $\langle \hat{A}(t) \rangle \approx \mathrm{Tr}(\hat{A} \hat{\rho}_\beta)$ of a corresponding thermal state $\hat{\rho}_\beta$. 
In particular, the thermal state with infinite effective temperature $\hat{\rho}_{\beta = 0}$ gives the expectation value $\langle \hat{A}(t) \rangle \approx \mathrm{Tr}(\hat{A})/D$. 
Here, we extend the above expectation beyond local quantities to the time-average of a global quantity $\overline{p}_z$ and expect this normalization factor $\overline{p}_z$ to be approximately constant since our initial state and measurement states are at infinite temperature. 
Consequently, we expect the raw, unnormalized $\{p_z\}$ to follow the PT distribution at intermediate evolution times, as shown in the middle panel of Fig.~4b in the main text. 

\subsection{Additional results of bit-string distribution}
In this section, we provide additional results of numerical simulations and experimental data as a supplement to the bit-string distributions shown in Fig.~\ref{fig:FigureN4}b of the main text.

\subsubsection{Numerical simulations of bit-string distributions}
In addition to the experimental bit-string distribution shown in Fig.~4b, we present the results from numerical simulations of the system Hamiltonian in Eq.~\eqref{eq:Ham} of the main text and the integrable (nearest-neighbor and hard-core) Bose-Hubbard Hamiltonian in Fig.~\ref{fig:ergodic}b and c, respectively.
After a short evolution time of $\tau=16\,$ns, the distribution  is similar in the three cases of experiment (Fig.~\ref{fig:FigureN4}b of the main text, left), theory (Fig.~\ref{fig:ergodic}b, left), and integrable theory (Fig.~\ref{fig:ergodic}c, left), exhibiting a large probability $p_z$ of bit-strings close to $z_\mathrm{init}$.
At intermediate evolution times (e.g. $\tau=360\,$ns), both the experimental (Fig.~4b in the main text, middle) and the theoretical (Fig.~\ref{fig:ergodic}b, middle) bit-string distributions follow the PT distribution while the integrable theory predicts more bitstrings with large $p_z$, resulting in a larger second moment $\mu_2$.
Due to the absence of decoherence processes, the distributions stay the same for the numerical simulations at long evolution times, e.g. $\tau=5.4\,\upmu$s  (right panels of Fig.~\ref{fig:ergodic}b and c).

\subsubsection{Experimental bit-string distribution from a single, two-excitation initial state}
The probability distributions shown in Fig.~\ref{fig:ergodic}b and c and in Fig.~\ref{fig:FigureN4}b of the main text are obtained by combining the results from 20 randomly chosen initial states $z_\mathrm{init}$ to reduce statistical errors. 
Here, we show that the experimental results from a single initial state ($z_\mathrm{init}=0000110000$), displayed in Fig.~\ref{fig:ergodic}d, obey the same trend as predicted by the ergodic evolution.
Due to the limited counts $N(p_z)=D=45$ of $p_z$'s, the tail of PT distribution is not very clear in the middle panel of Fig.~\ref{fig:ergodic}d. 
Nonetheless, the distinction among the three panels in Fig.~\ref{fig:ergodic}d is obvious, with $p_z$ aggregating towards $1/D$ in the right panel due to decoherence.

\subsubsection{Experimental bit-string distribution from a five-excitation initial state}
To illustrate the generality of the bit-string distribution during ergodic evolution, we show the experimental bit-string distribution obtained from preparing a randomly chosen five-excitation initial state $z_\mathrm{init}=0100111010$ in Fig.~\ref{fig:ergodic}e.
The left panel displays the initial evolution stage at $\tau=16\,$ns and the right panel shows a histogram at $\tau=360\,$ns that is closer to the PT distribution than the middle panel of Fig.~\ref{fig:ergodic}d owing to the larger Hilbert space dimension $D=252$.

\subsection{Effect of decoherence}

\begin{figure}[b!]
\begin{center}
\includegraphics[width=0.48\textwidth]{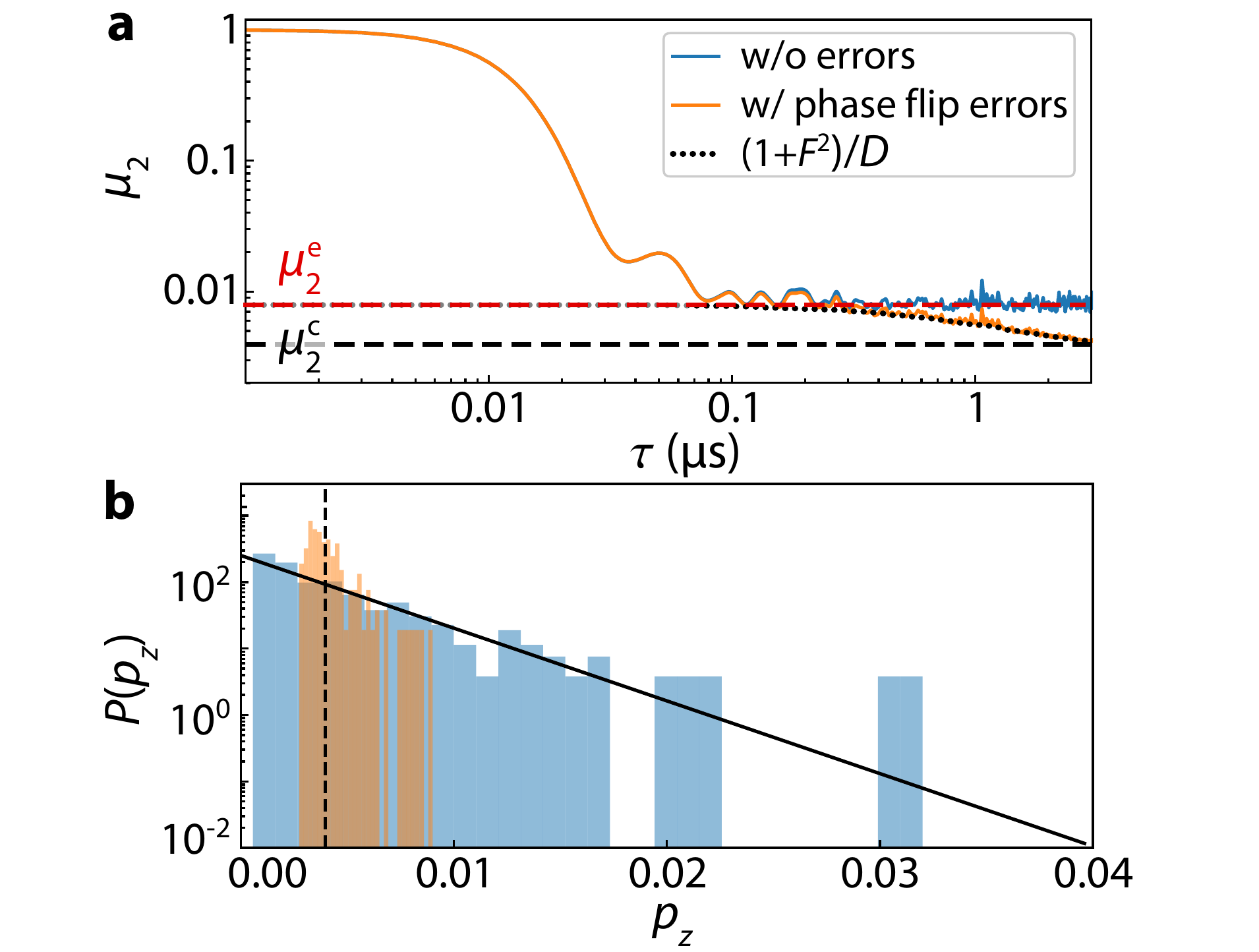}
\caption{\textbf{Numerical simulation of bit-string statistics under dephasing errors.} 
\textbf{a}, The evolution of second moment $\mu_2$ of a five-excitation initial state $z_\mathrm{init}=1100100101$ assuming the error-free model (blue) and the model with added phase-flip errors (orange). 
The red (black) dashed line shows the ergodic (classical) limit $\mu_2^\mathrm{e}$ ($\mu_2^\mathrm{c}$) of the second moment and the dotted curve shows the decay of the second moment predicted by the many-body fidelity as $\mu_2 \approx (1+F^2)/D$. 
\textbf{b}, The histogram of bit-string probability distribution $\{p_z\}$ obtained at evolution time $\tau=3\,\upmu$s in panel \textbf{a}. The error-free result and the result with dephasing are shown in blue and orange, respectively. The solid line represents the PT distribution and the dashed line shows the uniform classical distribution $p_z=1/D$.}
\refstepcounter{SIfig}\label{fig:decohered_mu2} 
\end{center}
\end{figure}

In the measurement and data processing of bit-string statistics, we use the same pulse sequence and post-select the sector that conserves the excitation number discussed in Sec.~\ref{sec:Fd_decohere}, which mitigates the effects of population decay.
To illustrate the effect of dephasing on bit-string statistics, we calculate the second moment $\mu_2$ and the histogram of bit-string probability distribution $\{p_z\}$ using data from the numerical simulation of the system Hamiltonian without errors and with added phase-flip errors that generates Fig.~\ref{fig:decoheredFd}.
Here, we find that the second moment of the error-free simulation converges to the ergodic limit $\mu_2^\mathrm{e}$ after the quench while $\mu_2$ obtained from the simulation with phase-flip errors keeps decreasing (see Fig.~\ref{fig:decohered_mu2}a). 
The second moment eventually converges to the classical limit
\begin{equation}
    \mu_2^\mathrm{c} = \sum_z p_z^2 = \sum_z \frac{1}{D^2} = \frac{1}{D}.
\end{equation}
The corresponding histogram at a long evolution time $\tau=3\,\upmu$s, displayed in Fig.~\ref{fig:decohered_mu2}b, follows the PT distribution for the error-free simulation and approaches the classical distribution for the simulation with dephasing.
The above numerical simulations qualitatively reproduce the experimental results shown in Fig.~\ref{fig:FigureN4}a and b of the main text and in Fig.~\ref{fig:ergodic}a.

We also provide a heuristic argument to show that the decay of the second moment $\mu_2$ can be predicted by a quantity $ (1+F^2)/D$ involving the many-body fidelity $F$ under certain error models. This can be understood by utilizing an ansatz 
\begin{equation}
    p_z = F q_z + (1-F) q^\perp_z
\end{equation}
introduced in Refs.~\cite{dalzell2021random, mark2022inprep}, which relates the empirical distribution of $p_z$ in the presence of noise with the ideal PT distribution of $q_z$ and a classical distribution of $q^\perp_z\approx 1/D$, uncorrelated with $q_z$, with a second moment close to $1/D$, i.e., $\sum_z (q^\perp_z)^2 \approx \mu_2^\mathrm{c}=1/D$.
Using this ansatz, the second moment is predicted to be $\mu_2=\sum_z p_z^2 \approx (1+ F^2)/D$, which is supported by our numerical simulations with dephasing errors\footnote{However, we note that this result depends on the specific error model. For example, if the dominant error source is qubit decay from the $\ket{1}$ to $\ket{0}$ state, the term $q^\perp_z$ will favor bit-strings with 0's, converging over time to a single bit-string $00\cdots 0$. This limiting distribution has a large second moment $\sum_z \delta_{z=0\cdots 0}^2 = 1$, hence the overall second moment $\mu_2$ grows over time.} in Fig.~\ref{fig:decohered_mu2}a.

\end{document}